\documentclass[preprint,review]{elsarticle}

\usepackage{amssymb, upgreek, amsmath,fp,color}
\renewcommand{\vec}[1]{{\bf #1}}
\usepackage{url}

\usepackage{graphicx}
\DeclareGraphicsExtensions{.jpg}

\journal{Journal of Computational Physics}
\begin{document}

\begin{frontmatter}

\title{Mixed-Hybrid and Vertex-Discontinuous-Galerkin Finite Element Modeling of Multiphase Compositional Flow on 3D Unstructured Grids\footnote{\copyright ~2016. This manuscript version is made available under the CC-BY-NC-ND 4.0 license 
\url{http://creativecommons.org/licenses/by-nc-nd/4.0/}}}

\author{Joachim Moortgat}
\ead{moortgat.1@osu.edu}
\address{School of Earth Sciences, the Ohio State University,
Columbus, Ohio, USA.}

\author{Abbas Firoozabadi}
\address{Reservoir Engineering Research Institute, Palo Alto, CA 94301, USA.}
\address{Department of Chemical and Environmental Engineering,Yale University, New Haven, CT}

\begin{abstract}
Problems of interest in hydrogeology and hydrocarbon resources involve complex heterogeneous geological formations. Such domains are most accurately represented in reservoir simulations by unstructured computational grids. {\color{black}Finite element 
methods accurately describe flow on unstructured meshes with complex geometries, and their flexible formulation allows implementation on different grid types.}
In this work, we consider for the first time the challenging problem of fully compositional three-phase flow in 3D unstructured grids, discretized by any combination of tetrahedra, prisms, and hexahedra. We employ a mass conserving mixed hybrid finite element (MHFE) method to solve for the pressure and flux fields. The transport equations are approximated with a higher-order {\color{black}vertex}-based discontinuous Galerkin (DG) discretization. We show that this approach outperforms a face-based implementation of the same polynomial order.
These methods are well suited for heterogeneous and fractured reservoirs, because they provide globally continuous pressure and flux fields, while allowing for sharp discontinuities in compositions and saturations. The higher-order accuracy improves the modeling of strongly non-linear flow, such as gravitational and viscous fingering. We review the literature on unstructured reservoir simulation models, and present many examples that consider gravity depletion, water flooding, and gas injection in oil saturated reservoirs. We study convergence rates, mesh sensitivity, and demonstrate the wide applicability of our chosen finite element methods for challenging multiphase flow problems in geometrically complex subsurface media.
\end{abstract}

\begin{keyword}
unstructured 3D grids \sep higher-order \sep compositional \sep compressible  \sep multiphase flow \sep Discontinuous Galerkin \sep Mixed Hybrid Finite Elements  \sep SPE 10 \sep Egg model 

\end{keyword}

\end{frontmatter}

\section{Introduction}
Subsurface geological formations generally have complex geometries that require highly flexible meshing for accurate representation.
Structured (or logically Cartesian) grids may not accurately describe many subsurface problems in hydrogeology and the recovery of hydrocarbon resources. They are also not well suited to model radial flow near wells, and results from commercial simulators may not converge in the near-well region \citep{larry2013}. 

The most commonly used numerical method to model flow on structured grids is the finite difference (FD) approach, while finite volume (FV) methods are usually adopted for unstructured grids. In their lowest-order form, both assume element-wise constant scalar variables (such as saturations) and use a two-point flux approximation (TPFA) to compute vectors (fluxes) from (pressure) gradients between two points.
It is well known that such lowest-order approximations suffer from excessive \textit{numerical dispersion}, and \textit{grid sensitivity}.
The former can be reduced through `brute force' by significantly refining the mesh, which is made more feasible by the development of massively parallelized simulators in the industry \citep{larry2008}. However, sufficient mesh refinement is often not feasible when modeling flow in field-scale hydrocarbon reservoirs or aquifers. 
Grid sensitivity cannot be resolved by mesh refinement. Specifically, it is well known that the TPFA may not converge {\color{black}unless the grid is $K$-orthogonal} \citep{Aavatsmark19962, wu2009}. 
Recently, significant improvements have been made to the FV approach, for instance to accommodate the full permeability tensor \citep{Aavatsmark1998, Edwards2002,Edwards2010, Edwards2011tensor, Edwards2011tensorb, Sadok2013,Sadok2010,Eymard4,Eymard5} and fractures \citep{karimifractures2004,Monteagudo2007,Geiger2013}. To improve FD flux computations on general grids and with tensor permeability, the multipoint flux approximation (MPFA) was introduced. In MPFA, fluxes are reconstructed from the pressures in multiple surrounding elements \citep{AavatsmarkIntro, kozdon2011, Sepehrnoori2013,Nordbotten2007, AavatsmarkMonot,Edwards2002,Younes2013, AavatsmarkConvergence}, similar to the stencil of a standard continuous Galerkin discretization.
Several flavors of MPFA have been proposed since the original version \citep{AavatsmarkConvergence, matringeMPFA, Edwards2011tensor, Edwards2011tensorb,matringeMPFA, Amgad,Eymard2}. MPFA has been compared to the Vertex Approximate Gradient scheme \citep{Eymard1} and to BDM$_{1}$ space under numerical quadrature \citep{matringeMPFA, Amgad}.

The last category of numerical flow models rely on finite elements (FE). 
FE are the method of choice in many disciplines in science and engineering that involve unstructured grids. 
The FE methods that we adopt in this work are motivated by two essential physical properties of flow through porous media: 1) pressures and fluxes are \textit{continuous}, even across layers and fractures, while 2) fluid properties are often \textit{discontinuous} across phase boundaries, fractures, and layers. 

In light of the latter realization, we {\color{black}adopt} the \textit{discontinuous} Galerkin (DG) method for the mass transport update. DG is strictly mass conserving at the element level. In higher-order DG, compositions or saturations can be updated at all vertices or faces and the values can be discontinuous across faces. This is particularly useful in fractured or layered reservoirs. 
In this work, we {\color{black}employ} a multi-linear DG approximation as a compromise between higher-order convergence versus the number of phase-split computations that have to be carried out at each degree-of-freedom.
Many flavors of DG have been analyzed in terms of error estimates and convergence properties, and it is hard to do justice to the full scope of this work (the following papers provide an overview of pioneering and recents efforts in the analysis of DG methods: \citep{peyman,arnold82,babuska73, babuskazlamal73, wheeler, brezzifortini91, cockburnshubook, dawsonsunwheeler, rivierewheeler, giraultshuuy, sunwheeler2005a, sunwheeler2005b, sunwheeler2005c, sunwheeler2006a, sunwheeler2006b, sunwheeler2006c, sunwheeler2007a, Arnoldreview, Persson2008, Nguyen2009, Nguyen2009a, Eymard3, Eymard6, persson2013,Scovazzi2013}). 

We use a mixed-hybrid finite element (MHFE) to satisfy the second aforementioned physical property: that both pressure and flux fields are continuous everywhere. MHFE simultaneously and to the same order {\color{black}of} accuracy solves for globally continuous pressure and flux fields \citep{chaventjaffre, mosemfe, brezzifortini91, hoteitreliability, wheeler1983, wheeler1984}. The high accuracy in the velocity field in highly heterogeneous and/or anisotropic domains is the main attraction of the MHFE method. Computing the pressures on element faces is also convenient when modeling capillarity and fractured reservoirs. Unlike some FE methods, the MHFE-DG combination is strictly mass conserving at the element level. 

A comparison between MFE and MPFA was presented in \citet{matringeMPFA} for single-phase incompressible flow without gravity. \citet{hoteit2005} and \cite{jiri} compared MHFE-DG to the traditional TPFA-FD approach in a commercial simulator, and to an equal-order MUSCL FV scheme, respectively. Both MPFA \citep{Edwards2006} and MHFE \citep{hoteit2008} flux approximations have been presented on unstructured 3D grids for two-phase incompressible flow. However, to the best of our knowledge, neither method has been investigated for unstructured 3D grids and (EOS-based) compositional and compressible multiphase flow with gravity, which is the subject of this work. We emphasize that the latter problem is governed by a different set of equations which involve the highly non-linear total compressibility and total partial molar volumes of multicomponent multiphase mixtures. 

Based on this discussion, we adopt an implicit-pressure-explicit-composition (IMPEC) scheme with a higher-order DG explicit mass transport update and a MHFE implicit pressure and flux update. 
This MHFE-DG scheme was explored for single-phase compressible compositional flow in fractured media in \citet{hoteit2005}, and generalized to two-phase compositional flow in homogenous \citep{hoteit2006a} and fractured domains \citep{hoteit2006b}, all on structured 2D grids; and to two-phase immiscible and incompressible flow with capillarity on 3D unstructured grids in \citep{hoteitcapillary}. More recently, MHFE-DG has been applied to problems of increasing complexity and non-linearity: three-phase flow with an immiscible aqueous phase \citep{moortgatII}, three fully compositional multicomponent hydrocarbon phases or two hydrocarbon phases and a compositional aqueous phase modeled by the cubic-plus-association (CPA) equation-of-state (EOS) \citep{moortgatIII, liCPA}. Fickian diffusion, three-phase capillarity, and discrete fractures were modeled in 3D in \citet{moortgatIV, moortgatV, moortgatVI}.

Our past studies of compositional multiphase flow have been restricted to structured grids. The objective of this work is to unleash the full potential of our FE methods by moving to unstructured grids and allowing for all types of commonly used elements: triangles and quadrilaterals in 2D, and hexahedra, prisms, and tetrahedra in 3D. One other improvement is that we consider {\color{black}vertex}-based DG discretizations rather than face-based (which requires a different slope-limiter \citep{chaventjaffre, hoteitcapillary}). The superiority of this approach is demonstrated in one of the numerical examples.

We briefly summarize our mathematical fractional flow formulation in Section~\ref{sec::mathmodel}{\color{black}. In} Section~\ref{sec::mhfedgimp} we discuss the MHFE-DG implementation on unstructured grids. 
The numerical experiments in Section~\ref{sec::exs} are a main focus of this work. First, we model the recovery of hydrocarbon energy resources by three important processes (gravity depletion, water flooding, and compositional $\mathrm{CO}_{2}$ injection) from a 3D reservoir discretized by $5$ different structured and unstructured hexahedral, prismatic and tetrahedra grids. This example shows that we obtain the same results irrespective of grid-types for a wide range of multiphase processes that exhibit counter-current flow and phase behavior. Other sets of examples investigate grid sensitivity, anisotropic domains, and the convergence properties of the MHFE-DG method on structured and unstructured grids. The last example considers realistic petrophysical properties from the benchmark `SPE~Tenth Comparative Solution Project' \citep{spe10}, and `Egg Model' \citep{eggmodel2, eggmodel}. We end with a brief summary of our findings.

\section{Mathematical Model}\label{sec::mathmodel}
\subsection{Fluid and Formation Description and Notation}
We consider multicomponent multiphase flow through porous media. The porous medium is characterized by an absolute permeability \textit{tensor} $\mathrm{K}$, porosity $\phi$, tortuosity $\tau$, and formation compressibility $C_r$. The fluid mixtures consist of $n_c$ components, labeled by index $i$, and each species can transfer between up to three phases $\alpha = \alpha_1, \alpha_2, \alpha_3$. The overall molar density of species $i$ in the mixture is $c_i = c z_i$ in terms of the overall molar density $c$ and overall molar composition $z_i$. Similarly, $c_{i,\alpha} = c_\alpha x_{i,\alpha}$ is the molar density of species $i$ in phase $\alpha$, and $c_\alpha$ and $x_{i,\alpha}$ are the phase molar density and composition. Molar and mass densities ($\rho_\alpha$) are related through the molar weights $M_i$: $\rho_\alpha = c_\alpha \sum_i M_i x_{i, \alpha} $.
Phase saturations are indicated by $S_\alpha$.

Phase compositions $x_{i,\alpha}$ and molar fractions are found from phase stability analyses and phase split computations, based on a given total composition $z_i$, temperature $T$ and pressure $p$. The phase-splits are based on the Peng-Robinson (PR) EOS \citep{PREOS} for hydrocarbon phases, and a cubic-plus-association (CPA) EOS \citep{liCPA} for mixtures that contain polar component such as water. The basic algorithm is described in Appendix~A.

\subsection{Governing Equations}
In terms of the above definitions of fluid and formation properties, we can write the species transport, or material balance, equations as
\begin{eqnarray}\label{eq::transfer}
\phi \frac{\partial c_i}{\partial t} + \nabla\cdot\vec{U}_i &=& F_i, \quad i = 1, \ldots, n_c,
\end{eqnarray}
in which $F_i$ represent sinks (production wells) or sources (injection wells) of species $i$.
$\vec{U}_i$ are the total molar velocities of each species $i$. $\vec{U}_i$ consists of \textit{convective} Darcy phase velocities $\vec{u}_\alpha$ and \textit{diffusive} (species) phase velocities $\vec{J}_{i, \alpha}$:
\begin{eqnarray}\label{eq::totmolflux}
\vec{U}_i &=& \sum_{\alpha} \left(c_{i,\alpha} \vec{u}_\alpha + f(\phi,\tau) S_\alpha \vec{J}_{i,\alpha}\right),\quad  i = 1, \ldots, n_c.
\end{eqnarray}
The factor $f(\phi,\tau)$ is a modification from open space diffusion to account for the reduction in available flow paths due to rock porosity and tortuosity. In this work, we focus on the MHFE discretization of the Darcy velocity on different types of 2D and 3D unstructured grids and omit diffusion from the equations for clarity of presentation. Fickian diffusion is modeled similar to \citet{moortgatI, moortgatVII}, and is included in the example in Section~\ref{sec::ex6b}. Similarly, we refer to \cite{moortgatV} for a discussion on capillarity and neglect the capillary pressure here in the Darcy relation for the convective phase velocities:
\begin{eqnarray}\label{darcy}
\vec{u}_\alpha &=& - \lambda_{\alpha}\mathrm{K} (\nabla p - \rho_\alpha \vec{g}), \quad \alpha = \alpha_1, \ldots, \alpha_3
\end{eqnarray}
in which $\vec{g}$ is the gravitational vector and $\lambda_{\alpha} = \lambda_\alpha (S_\alpha)$ is the phase mobility. 

We use an explicit equation for the pressure field \citep{acs, watts}:
\begin{eqnarray}\label{eq::acs}
\zeta \frac{\partial p}{\partial t} +  \sum_{i=1}^{n_c} \bar{v}_i (\nabla\cdot \vec{U}_i - F_i) &=& 0,
\end{eqnarray}
where $\zeta = \phi (C_r + C_f)$, and $C_f$ and $\bar{v}_i$ are, respectively, the total compressibility and total partial molar volumes of the three-phase fluid mixture (derived in \citep{moortgatIII}). 
We adopt a fractional flow formulation in terms of the \textit{total} velocity $\vec{u}_t$:
\begin{eqnarray}\label{eq::darcytot}
\vec{u}_t &=& - \lambda_t \mathrm{K} \left( \nabla p - \rho_t \vec{g}\right),
\end{eqnarray}
in which $\lambda_t = \sum_\alpha \lambda_\alpha$ and $\rho_t = \sum_{\alpha}\rho_{\alpha} f_\alpha$ with $f_\alpha =  \lambda_\alpha/\lambda_t$ the fractional flow functions. The main advantages of the fractional flow formulation are that \eqref{eq::darcytot} can be readily inverted in favor of the pressure ($\lambda_t > 0$, while $\lambda_\alpha \ge 0$), and that we only solve directly for one velocity field. More specifically, in the next section we describe our MHFE method to simultaneously solve \eqref{eq::darcytot} for the total velocity and \eqref{eq::acs} for the pressure. 

The {phase} velocities are reconstructed from $\vec{u}_{t}$ by
\begin{eqnarray}\label{eq::vafromvt}
 \vec{u}_{\alpha} &=& f_{\alpha} \left(\vec{u}_{t} + \vec{G}_\alpha \right),\\\label{eq::ttt}
\vec{G}_\alpha &=& \sum_{\alpha^\prime} \lambda_{\alpha^\prime}\mathrm{K} (\rho_{\alpha} - \rho_{\alpha^\prime}) \vec{g},
 \end{eqnarray}
 in which the prime notation denotes independent phase indices $\alpha$ and $\alpha^{\prime}$.
  Gravity (and capillarity) can drive counter-current flow, which causes complications in finding the upwind values of $\lambda_{\alpha^\prime}$, which were resolved in earlier work \citep{moortgatIII}.

\section{MHFE-DG on 3D Unstructured Grids}\label{sec::mhfedgimp}
In this section we discuss the IMPEC discretization of our mathematical model through higher-order finite element methods. First, we provide the FE basis (vector) functions for scalar and vector quantities (Section~\ref{sec::bases}) and then we summarize the MHFE discretization of the velocity \eqref{eq::darcytot} and pressure \eqref{eq::acs} equations (Section~\ref{sec::mhfe}) and species transport \eqref{eq::transfer} (Section~\ref{sec::dg}). Details have been presented in our earlier work \citep{moortgatVI}. In these sections we mainly emphasize the coefficient matrices that depend on mesh geometry. {Appendix~B provides the complete expressions for the DG discretization on all grid types.}

 The domain is discretized by elements $K$ with volume $|K|$, which have faces $E$ with surface area $|E|$, such that the boundary of $K$ is $\partial K = \sum_E |E| $. We label the {(local)} vertices {inside} each element by $N$, and denote the total number of elements, faces, and nodes by $N_K$, $N_E$, and $N_N$, respectively. The coordinate vector is defined as $\vec{x} = (x, y, z)^T$.

\subsection{Finite Element Bases}\label{sec::bases}
In FE discretizations, scalars and vectors are decomposed in terms of appropriately defined basis functions $\varphi (\vec{x})$ and basis vector fields $\vec{w} (\vec{x})$, respectively. Differential equations are put in the weak form by integrating over each grid cell $K$. The number and choice of basis (vector) functions depend on the number of degrees-of-freedom {(DOF) associated with the order of the method. In our MHFE-DG approach, the scalar state variables are evaluated at the \textit{vertices} (nodes) by a multi-linear DG approximation, while the velocity vector field is evaluated across element faces in the MHFE discretization. The corresponding DOF are illustrated in Figures~\ref{fig::doftrian}-\ref{fig::dofcube}.

Note that while for a \textit{continuous} higher-order method, such as continuous Galerkin, a single value is updated for each {\color{black}vertex} in the domain, for our \textit{discontinuous} Galerkin method the DOF are the nodes \textit{inside} each element and properties can be discontinuous across edges. This means that for hexahedra, for instance, eight values (e.g.~for compositions) are updated at each {\color{black}vertex}, one for each of the eight elements sharing that {\color{black}vertex}. Physically, this is a desirable feature when discontinuities occur on grid faces, e.g. at layer or fracture-matrix interfaces, while mathematically this gives one the freedom to use different orders of approximations in each grid cell (because the interpolations do not have to conform at the grid boundaries). Pressures and fluxes, on the other hand, are required to be continuous across edges to satisfy the physical constraints of the problem.}

{We expand $c_i (t, \vec{x})$ (and $c_{i, \alpha} (t, \vec{x})$) and $\vec{u}_t (t, \vec{x})$ (and $\vec{u}_\alpha (t, \vec{x})$ and $\vec{g} (\vec{x})$) as:}
\begin{eqnarray}\label{eq::expansionscalar}
c_i (t, \vec{x}) &=& \sum_N (c_{i})_{N} (t) \varphi_N (\vec{x}),\\\label{eq::expansionvector}
\vec{u}_t (t, \vec{x}) &=& \sum_E q_{E} (t) \vec{w}_E (\vec{x}),
 \end{eqnarray}
 where the coefficients $(c_{i})_{N}$ are the species molar densities at the vertices $N$, and $q_{E}$ is the total flux across faces $E$.

 \subsection{MHFE Pressures and Fluxes}\label{sec::mhfe}
 To find the MHFE discretization of the Darcy law for the total velocity, we expand $\vec{u}_t$ as in \eqref{eq::expansionvector}, multiply \eqref{eq::darcytot} by a test function, which we choose to be the same as the basis vector fields $\vec{w}_E$, and integrate over elements $K$. The result is
 \begin{equation}\label{eq::qmatrix}
q_{K, E} = \theta_{K, E} p_K - \sum_{E^\prime \in\partial K} \beta_{K, E, E^\prime} tp_{K, E^\prime} - \gamma_{K, E}, \quad E\in \partial K,
\end{equation}
in which $p_{K} = \int_{K} p$ and $tp_{K,E} = \int_{E\in\partial K} p$ follow from partial integration of the divergence term by Gauss' theorem, and the coefficients are defined (with $\mathrm{d} \mathbf{x}$ the volume increment) as:
\begin{eqnarray}\label{eq::coeffsqke1}
\beta_{K, E, E^\prime} &=&  \lambda_{t,K}\left[\int_{K} \vec{w}_{K,E} \mathrm{K}_{K}^{-1} \vec{w}_{K,E^{\prime}}{\color{black}\ \mathrm{d} \mathbf{x} }\right]^{-1},\\\label{eq::coeffsqke2}
\theta_{K, E} &=& \sum_{E^\prime}  \beta_{K, E, E^\prime},\\\label{eq::coeffsqke3}
 \gamma_{K, E} &=&  
  -   \mathrm{\lambda}_{t, K} \rho_{t,K} \mathrm{K} (\vec{g}\cdot \vec{n}_E) |E|.
\end{eqnarray}
Detailed derivations can be found in \cite{moortgatVI}. The MHFE discretization of the pressure equation proceeds along the same lines. We multiply \eqref{eq::acs} by $\vec{w}_E$, expand the phase velocities in $\vec{U}_{i}$ in terms of the total velocity using \eqref{eq::vafromvt}, integrate over each element $K$, and adopt a backward Euler discretization of the time derivative (indexed by superscript $n$). The final expression for the pressure evolution is:
 \begin{eqnarray}\nonumber
&&p_K^{n+1} =\frac{\Delta t}{\tilde{\alpha}_K\Delta t + \zeta_K} \times \\\label{eq::pfinal}
&& \times\left\{\frac{\zeta_K}{\Delta t}p^n_K + \!\sum_{E\in \partial K}\tilde{\beta}_{K, E}  tp_{K, E}^{n+1} +  \tilde{\gamma}_K + 
\! \sum_i \bar{\nu}_{i,K}  F_{i,K}  \right\}
\end{eqnarray}
in which $\tilde{\alpha}_K$, $\tilde{\beta}_{K, E}$ and $\tilde{\gamma}_K$ are essentially multiplications of \eqref{eq::coeffsqke1}--(\ref{eq::coeffsqke3}) by the partial molar volumes and phase mobilities and densities, without additional dependencies on the geometry. {In our hybridized MHFE approach, a Schur decomposition is performed that leaves the pressure traces on faces as primary variables. {\color{black}The sparsity pattern involves for each row (corresponding to a face) all the faces of the neighboring two elements}. Cell-centered pressures and fluxes are obtained by computationally inexpensive back-substitution.}
Details of the derivations are provided in \cite{moortgatVI,shahraeeni}.

The purpose of this brief reiteration of our MHFE method is to demonstrate the elegance with which the method can be generalized from Cartesian grids to any type of unstructured partitions. The only difference between different grid types is in the integral over permeability in \eqref{eq::coeffsqke1}, and the normal components of gravity with respect to element faces in \eqref{eq::coeffsqke3}. Both these terms are time-independent and only have to be evaluated once in the initialization of a simulation on a particular grid. Once the geometrical coefficients in \eqref{eq::coeffsqke1} and (\ref{eq::coeffsqke3}) are given, the MHFE computation of the pressures and fluxes at each time-step is identical for all grid types.  

\subsection{DG Species Transport}\label{sec::dg}
The DG discretization of the transport equation is similar: \eqref{eq::transfer} is put into weak form by multiplying by test functions $\varphi_N$ and integrating by parts over each element $K$. {The time derivative is discretized by a forward Euler method. Using standard calculus, the derivatives over the unknown phase properties are changed into derivatives of the known basis functions, which results in the following discretized transport equation {\color{black}(with $\mathrm{d} \mathbf{s}$ the surface increment)}: 
\begin{eqnarray}\nonumber 
&&  \sum_{N} \frac{(c_i)^{n+1}_{K,N}-(c_i)^{n}_{K,N}}{\Delta t/\phi_K}
 \int_K  \varphi_{N} \varphi_{N^\prime}\ \mathrm{d} \mathbf{x} =  \sum_{N}\sum_\alpha \sum_{ E} q_{\alpha, K, E}\times \\\nonumber
&& \left[  (c_{i,\alpha})^n_{K, N} \int_K   \varphi_{N} \vec{w}_E \cdot \nabla  \varphi_{N^\prime}{\color{black}\ \mathrm{d} \mathbf{x}} - \frac{(\widetilde{c_{i,\alpha}})^n_{K, E, N}}{|E|} \int_{E \in \partial K}  \varphi_{ N} \varphi_{N^\prime}{\color{black}\ \mathrm{d} \mathbf{s}}
 \right]\\\label{eq::DGdiscr}
&& + (F_i)_K^n \int_K  \varphi_{N^\prime}{\color{black} \ \mathrm{d} \mathbf{x}}
\end{eqnarray}
in which $(\widetilde{c_{i,\alpha}})_{K, E, N}$ are the \textit{upstream} values of $c_{i,\alpha}$ at all the vertices of face $E$ with respect to the phase flux $q_{\alpha, K, E}$ through $E$ {\color{black}(this defines the numerical flux in our DG discretization and guarantees its continuity)}. 
Appendix~B provides the explicit algebraic expressions from working out \eqref{eq::DGdiscr} for all grid types considered in this work, i.e.~triangles, rectangles, tetrahedra, prisms, and hexahedra. These expressions can be readily implemented in any reservoir simulator and should facilitate the popularization of the Discontinuous Galerkin mass transport update for non-trivial unstructured 3D grids.}

An important advantage of this \textit{{\color{black}vertex}}-based DG discretization is that up to $4$ {\color{black}vertex} values are upwinded for each face $E$, rather than just one value in a face-centered DG approach. This means that not only the magnitude of compositions and densities are communicated between elements, but also the gradients in those variables along faces. This further reduces numerical dispersion and grid sensitivity.

Similar to our discussion of the MHFE method on unstructured grids, we only summarize the DG-discretized mass transport equation to highlight the ease with which complex geometries can be accommodated. The only grid dependencies are captured in the matrices {\color{black}$\int_K \varphi_N \varphi_N^\prime \mathrm{d}\mathbf{x}$, $\int_K   \varphi_{N} \vec{w}_E \cdot \nabla  \varphi_{N^\prime}\mathrm{d}\mathbf{x}$, and $\int_E  \varphi_{N} \varphi_{N^\prime}\mathrm{d}\mathbf{s}$}, which are again time-independent and can be computed as part of an initialization routine once the grid is specified. 

We note that while Equation~\ref{eq::DGdiscr} is expressed as a first-order time-discretization, we have also implemented a second-order Runge-Kutta scheme. However, it appears that the dependence of numerical dispersion on the time-step size is weak, partly due to the small time-steps enforced by the CFL condition in our IMPEC scheme {\color{black}\citep{moortgatVI}}.

\subsection{Slope Limiter}
Higher-order methods generally require a slope limiter to avoid spurious oscillations around sharp front, and our DG implementation is no exception. We adopt the limiter developed by \citep{chaventjaffre, hoteitcapillary}. Consider {\color{black}vertex values} $(c_{i})_{K,N}$ for an element $K$ with local node number $N$. In the full grid, this node is shared with multiple neighboring element, e.g.~eight for a hexahedral grid and often even more in fully unstructured tetrahedral grids. Denote $\vec{\bar{c}}$ as the vector of averages
\begin{eqnarray}
&&[\vec{\bar{c}}]_{K^{\prime}} =  \frac{1}{|K^{\prime}| }\int_{K^{\prime}}c_{i} =   \frac{1}{N^{\prime} }\sum_{N^{\prime}}(c_{i})_{K^{\prime},N^{\prime}}
\end{eqnarray}
in all the elements $K^{\prime}$ sharing node $N$.

The basic idea is to limit $(c_{i})_{K,N}$ to lie between the minimum and maximum of the average state variables in all the elements surrounding a given node. This is done by minimizing the $L_{2}$ norm of the limited, or reconstructed, $(c_{i}^{l})_{K,N}$ with respect to the original {\color{black}vertex} values $(c_{i})_{K,N}$ under the physical constraints of mass conservation
\begin{eqnarray}
&& \min_{(c_{i}^{l})_{K,N}} || (c_{i}^{l})_{K,N}- (c_{i})_{K,N} ||_{2},\quad \mbox{satisfying}\\
&&\min(\vec{\bar{c}})<(c_{i}^{l})_{K,N} < \max(\vec{\bar{c}}),\quad \mbox{and}\\
&& \frac{1}{N}\sum_{N}(c^{l}_{i})_{K,N} = \frac{1}{N}\sum_{N}(c_{i})_{K,N}.
\end{eqnarray}
The mesh dependence of this slope limiter lies only in keeping track of which elements share which (global) vertex, and the (possibly) different numbers of nodes in each type of grid cell.

{\color{black}It has been shown that this slope limiter does not enforce a maximum principle on the \textit{face-averaged} values of state variables \citep{hoteitslope}, which could theoretically cause spurious oscillations, e.g.~if phase fluxes are computed with face-centered fractional flow functions. Two-phase immiscible flow is particularly sensitive because mobilities depend directly on the primary saturation variable. Nevertheless, this slope limiter was applied successfully in MHFE-DG modeling of two-phase immiscible flow on unstructured 3D grids by \citet{hoteitcapillary}.
In compositional flow, species molar densities are the primary variables that are updated at vertices by DG for given phase fluxes. After slope limiting, molar densities and overall compositions ($0<z_{i}<1$) are computed and used to update the phase split (flash) computations. Phase saturations (and mobilities) are derived quantities, with $0<S_{\alpha}<1$ guaranteed by the flash algorithms. In two-phase flow, face-centered mobilities can be computed from face-averaged saturations (which are assumed to vary linearly in each dimension), but in compositional flow, saturations are a strongly non-linear function of compositions. Phase properties (compositions and saturations) cannot be extrapolated away from the points where flash calculations were performed.  Instead, motivated by the first-order accuracy of the MHFE flux update, we perform a cell-centered flash computation to update saturations and mobilities. With this approach, we avoid any spurious oscillations in our solution. In future work, we may explore methods to further improve accuracy by coupling DG to a higher-order flux update (e.g., BDM$_{1}$) and developing a slope limiter with a formal maximum principle.}

\subsection{Convergence}\label{sec::norms}
Formally, the order of convergence for our multi-linear DG approximation should be quadratic. However, these formal rates are derived for highly simplified problems, and observed rates are generally lower for highly non-linear problems. Nevertheless, the convergence rate scales with the order of approximation. In \citep{moortgatIII}, we found that the convergence rate of a 2D face-based bilinear DG mass transport update was twice that of an element-wise constant (FV) approximation. The order of convergence is particularly important for IMPEC schemes in 3D. The CLF constraint on the time-steps scales as $1/h$ (where $h$ is a characteristic length-scale of the elements, i.e.~$\Delta x$, $\Delta y$ or $\Delta z$ on structured grids, or, say, $V^{1/3}$ for an unstructured element with volume $V$). Therefore, {\color{black}for explicit methods} the increase in CPU time for a mesh refinement $\delta f = h_\mathrm{coarse}/h_\mathrm{fine}$ in each direction scales with at best $\delta f^4$! Our DG approximation on tetrahedra requires the update of $4$ $c_i$ at {\color{black}vertices} instead of an element-wise constant value, but the associated additional CPU cost is negligible compared to the $\propto \delta f^4$ cost of mesh refinement to reduce numerical dispersion. 

{\color{black}Note that, when flow instabilities (gravitational or viscous fingering) or strong heterogeneity occur at small spatial scales, such features cannot be resolved on coarse grids and mesh refinement may be unavoidable \citep{moortgatFinger}. Up- and down-scaling static reservoir models for permeability and porosity heterogeneity on unstructured grids adds further complexity, but geostatistical facies modeling can be applied to unstructured grids as well as to structured ones.
When a formation is layered or fractured, grids should conform to the layer and matrix block sizes, but within a layer or matrix block, less mesh refinement is required, e.g., to resolve flow instabilities, when higher-order methods are used. Moreover, the \textit{discontinuous} Galerkin method preserves sharp changes in fluid properties at layer and fracture-matrix interfaces.}

We also note that it is not straightforward to compare the CPU times and degree of mesh refinement for the different types of elements: time-steps are governed by the element volumes and will be similar for any grid types, provided they have the same number of elements. However, the computational cost of the DG transport update scales with the number of nodes per element. 
Finally, the MHFE pressure update is proportional to the number of faces, which again scales differently for each type of element. To summarize: in order to make a `fair' comparison between simulations on different grids we have to balance 1) the number of elements, which governs the time-steps, 2) the number of nodes, which determines the cost of the DG update and associated phase-split computations, and 3) the number of faces, which determines the size of the matrix that need to be inverted in the implicit pressure solve.

\section{Numerical Experiments}\label{sec::exs}
We present the results of nearly 50 simulations that explore key challenges in the modeling of subsurface flow.
Grid independence is the first such challenge. In certain numerical models (e.g.~TPFA), simulated results are sensitive to the orientation and quality of the mesh. We claim that MHFE shows virtually no mesh dependence. We demonstrate this in the first set of examples by simulating gravity depletion, water flooding, and $\mathrm{CO}_{2}$ injection on structured and unstructured tetrahedral, prismatic, and hexahedral grids. The fluid characterization, rock properties, temperature and pressure conditions are representative of a North Sea light-oil reservoir.
 The second example further explores gridding by verifying that in the absence of gravity the expansion front from a point-source in a homogeneous domain is spherical. For unstable flow, which exhibits viscous fingering, we demonstrate the advantage of our {\color{black}vertex}-based DG formulation over a face-based approach.
Anisotropic formations with a full permeability tensor are considered in the third example. The convergence rate upon grid refinement of our MHFE-DG method on unstructured grids is analyzed in the fourth example. In that example, we compare the CPU times for a range of grid refinements and different grid types. In the other examples, the bulk of the CPU time may be consumed by phase-split computations, and comparisons between different grid types is less informative of the efficiency of the FE methods. In the fifth set of examples, we demonstrate the flexibility of our methods in dealing with complicated domain geometries and mixing different elements types in a single mesh. 

The first five examples are not necessarily representative of realistic reservoir geometries. Rather, each example is constructed to demonstrate a \textit{specific} powerful feature of our finite element approach. In the sixth example, we consider two benchmark reservoirs from the literature: Model 2 of the `SPE~Tenth Comparative Solution Project' \citep{spe10}, and the `Egg Model' developed at the Technical University of Delft \citep{eggmodel2, eggmodel}. These two models use logically Cartesian grids with low geometrical complexity, but they incorporate realistic petrophysical properties with many orders of magnitude variations in permeabilities and porosities. 

In all the examples we determined that the maximum mass balance error of a species for the entire simulations is of the order $\sim 10^{-13}$. All simulations were carried out in serial mode on a 2.8 GHz Intel Core i7 with 12 GB RAM. 

\subsection{Example~1: Depletion, Water Flooding, and $\mathrm{CO}_{2}$ Injection}
We consider a $0.3\ \mathrm{km}\times 0.1\ \mathrm{km}\times 0.05\ \mathrm{km}$ section of a North Sea reservoir, discretized by structured hexahedra, unstructured prisms, and various tetrahedralizations. We consider the 7-(pseudo)-component fluid characterization and formation properties (homogeneous rock permeability of $100\ \mathrm{mD}$ and 20\% porosity) from Example~2 in \citet{moortgatVI}. The temperature is $400\ \mathrm{K}$, and the initial bottom-hole pressure of $337.5$ bar is just below the saturation pressure ($338$ bar) of the oil. We assume Stone \citep{stone1, stone2} relative permeabilities, with quadratic exponents and end-points of $k^{0}_{rg} = k^{0}_{ro} =0.4$ for gas and oil. The aqueous phase has an end-point relative permeability of $0.3$ and exponent of $3$. The residual oil saturations are $50\%$ to water and $0\%$ to gas (due to phase behavior).

We test our unstructured higher-order FE methods for three important oil recovery processes: gravity depletion, water flooding, and compositional $\mathrm{CO}_{2}$ gas injection. 

To investigate whether our method suffers from grid dependency, we consider $5$ different grids. The first is a structured mesh with $50\times18\times10 = 9,000$ elements, while the other $4$ are unstructured. The second grid has $10$ layers of prisms, the third (tetrahedra 1) consists of $9,000$ equal size tetrahedra constructed from a $1,500$ element structured grid (5 layers) by dividing each cuboid into 6 tetrahedra, the fourth grid (tetrahedra~2) is an irregular CDT with $10,467$ elements of widely varying sizes, and the fifth (tetrahedra~3) is a $36,000$ element refinement (10 layers) of the tetrahedra~1 grid. The number of elements, vertices and faces for each grid are summarized in Table~\ref{table::NNNMNF}. Wells are placed at the origin (bottom) and the diagonally opposite corner (top) with each well's function (injection or production) determined by the production scenario.
For the depletion case, oil is produced at a constant rate of $6\%$ pore volume (PV) per year from the bottom well, water flooding and gas injection are at a constant rate of $5\%$ PV/yr from the bottom well with constant production pressure from the top. 

\subsubsection{Depletion}
\begin{figure}
 \centering
\includegraphics[width=\textwidth]{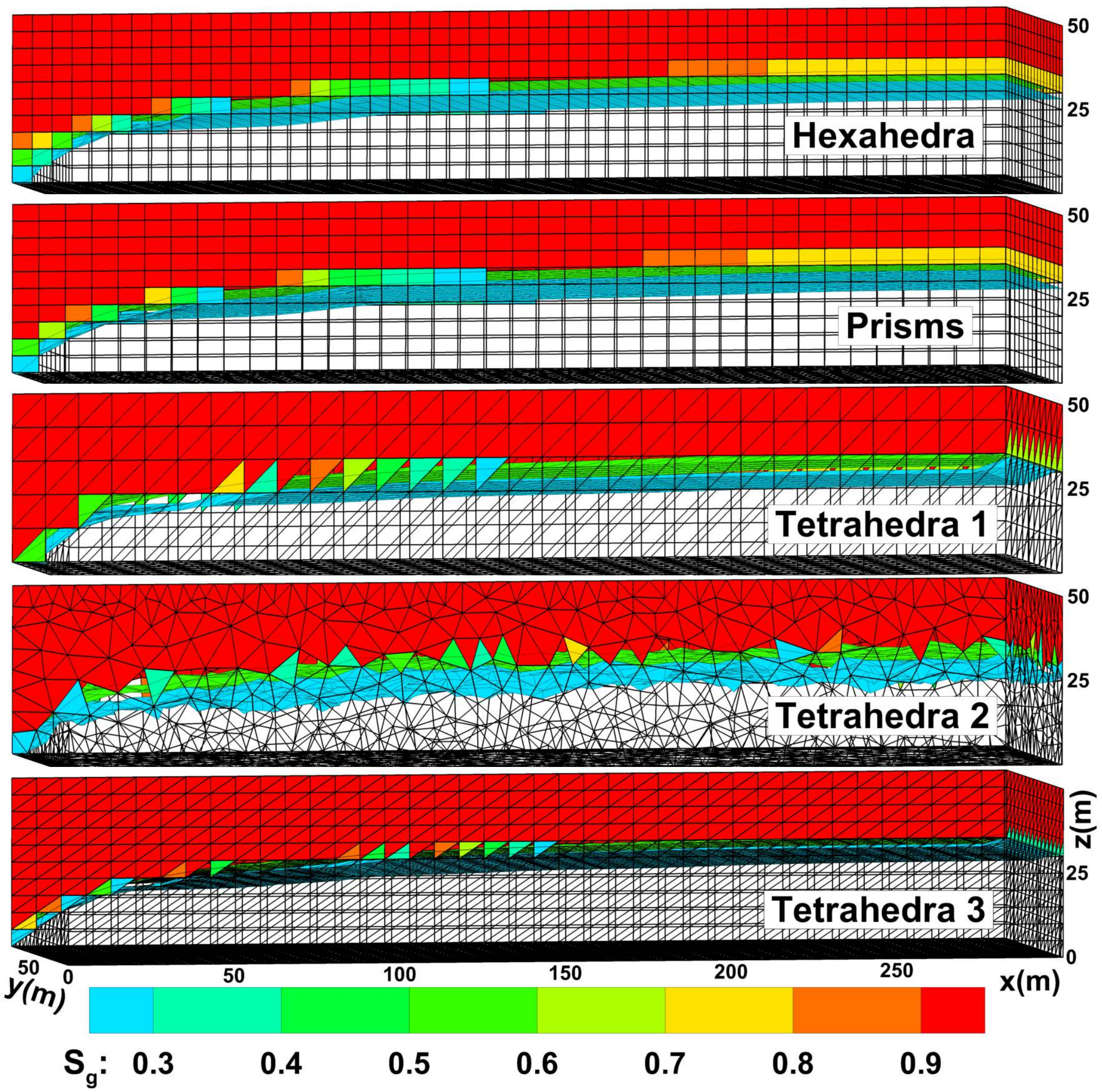} 
\caption{Example 1: Gas saturation after $5$ years of depletion, simulated on $5$ different grids.}
 \label{fig::deplex1}
 \end{figure}

\begin{figure}
\centering
 \includegraphics[width=0.5\textwidth]{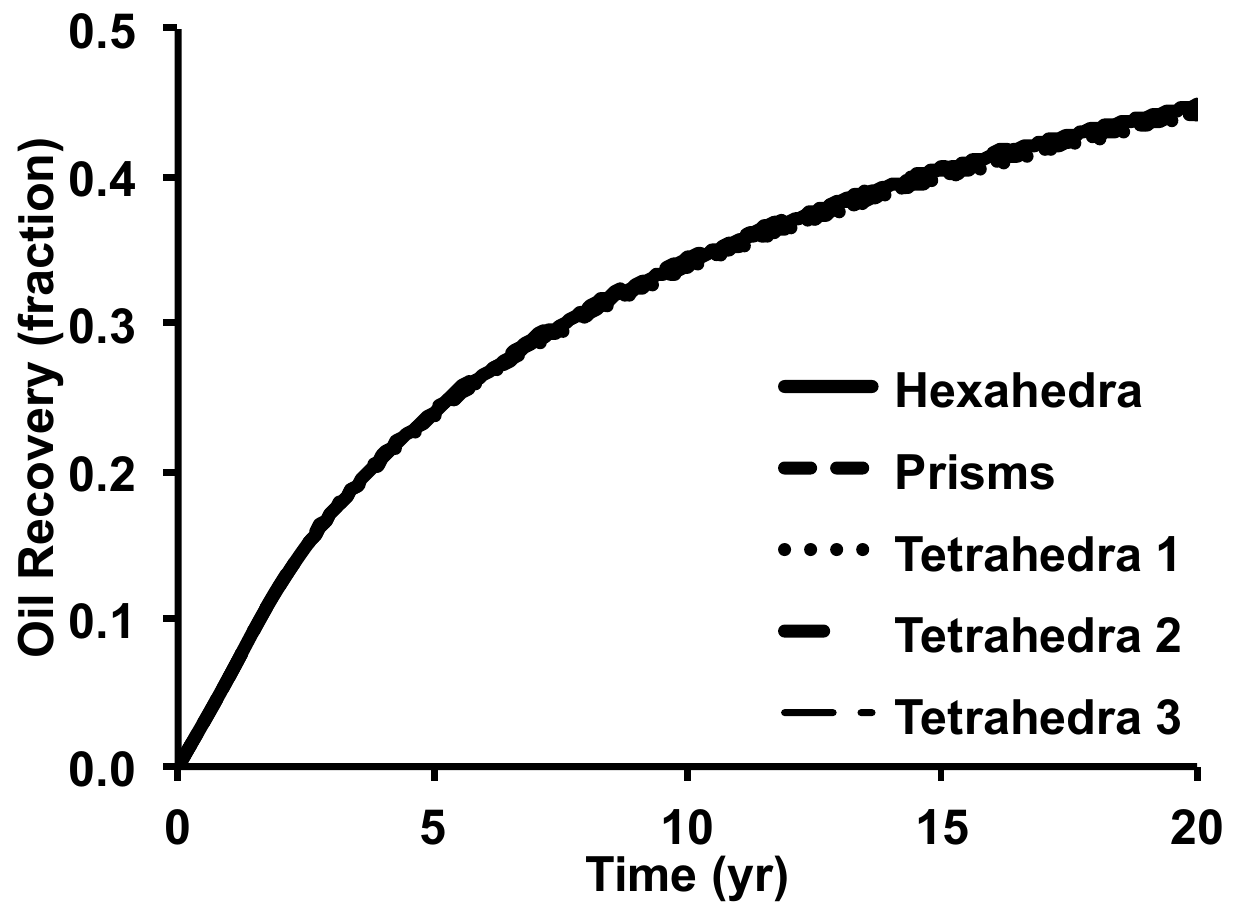}
\caption{Example 1: Oil recovery from $20$ years of gravity depletion, computed on $5$ different structured and unstructured grids.}
 \label{fig::deplex1recovery}
 \end{figure}

Figure~\ref{fig::deplex1} shows the gas saturation throughout the domain on all 5 grids after 5 years of gravity depletion. The results are in good agreement on all types of grids. The complete lack of mesh dependence for these simulations is even more apparent in Figure~\ref{fig::deplex1recovery}, which shows the oil recovery over a simulation time of 20 years. The results are indistinguishable. Depletion is a somewhat challenging test-case for multiphase flow simulations, because the flow exhibits gravitational counter-current flow with gas buoyantly rising to the top while oil drains to the bottom. 

\subsubsection{Water flooding}
\begin{figure}
 \includegraphics[width=\textwidth]{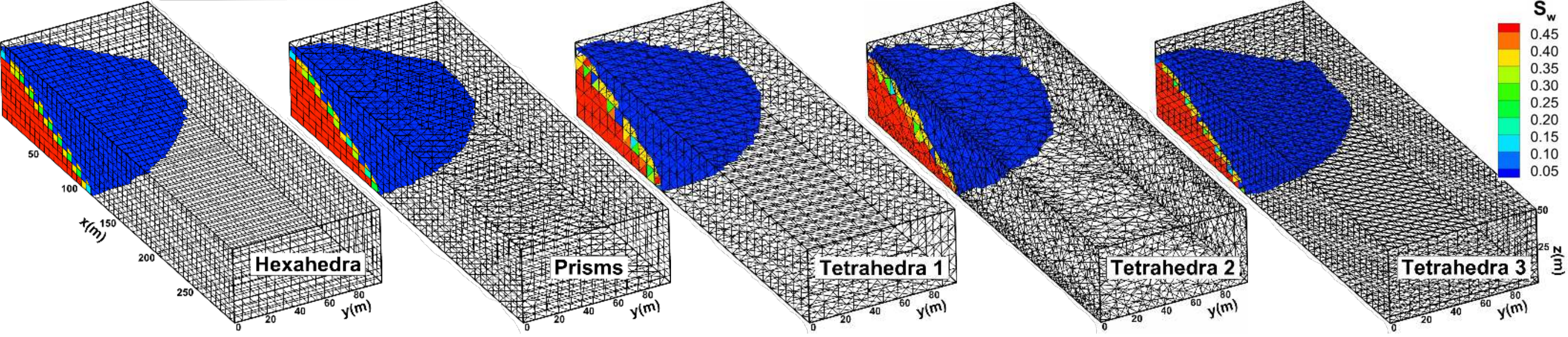}
\caption{Example 1: Water saturation after $1$ year of flooding, simulated on $5$ different grids.}
 \label{fig::H2Oex1}
 \end{figure}

\begin{figure}
 \centering
 \includegraphics[width=0.5\textwidth]{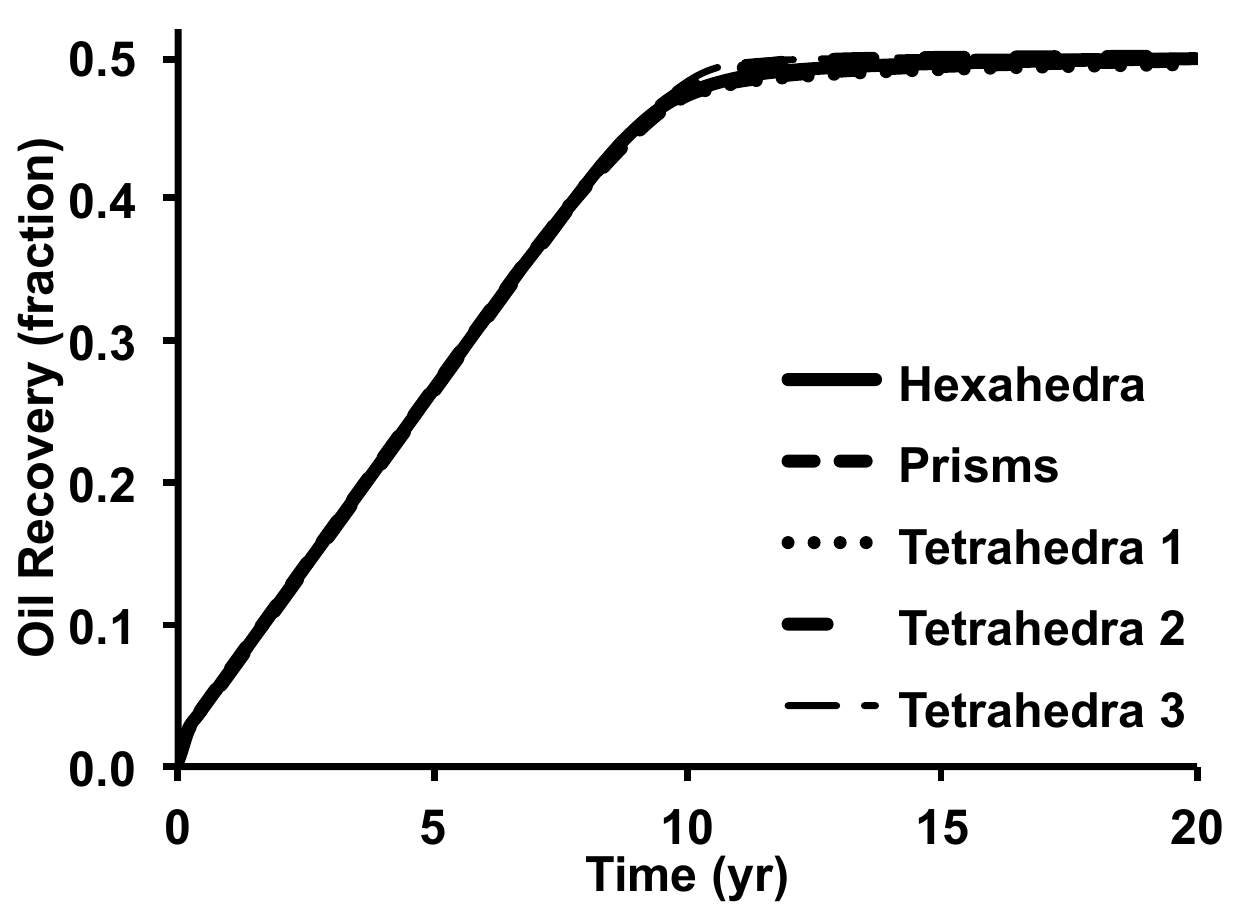}
\caption{Example 1: Oil recovery from $20$ years of water flooding, computed on $5$ different structured and unstructured grids.}
 \label{fig::H2Oex1recovery}
 \end{figure}

The water saturation after 1 year of $5\%$ PV/yr injection is shown in Figure~\ref{fig::H2Oex1} on all 5 grids. Due to the $50\%$ residual oil saturation, water has invaded about $10\%$ of the domain. Again, we find very good agreement in the front locations on all meshes and nearly identical oil recoveries over 20 year in Figure~\ref{fig::H2Oex1recovery}. We do see a slightly sharper kink in the oil recovery around the time of water breakthrough for the finest tetrahedral grid, suggesting that the $\sim 9,000$ element grids still have a small degree of numerical dispersion. 

\subsubsection{$\mathrm{CO}_{2}$ Injection} 
\begin{figure}
\centering
 \includegraphics[width=\textwidth]{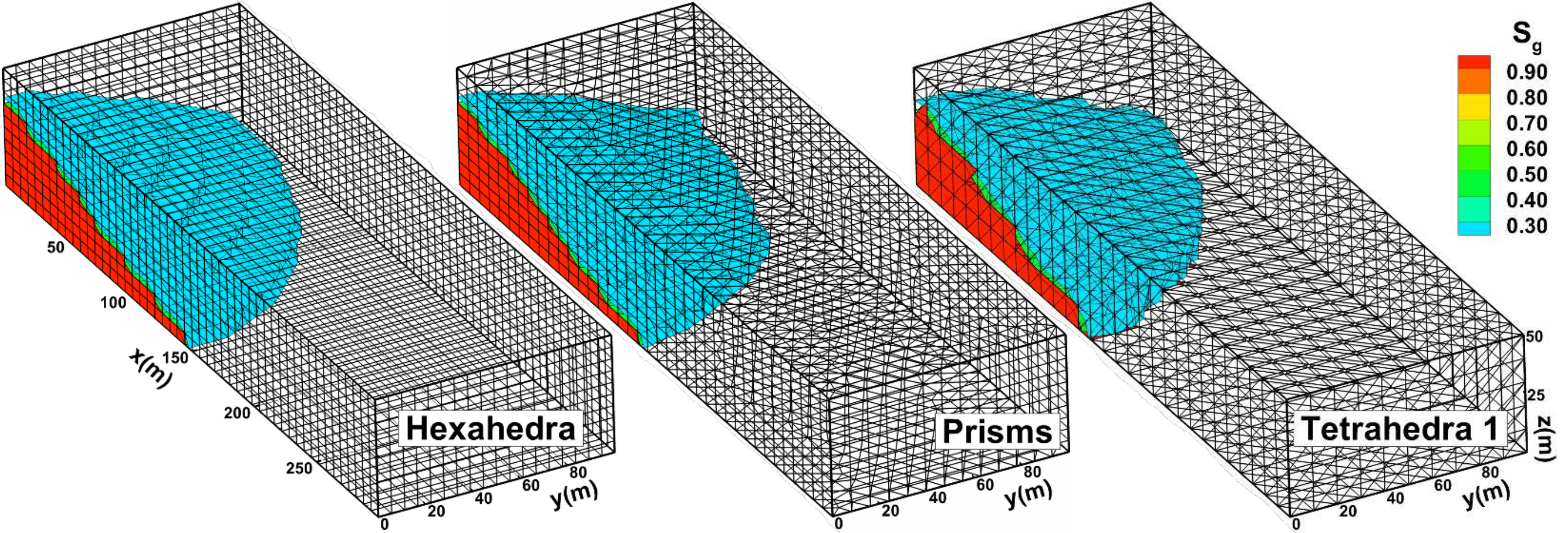}
  \includegraphics[width=\textwidth]{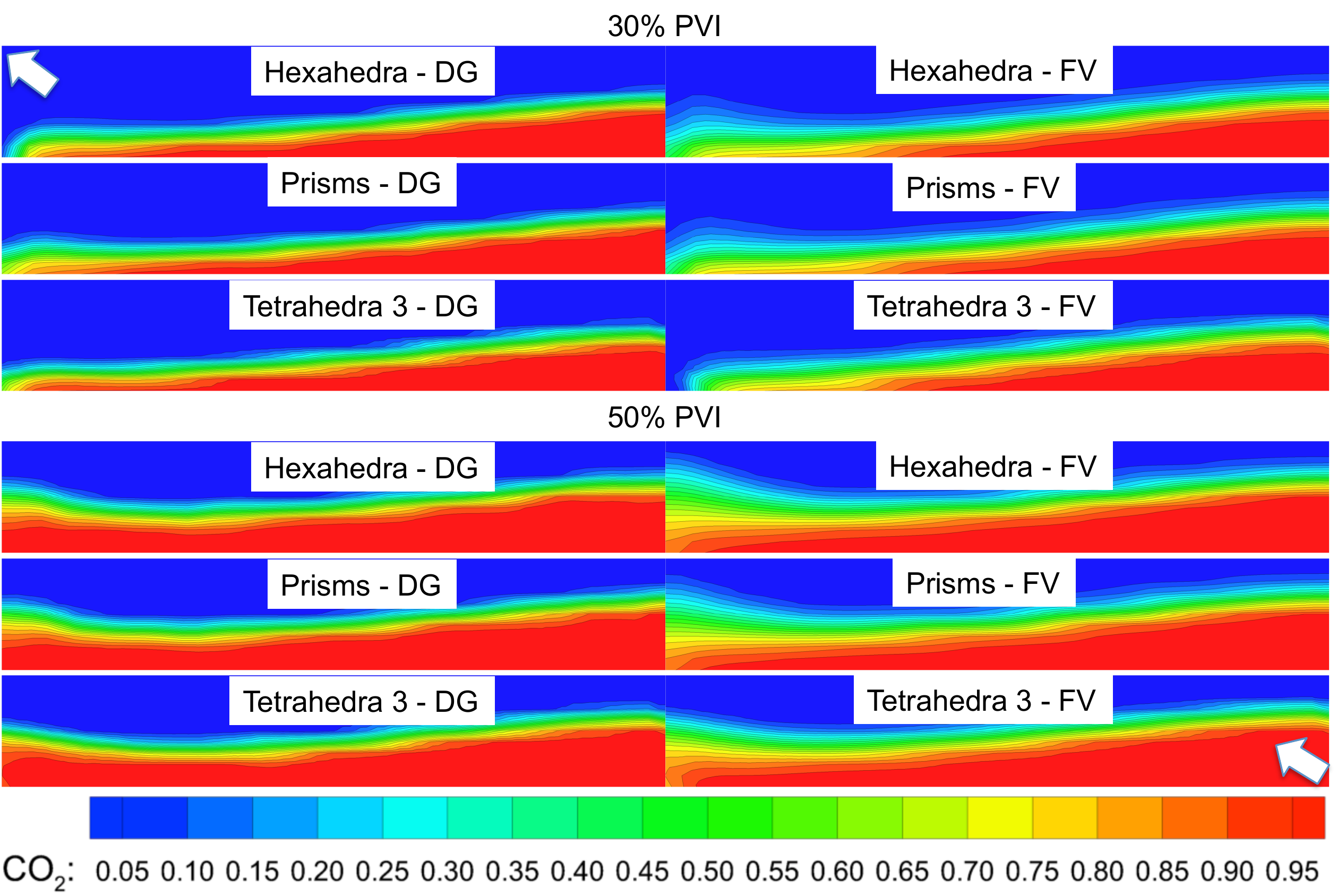}
\caption{Example 1: Gas saturation after 10\% PV of $\mathrm{CO}_{2}$ injection from the bottom, MHFE-DG simulations on $3$ different grids. Overall CO$_{2}$ composition is also shown at 30\% and 50\% PVI on a cross-section at $y=70$ m for both MHFE-DG and lower order MHFE-FV simulations, with arrows indicating the (projected) injection (top) and production (bottom) well locations.}
 \label{fig::CO2bottomex1}
 \end{figure}

\begin{figure}
 \centering
\includegraphics[width=.7\textwidth]{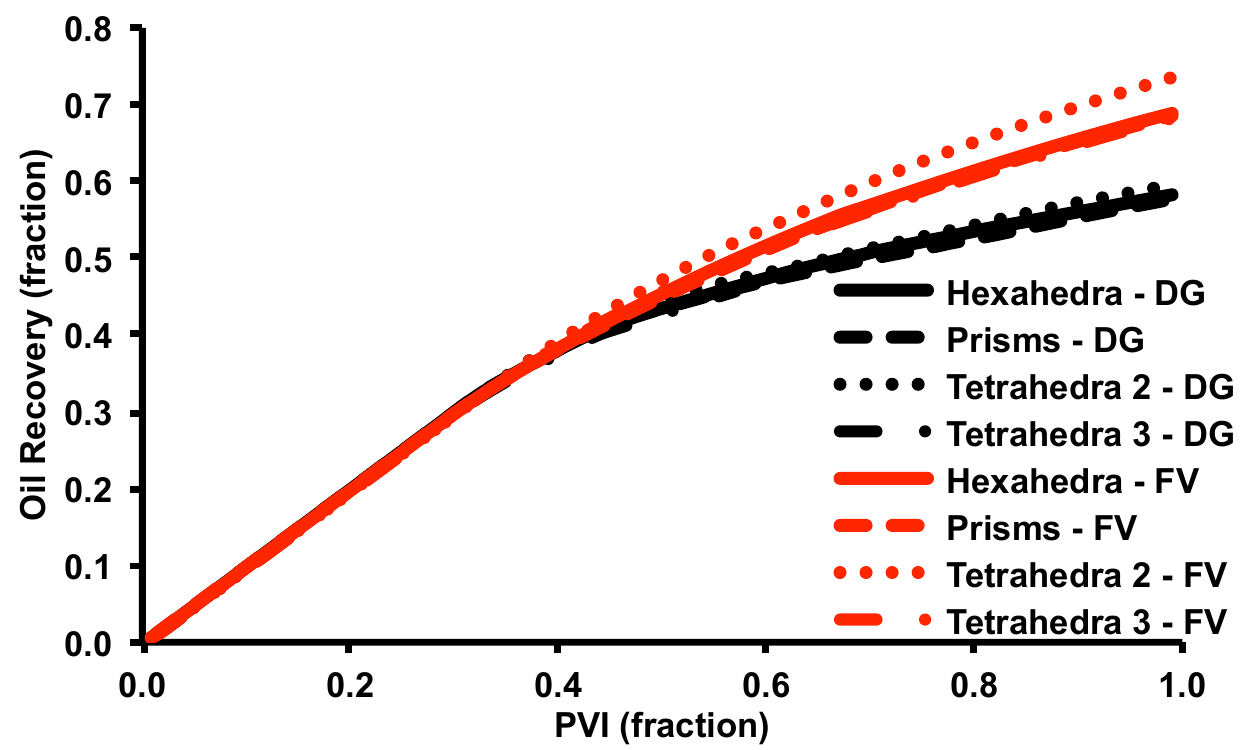}
\caption{Example I: Oil recovery from $20$ years, or 1 PV, of $\mathrm{CO}_{2}$ injection from the bottom, computed on $4$ different structured and unstructured grids with MHFE-DG and MHFE-FV.}
 \label{fig::CO2ex1recovery}
 \end{figure}

{\color{black}The injected $\mathrm{CO}_{2}$ has a higher density ($614\ \mathrm{kg}/\mathrm{m}^3$) than the oil ($543\ \mathrm{kg}/\mathrm{m}^3$), so $\mathrm{CO}_{2}$ will preferentially flow below the oil. Figure~\ref{fig::CO2bottomex1} illustrates that the gas saturation after $10$ years of injection is nearly identical for MHFE-DG simulations on hexahedral, prismatic and tetrahedral~1 grids. Also shown is a comparison between MHFE-DG and a lower-order transport update (MHFE-FV) on a 2D cross-section at $y=70\ \mathrm{m}$ for 30\% and 50\% PVI and for 3 different grid types. The MHFE-FV results have not converged and are therefore more dispersed. The implications of this numerical dispersion are more apparent in Figure~\ref{fig::CO2ex1recovery}. It shows that while the converged MHFE-DG simulations predict the same oil recovery on all grid types (with different levels of refinement as given in Table~\ref{table::NNNMNF}), the MHFE-FV results have different oil recoveries on different grids and the recovery predictions are overestimated by 15 -- 25\%.}

\subsection{Example 2: Grid Sensitivity Studies}
We carry out a few more targeted grid sensitivity studies. 
\subsubsection{Stable Waterflooding}\label{sec::sphere}
First, water is injected (at $10\%$ PV/yr) from the center of a $3\ \mathrm{m}\times3\ \mathrm{m}\times3\ \mathrm{m}$ cube, with constant pressure production from all 8 vertices. All rock and fluid parameters are the same as in Example~1. 
We set the gravitational acceleration to zero to make the problem fully symmetric. The water front should therefore expand as a sphere. Sensitivity to grid orientation manifests itself as enhanced flow along the prevalent grid directions. On 2D structured grids (below), this would result in a diamond-shaped distortion of the (real) circular outflow. In 3D, grid sensitivity would lead to deviations from the sphere, either along the coordinate axis for structured cuboid grids, or along preferred grid lines in unstructured prismatic and tetrahedral grids. 

\begin{figure}
 \centering
\includegraphics[width=\textwidth]{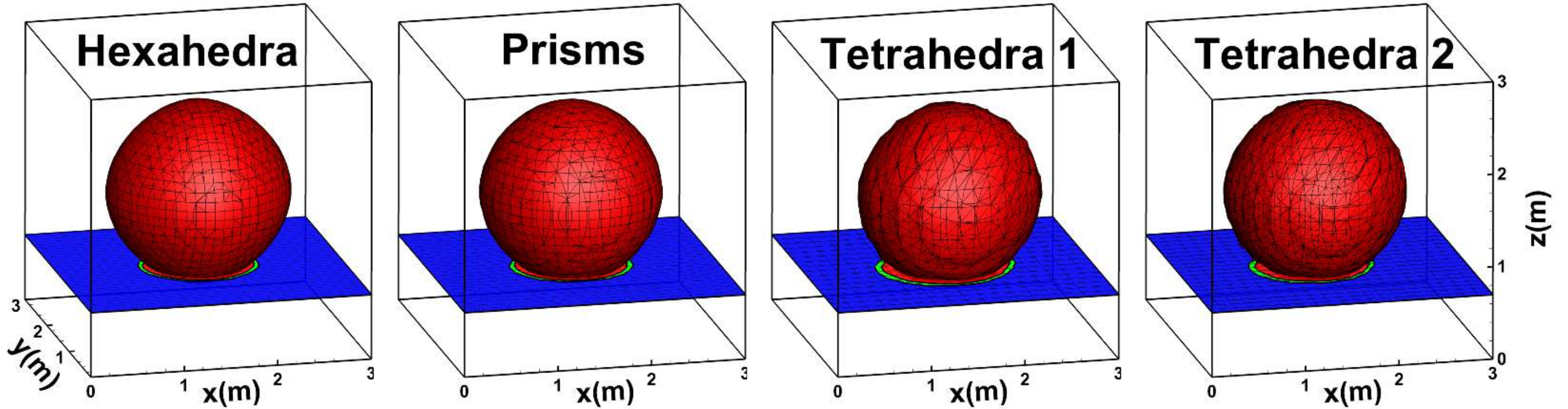}
\caption{Example 2: Water saturation isosurface ($S_w = 0.2$) and projected contours ($S_w = 0.05$ and $S_w = 0.1$) at $6\%$ PVI on $N_K = 29,791$ hexahedral, $N_K = 35,030$ prismatic,  $N_K = 29,478$ tetrahedra 1, and $N_K = 55,566$ tetrahedra 2 grids.} 
 \label{fig::ex4gridding}
 \end{figure}

Figure~\ref{fig::ex4gridding} shows the water saturation profile at $6\%$ PVI, as well as a projection onto a slice at $z = 0.7\ \mathrm{m}$, simulated on a hexahedral, a prismatic, and two tetrahedral grids. We see no discernible deviations from the sphere (or projected circle) due to grid sensitivity.

We also simulated this problem (and others) with our previous face-based DG formulation and found equally low grid sensitivity.
However, we expect that the {\color{black}vertex}-based DG formulation should be superior in some applications, as discussed in Section~\ref{sec::dg}.
To test this hypothesis we consider a pathological case of an \textit{unstable} displacement front next.

\subsubsection{Unstable Gas Injection with Viscous Fingering}
We consider a 2D horizontal $50\ \mathrm{m}\ \times\ 50\ \mathrm{m}$ five-spot pattern (source in the center, constant pressure production wells in all four corners), and a fine $99\ \times\ 99$ element mesh. All other parameters are the same as above. To simulate unstable flow, we inject $\mathrm{CO}_{2}$ and increase the gas/oil mobility ratio through the relative permeability by changing the gas relative permeability to linear, and setting $k^{0}_{rg} = 10\times k^{0}_{ro} =1$. The oil and gas viscosities are $0.14\ \mathrm{cp}$ and  $0.03\ \mathrm{cp}$, respectively. 

\begin{figure}
 \centering
\includegraphics[width=\textwidth]{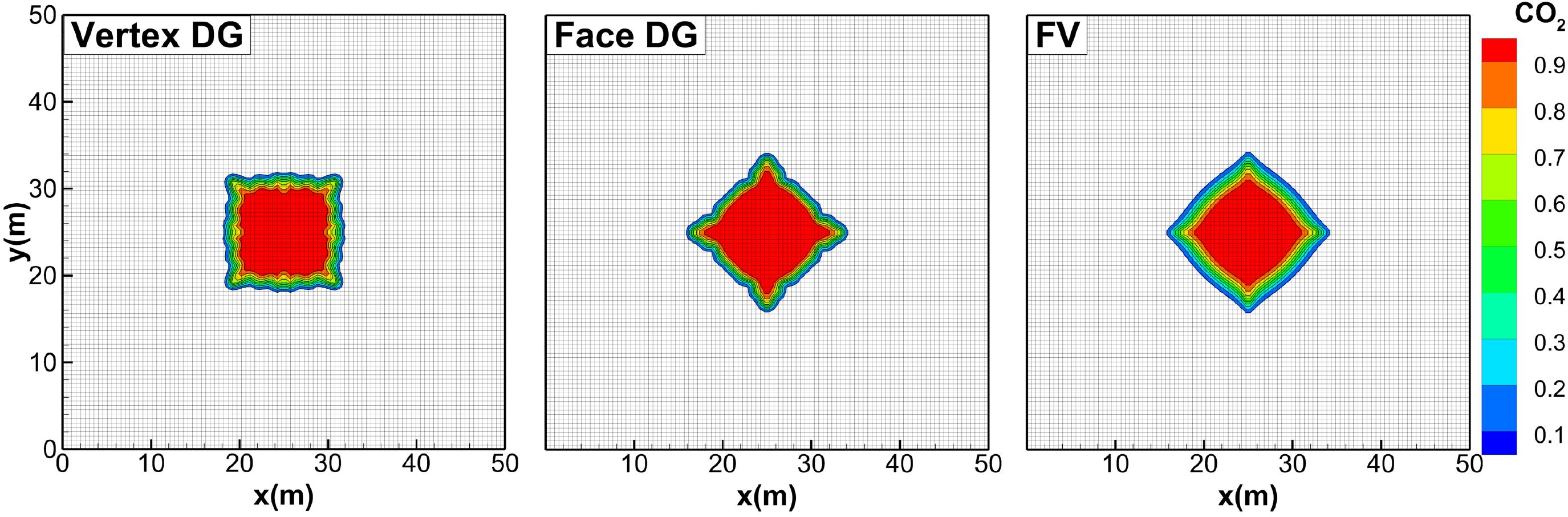}
\caption{Example 2: Overall $\mathrm{CO}_{2}$ molar fraction at $5\%$ PVI $\mathrm{CO}_{2}$ injection from the center for {\color{black} MHFE-FV and MHFE plus vertex- and face-based DG simulations.} Production is from the four corners. } 
 \label{fig::ex4griddingCO2}
 \end{figure}

Figure~\ref{fig::ex4griddingCO2} shows the overall $\mathrm{CO}_{2}$ composition at $5\%$ PVI for simulations with {\color{black}vertex}- versus face-based DG formulations. We present these early-time results, because the \textit{onset} of a viscous fingering instability is of interest. Due to the adverse mobility ratio, the flow is inherently unstable. However, the instability needs to be triggered by some deviation from complete symmetry (e.g.~numerical precision). What we see in Figure~\ref{fig::ex4griddingCO2} is that for the {\color{black}vertex}-based DG, viscous fingers emerge first along the \textit{physical} deviation from symmetry caused by the four production wells in the corners. For the face-based DG, on the other hand, we find that the most pronounced fingers prefer to flow along the coordinate axis, which is a typical \textit{unphysical} gridding effect. We conclude that for certain problems the {\color{black}vertex}-based DG formulation presented in this work performs better than the face-based discretization with the same formal convergence properties. {\color{black}Simulations results are also presented for a lowest-order FV transport update, which shows similar grid orientation effects as in the face-based DG approach, but with increased numerical dispersion.}

\subsubsection{General Quadrilateral and Hexahedral Grids}
\begin{figure}
 \centering
\includegraphics[width=\textwidth]{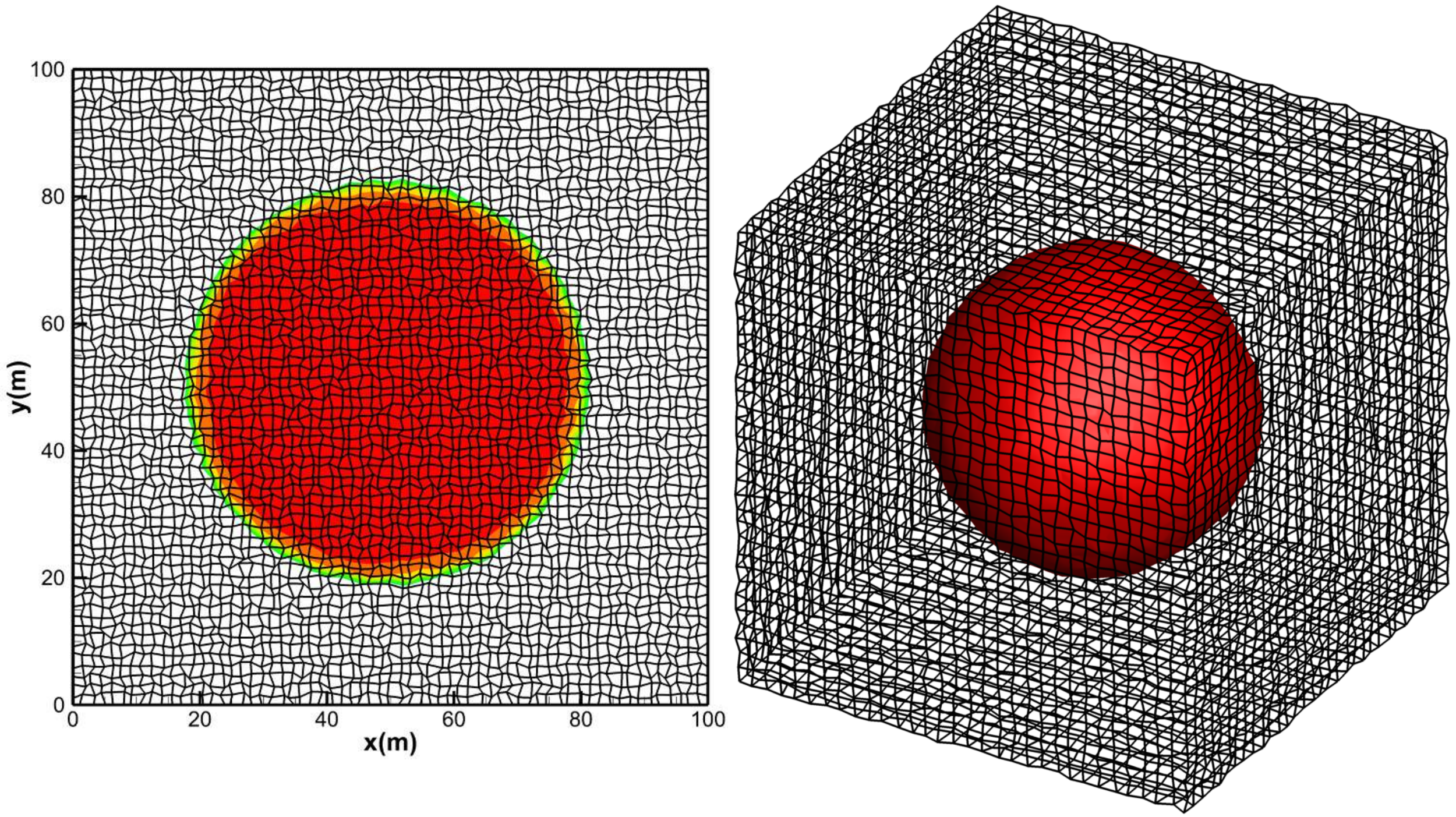}
\caption{Example 2: Water saturation for injection from the center with production from all corners. Left figure is at 15\% PVI for $N_K = 64\times 64$ distorted quadrilateral grid. Right figure is same $N_K = 29,791$ hexahedral grid and $S_{w}=20\%$ iso-surface at 6\% PVI as in Figure~\ref{fig::ex4gridding} but with nodes randomly distorted.} 
 \label{fig::ex4griddingwat2}
 \end{figure}

As a final test of mesh sensitivity, we consider water-flooding from the center of extremely poor quality distorted quadrilateral and hexahedral grids as shown in Figure~\ref{fig::ex4griddingwat2}. Production is again from all the corners. The nodes of structured 2D and 3D grids are randomly perturbed to the degree that, for instance,  many of the quadrilaterals look like triangles with a hanging node on one of the edges (which in reality is the fourth node of the quadrilateral). The hexahedra have non-parallel faces and many sharp angles. For the 2D grid, water is injected from the center at 10\% PV/yr. The 3D simulation is as in Section~\ref{sec::sphere}. We can still not detect any mesh sensitivity or other numerical artifact, and the total mass balance error (integrated in time, and over the domain) in each of the 6 components after 100\% PVI is $<10^{-14}$.

\subsection{Example 3: Water Flooding of Anisotropic Reservoir}
One of the strengths of our FE formulation is that we can readily accommodate full permeability tensors. On unstructured grids we effectively always consider anisotropic permeability, in the sense that even for a matrix-diagonal permeability on a \textit{real} mesh element, the permeability mapped onto the \textit{reference} finite element will look like an anisotropic permeability tensor. 

In this example we consider two tensors in which the vertical permeability is much lower than the horizontal permeability, and the permeability in the $x-y$ direction is either enhanced or reduced:
\begin{eqnarray}\label{eq::tensplus}
\mathrm{K}_\pm &=& 10\ \mathrm{mD} \left(\begin{array}{ccc} 50 & \pm 40 & 0 \\ \pm 40 & 50 & 0 \\ 0 & 0 & 1\end{array} \right)
\end{eqnarray}
such that the determinants of $\mathrm{K}_+$ and $\mathrm{K}_-$ are again $100\ \mathrm{mD}$. 
We repeat the water flooding case from Example~1 with this permeability, and on all $5$ (un)structured grids.  

\begin{figure}
\centering
\includegraphics[width=\textwidth]{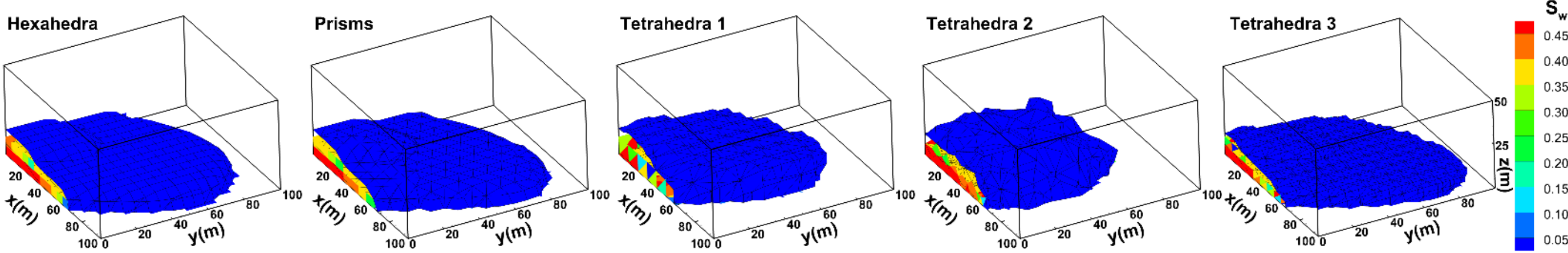}
\caption{Example 3: Water saturation after $1\%$ PVI of water, simulated on $5$ different grids with permeability tensor $\mathrm{K}_+$ given in \eqref{eq::tensplus}.}
 \label{fig::watex2pos}
 \end{figure}

\begin{figure}
\centering
\includegraphics[width=\textwidth]{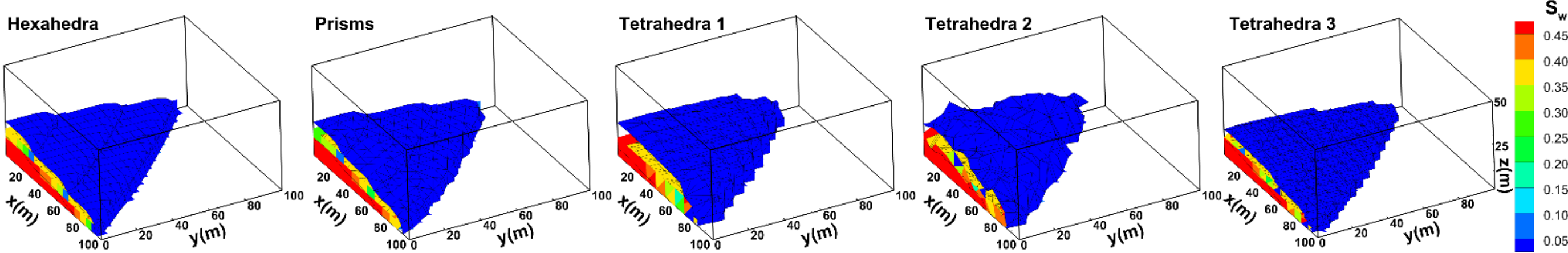}
\caption{Example 3: Water saturation after $1\%$ PVI of water, simulated on $5$ different grids with permeability tensor $\mathrm{K}_-$ given in \eqref{eq::tensplus}.}
 \label{fig::watex2neg}
 \end{figure}

The effect of the non-zero $K_{xy}$ component is most apparent before the water front reaches the back $y=100\ \mathrm{m}$ of the domain, so we plot the water saturation at $1\%$ PVI for $\mathrm{K}_+$ and $\mathrm{K}_-$ in Figures~\ref{fig::watex2pos} and \ref{fig::watex2neg}, respectively. Only $0< x < 100\ \mathrm{m}$ is shown for clarity. 

\begin{figure}
 \centering
\includegraphics[width=0.8\textwidth]{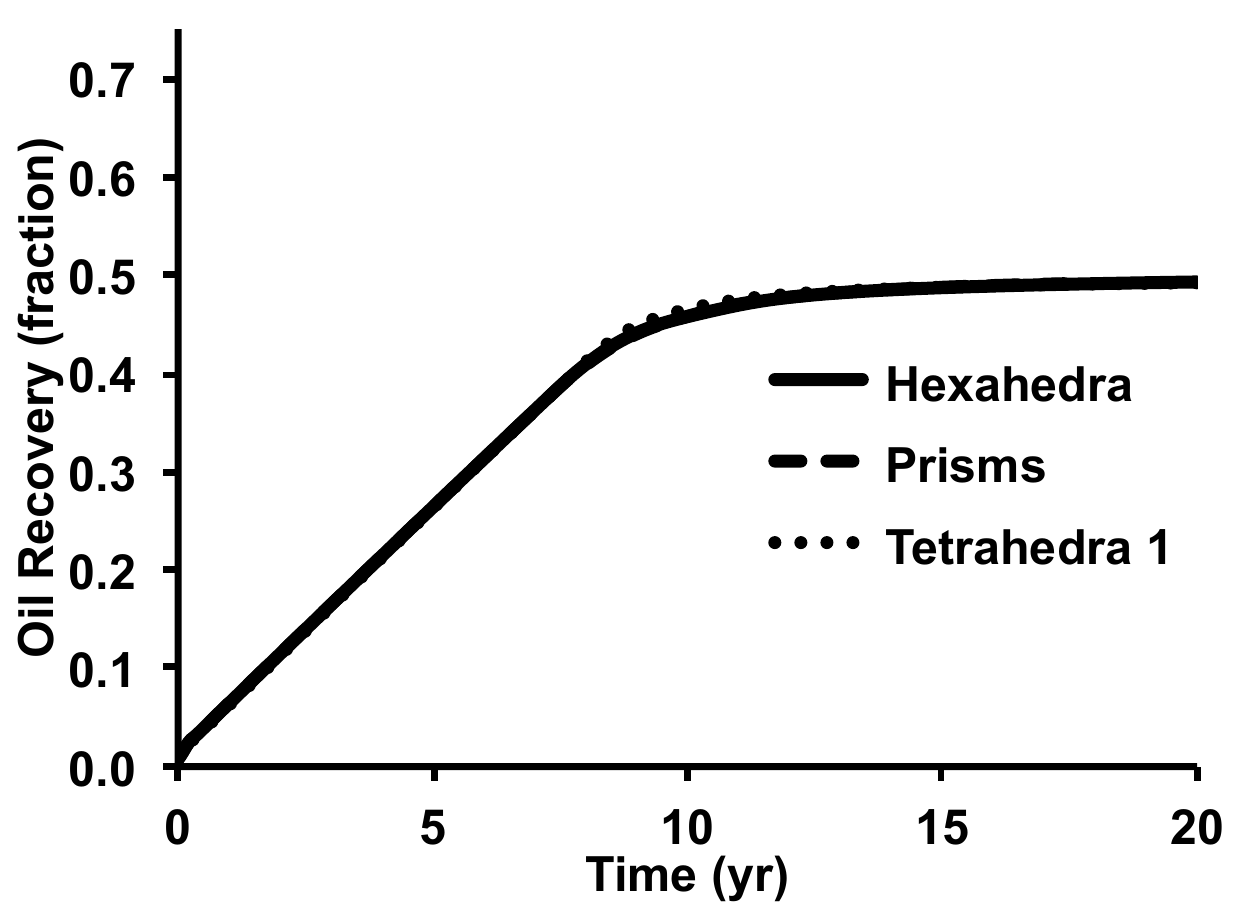}
\caption{Example 3: Oil recovery from $20$ years of water injection from the bottom of a reservoir with permeability tensor $\mathrm{K}_+$ given in \eqref{eq::tensplus}, computed on $3$ different grids.}
 \label{fig::ex2recovery}
 \end{figure}

We see how the positive $K_{xy}$ significantly enhances the (early) flow along the $x=y$ direction in Figure~\ref{fig::watex2pos}, while a negative $K_{xy}$ has the opposite effect in Figures~\ref{fig::watex2neg}. A similar example for triangular 2D grids was presented in \citet{moortgatI}. These observations are as expected and simply demonstrate the applicability of our MHFE-DG approach to non-grid-orthogonal tensor permeabilities.
The initial flow is mainly in the $x$-$y$ plane, due to the high horizontal to vertical permeability ratio. The fronts on the coarsest two tetrahedral grids (1 and 2) are slightly more dispersed, because they have only $\sim 5$ layers in $z$, as compared to $10$ layers for the hexahedral and prismatic grids. On the finer tetrahedral 3 grid the results appear to be converged. In terms of the oil recovery predictions, the results from the coarsest tetrahedral 1 grid converge to those on the finer hexahedral and prismatic grids, as shown in Figure~\ref{fig::ex2recovery} for $20$ years of water injection for $K_{+}$ (the figure for $K_{-}$ is omitted due to its similarity).

\subsection{Example 4: Convergence Analysis}\label{sec::convergg}
We analyze the convergence properties of our FE methods by simulating a simple problem on a wide range of grid refinements. Water is injected into a $4\ \mathrm{cm}\times 4\ \mathrm{cm} \times 30 \ \mathrm{cm}$ core saturated with nC$_{10}$. The temperature and pressure are $310\ \mathrm{K}$ and $100\ \mathrm{bar}$, respectively, the permeability is $10\ \mathrm{mD}$, and porosity is $20\%$. The relative permeabilities are linear with unit end-point for oil and $0.3$ for water, and the residual oil saturation is $30\%$. To simulate gravitationally stable displacement, we inject one PV/day from the bottom and produce from the top at a constant pressure. 

We perform simulations on $9$ different grids. The coarsest grid has $N_{1} = 2\times 2\times 15$ hexahedral elements, which we refine by a factor $2$ in each direction 4 times, such that the finest grid has $2^{12} \times N_{1}$ elements. Tetrahedral grids are created by dividing each hexahedron into $6$ tetrahedra. The number of elements, vertices, and faces for the $5$ hexahedral and $4$ tetrahedral grids are provided in Tables~\ref{table::ex3wathexa} and \ref{table::ex3wattetra}, respectively. The finest grid has $245,760$ elements, $262,449$ faces, and $753,664$ nodes, such that we update well over one million values for compositions, pressures and fluxes. 

\begin{figure}
 \centering
\includegraphics[width=0.8\textwidth]{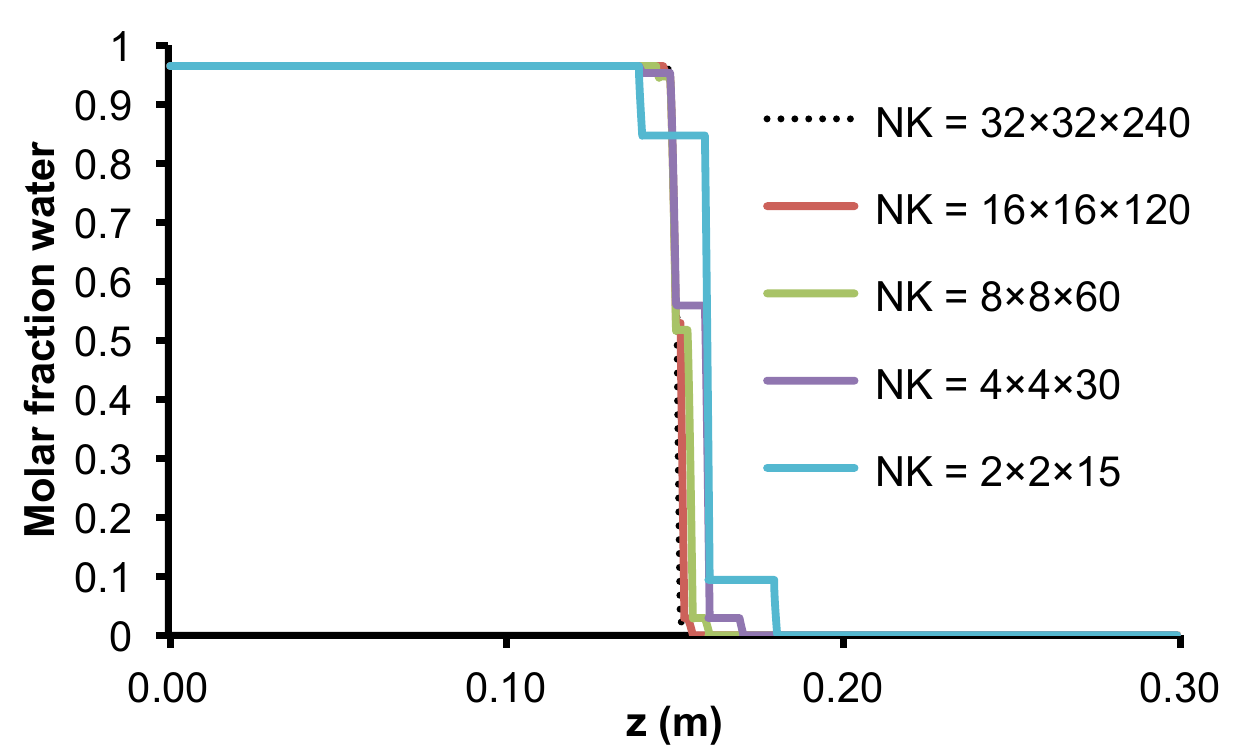}
\caption{Example 4: Water composition (molar fraction) at $50\%$ PVI on all $5$ hexahedral grids as provided in Table~\ref{table::ex3wathexa}.}
 \label{fig::ex3waterprofiles}
 \end{figure}

The flow of water through the core is piston-like displacement. The exact water composition profile is expected to be a 1D step-function propagating through the core, with increasing numerical dispersion exhibited on coarser grids. 3D plots of the results on all $9$ meshes are not very informative. Instead, Figure~\ref{fig::ex3waterprofiles} shows a projection onto the $z$-axis of all the hexahedral simulations at $50\%$ PVI.

The important results are summarized in Tables~\ref{table::ex3wathexa} and \ref{table::ex3wattetra}, which provide the $L_1$ errors, convergence rates, and CPU times of all the simulations. The order of convergence is lower than the formal order of $2$, but is still well above linear. In fact, it is higher than for the Buckley-Leverett problem simulated with a MHFE-DG approach in \citet{hoteitcapillary}, and our convergence study in \citet{moortgatIII}. The latter may be due to the {\color{black}vertex}, rather than face, based formulation. Figure~\ref{fig::ex3waterprofiles} illustrates the gain from the higher order of convergence: the simulation results on the $8 \times 8 \times 60$ grid have mostly converged to the same accuracy as on the $32 \times 32 \times 240$ grid, while requiring two orders of magnitude less CPU time. 

\subsection{Example 5: Complex Domains and Grids}

\subsubsection{Mixed Elements Grid}
\begin{figure}
 \centering
 \includegraphics[width=\textwidth]{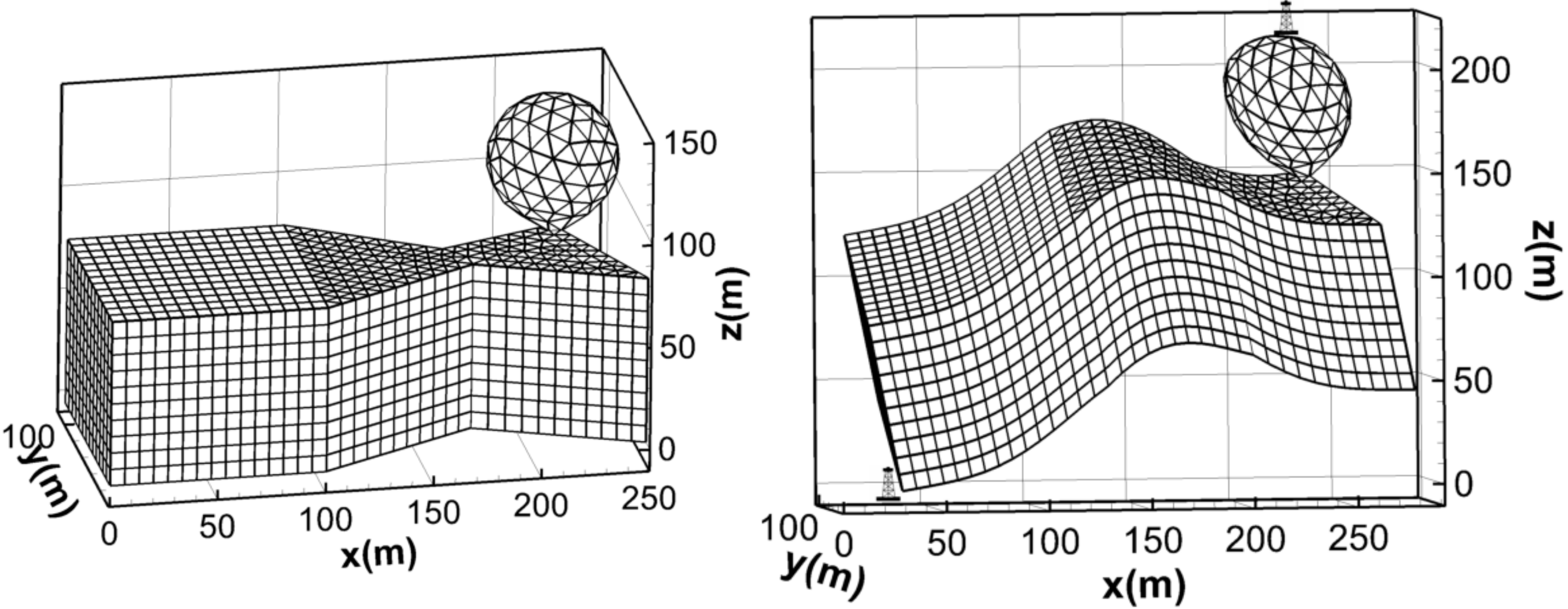}
\caption{Example 5: Compounded mesh before (left) and after applying the coordinate transformation in \eqref{eq::coordtransf1}-\eqref{eq::coordtransf2} (right), and location of injection and production wells in the final mesh. The original mesh on the left consists of 1) a $0.1\ \mathrm{km}\times 0.1\ \mathrm{km}\times 0.08\ \mathrm{km}$ cube discretized by $10\times 13\times 10$ hexahedral elements, 2) an hour-glass shaped domain consisting of 10 layers, with the horizontal cross-sections defined by the points $(100\ \mathrm{m}, 0\ \mathrm{m})$, $(175\ \mathrm{m}, 40\ \mathrm{m})$, $(250\ \mathrm{m}, 10\ \mathrm{m})$, $(220\ \mathrm{m}, 80\ \mathrm{m})$, $(165\ \mathrm{m}, 60\ \mathrm{m})$, and $(100\ \mathrm{m}, 100\ \mathrm{m})$, discredited by $2,870$ prisms, and 3) a $30\ \mathrm{m}$-radius sphere, discretized by $676$ tetrahedra (CDT).} 
 \label{fig::ex4combogrid}
 \end{figure}

To illustrate our capability to mix different element types, we start with $3$ sub-domains, which are discretized by hexahedra, prisms, and tetrahedra, respectively, and then form a compounded grid by `glueing' the three grids together, as shown in Figure~\ref{fig::ex4combogrid}. Because neighboring elements cannot have intersecting edges, we join the rectangular faces of the right-boundary of the cuboid with those of the left-boundary of the prismatic grid, and one of the triangular faces of the top-boundary of the latter grid with one of the triangular faces of the tetrahedral sphere.
We subsequently transform the $x$ and $z$ coordinates to $x^\prime$ and $z^\prime$ (with $L_x = 250\ \mathrm{m}$ and $L_z=143\ \mathrm{m}$) by 
\begin{eqnarray}\label{eq::coordtransf1}
x^\prime &=& x -  (z + L_z)/5, \\\label{eq::coordtransf2}
z^\prime &=& z + x/6 + 40 \sin (x \pi/L_x)^4 + y/2.5,
\end{eqnarray}
to create more general unstructured hexahedral and prismatic elements. The geometrical slopes and domes result in interesting multiphase flow, particularly when driven by gravity. The grid and well locations are shown before and after applying \eqref{eq::coordtransf1}-\eqref{eq::coordtransf2} in Figure~\ref{fig::ex4combogrid}. 

For the simulations, we consider the same fluid and rock parameters as in Example~1 and again simulate depletion, water flooding, and gas injection. 

\begin{figure}
 \centering
 \includegraphics[width=.5\textwidth]{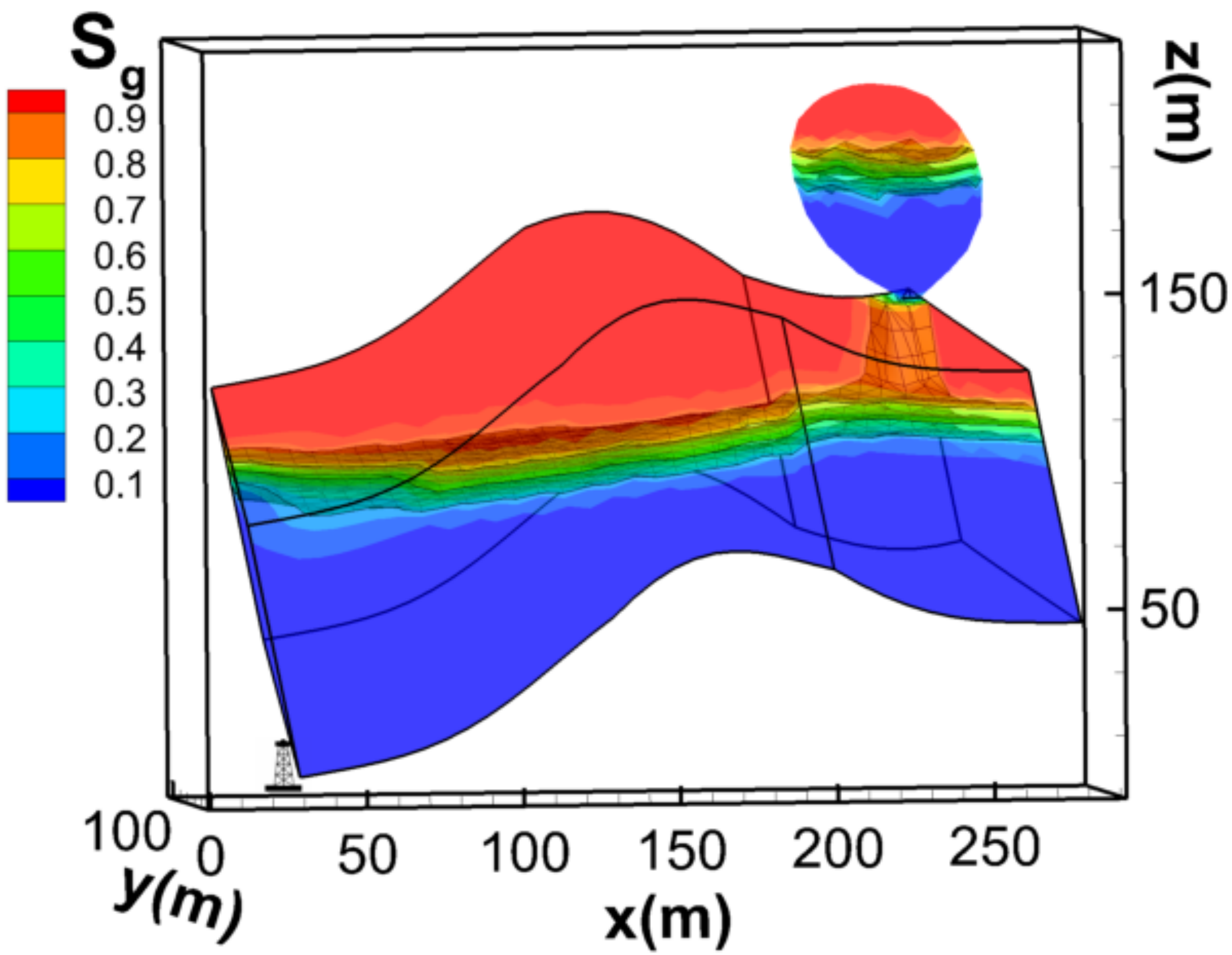}
\caption{Example 5: Gas saturation after $4$ years depletion at $5\%$ PV/yr from the bottom.} 
 \label{fig::ex4depl4yr}
 \end{figure}

\paragraph{Gravity Depletion}
We deplete at a constant rate of $5\%$ PV/yr from the lowest point in the domain. The initial bottom-hole pressure is $375$ bar. The saturation pressure of the reservoir oil is not much lower ($338$ bar), and after a few months the pressure in the top start to drop below the saturation pressure. It takes only $6$ months before the entire reservoir is in a two-phase state. Figure~\ref{fig::ex4depl4yr} shows the gas saturation throughout the reservoir after $4$ years. Because of the relatively high permeability ($100\ \mathrm{mD}$), the liberated gas buoyantly rises to the top and accumulates in the trap in the prismatic grid section and inside the sphere, although the latter is hampered by the bottle-neck intersection. In terms of numerical performance, we see a smooth and nearly flat saturation profile throughout the grid without any artifacts at the intersections between different element-types. 

\begin{figure}
 \centering
 \includegraphics[width=\textwidth]{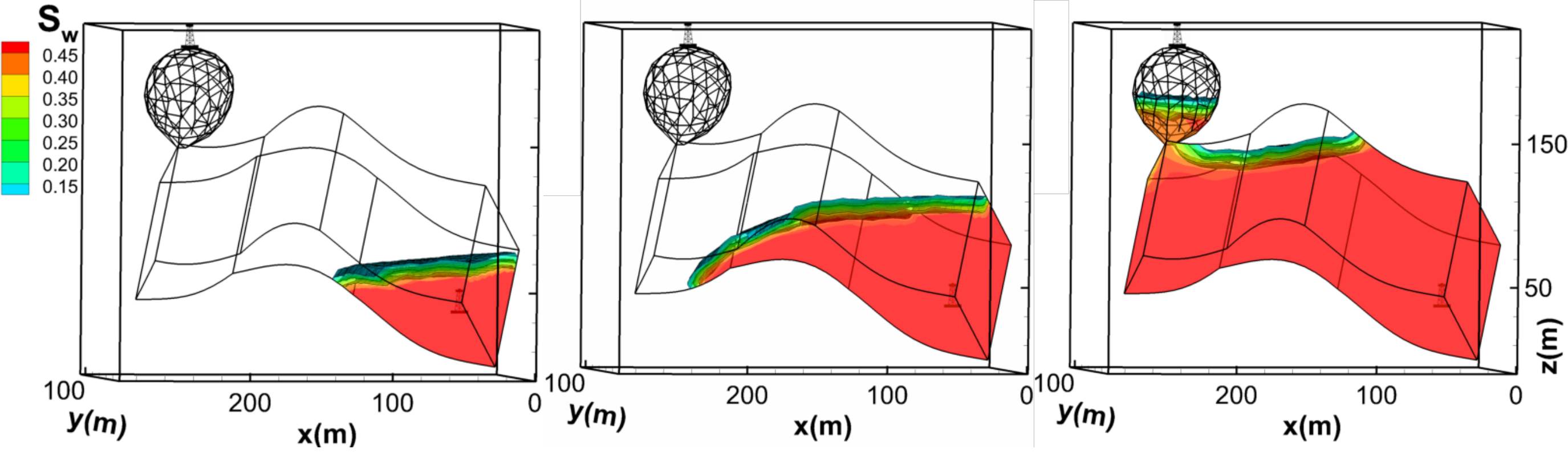}
\caption{Example 5: Water saturation at $10\%$ (left), $25\%$ (middle), and $45\%$ (right) PV of water flooding.} 
 \label{fig::ex4H2O}
 \end{figure}

\paragraph{Water Flooding}
Water is injected at $5\%$ PV/yr from the lowest point, and production is from the top of the sphere, resulting in immiscible compressible displacement of the reservoir oil. The water saturation at different times is shown in Figure~\ref{fig::ex4H2O}. Note that the mesh is rotated around the $z$-axis by $180^\circ$ to offer a better view. During the first $10\%$ PVI water mainly fills the lower trough of the hexahedral and prismatic sections, before spilling over into the second trough in the domain. After $42\%$ PVI the flooding front reaches the bottle-neck to the sphere. At later times, water fills the sphere, without recovering the oil in the dome section of the domain. Figure~\ref{fig::ex4H2O} nicely shows a level front due to gravitational segregation. 

\begin{figure}
 \centering
 \includegraphics[width=\textwidth]{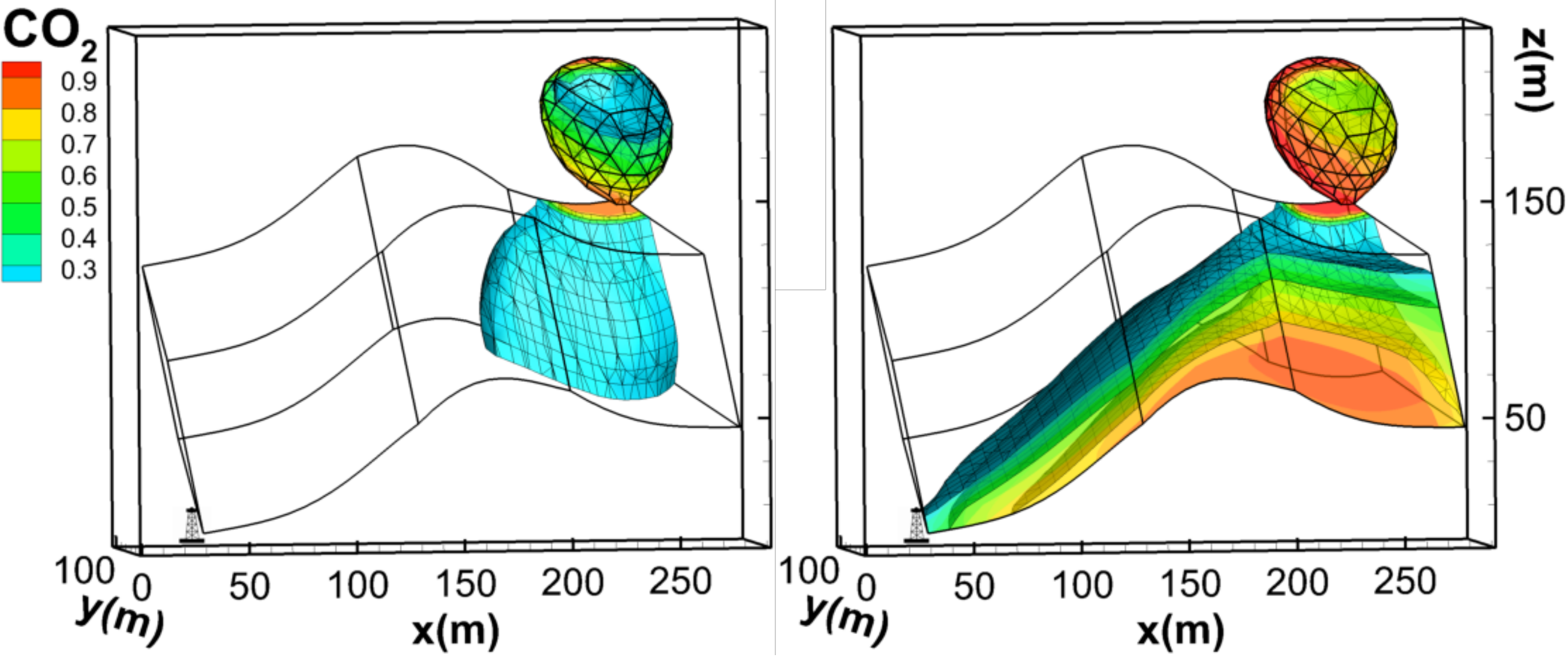}
\caption{Example 5: Overall $\mathrm{CO}_{2}$ composition (molar fraction) at $5\%$ (left) and $10\%$ (right) PV of $\mathrm{CO}_{2}$ injection.} 
 \label{fig::ex4CO2}
 \end{figure}

\paragraph{$\mathrm{CO}_{2}$ Injection}
As mentioned in Example~1, $\mathrm{CO}_{2}$ is denser than the oil at these reservoir conditions and should be injected from the bottom for the most efficient oil recovery. However, the flow in that case is similar to that of water in the previous example, so instead we simulate the more complicated scenario of $\mathrm{CO}_{2}$ injection from the top of the sphere, with production from the bottom. 
Figure~\ref{fig::ex4CO2} shows the overall molar fraction of $\mathrm{CO}_{2}$ after $10\%$ and $25\%$ PVI. We see how dense $\mathrm{CO}_{2}$ sinks to the bottom of the sphere, and then fills the right-side trough of the prismatic grid, before spilling over into the downward sloping prismatic and hexahedral sections of the domain. Again, we observe no sign of grid sensitivity within or across the transitions between the $3$ sub-grids.  

\subsection{Example 6: Benchmark Examples with Realistic Petrophysical Properties}

\subsubsection{SPE~Tenth Comparative Solution Project}
We consider the SPE 10 benchmark problem \citep{spe10} as an example of a realistic geo-statistical realization of petrophysical properties. This benchmark problem has more than eight orders of magnitude variation in permeabilities, as well as a wide range in porosities. We perform fine-grid simulations of two-phase flow in two different sectors of this domain, which is saturated with nC$_{10}$ at the same temperature and pressure as in Example~4.  C$_{1}$ is injected at $5\%$ PV/yr from one corner and production is at a constant pressure from the opposite corner.
The relative permeabilities are linear with an end-point relative permeability of $0.27$ for oil, and $1$ for gas.
	
\begin{figure}
 \centering
 \includegraphics[width=\textwidth]{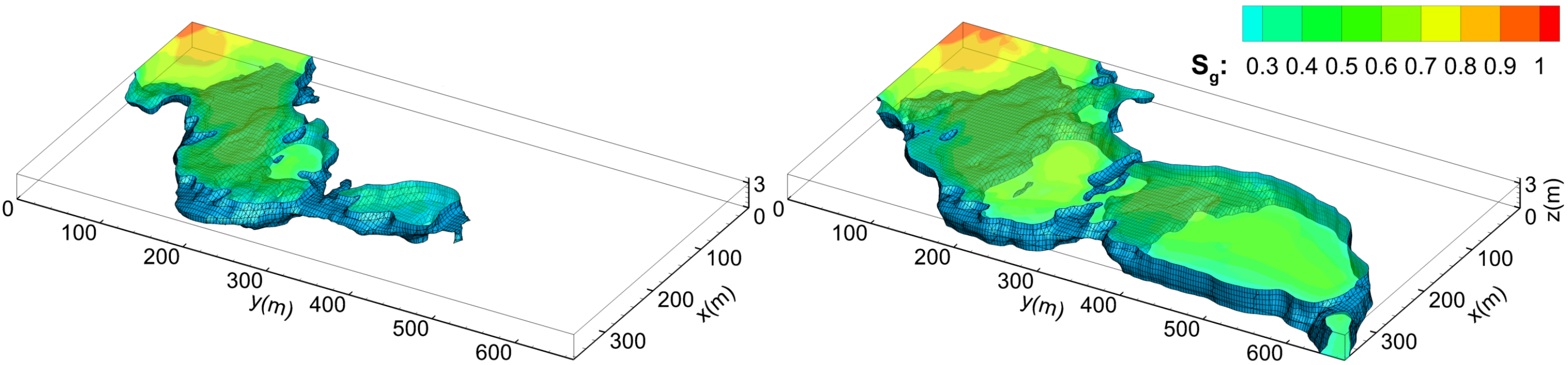}
 \caption{Example 6: Gas saturation for 5\% PV/yr methane injection into bottom five layers of SPE 10 domain at 25\% PVI (left) and 70\% PVI (right).} 
 \label{fig::ex6spebottom}
 \end{figure}

	The first section consists of the bottom five layers ($60\times 220\times 5$ grid cells). This is part of the fluvial Ness formation in the North Sea, which has well connected high-permeability sandstone channels, corresponding to river depositions with low permeability shale in between. 
	Figure~\ref{fig::ex6spebottom} shows the gas saturation at 25\% and 70\% PVI and clearly shows the channeling of injected methane through the high permeability sandstone with a poor sweep of the shale depositions.
	
\begin{figure}
 \centering
 \includegraphics[width=\textwidth]{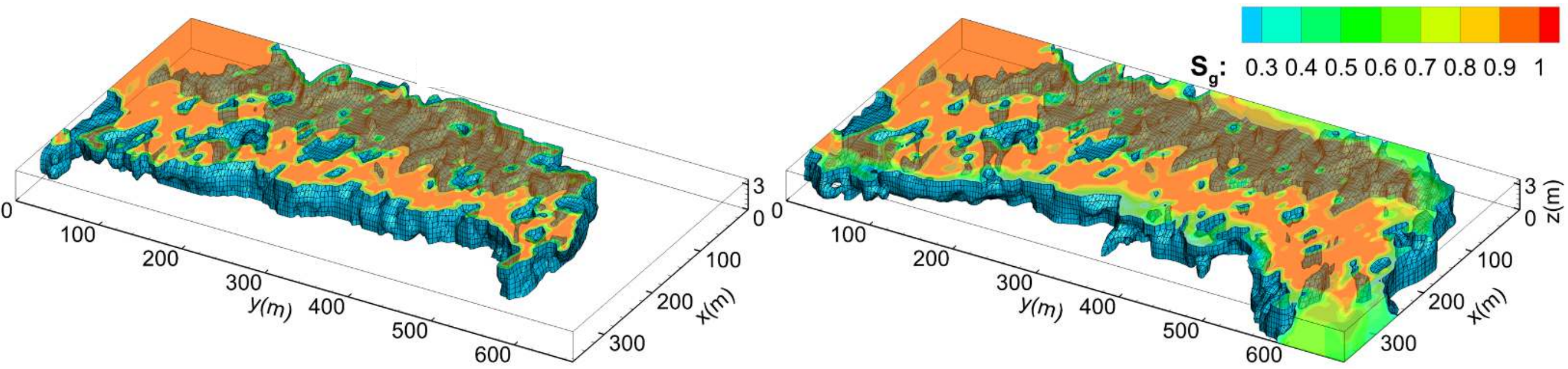}
 \caption{Example 6: Gas saturation for 5\% PV/yr methane injection into top five layers of SPE 10 domain at 50\% PVI (left) and 100\% PVI (right).} 
 \label{fig::ex6spetop}
 \end{figure}

	The second section has the top five layers from the Tarbert formation, which has a completely different shallow marine deposition history.
	 In Figure~\ref{fig::ex6spetop}, we see a better sweep for the Tarbert formation, in which the permeability variations still span 8 orders of magnitude, but with a shorter correlation length.
	
	\subsubsection{TU Delft Egg Domain}\label{sec::ex6b}
	This synthetic domain is a three-dimensional realization of a channelized reservoir, which was used to benchmark water flooding conditions with eight water injectors and four producers using $4$ different simulators \citep{eggmodel,eggmodel2}. We use the dataset adapted for MRST \citep{mrst}. An image of the channeled permeability field is also available on the SINTEF MRST website. The permeability varies between about 80 and $7,000$ milli-Darcy, while the porosity is uniform at 20\%.
	
\begin{figure}
 \centering
 \includegraphics[width=\textwidth]{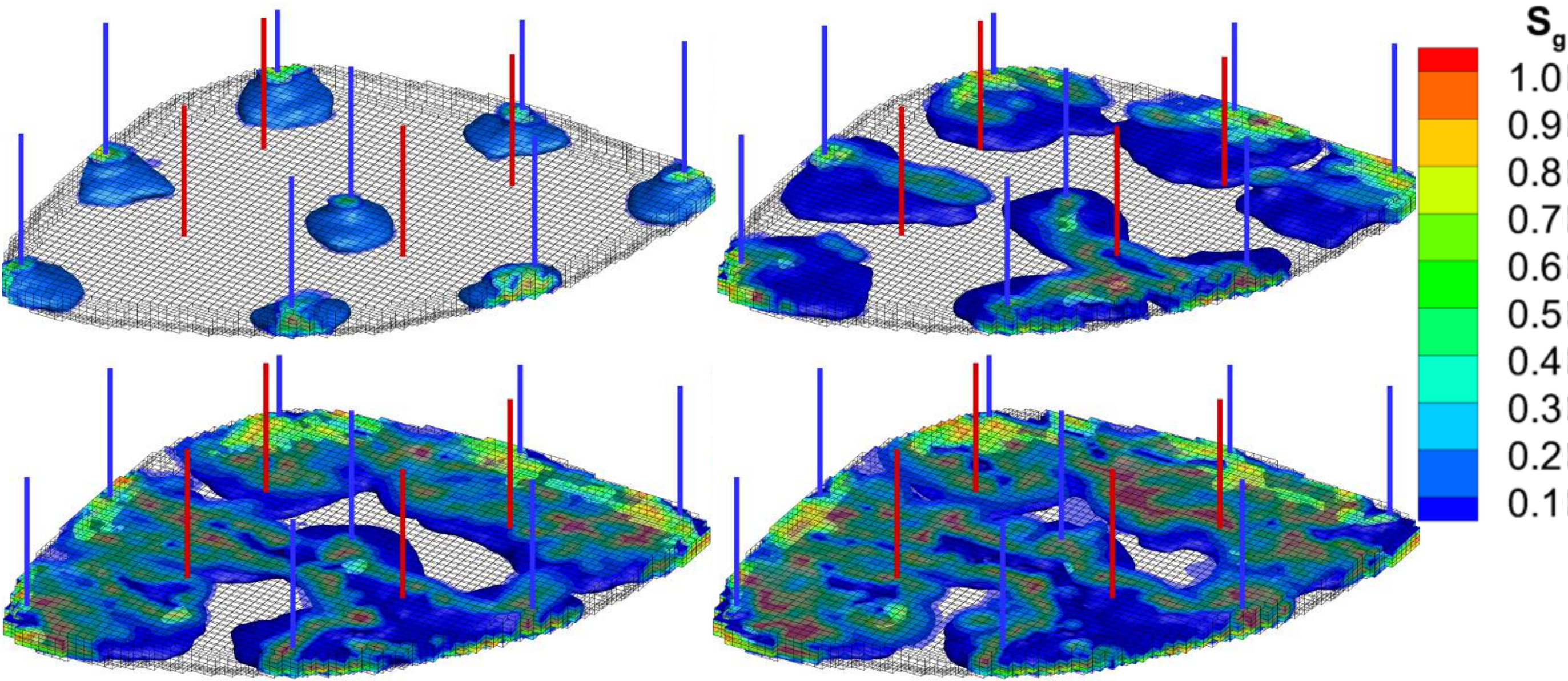}
 \caption{Example 6: Gas saturation for 5\% PV/yr CO$_{2}$ injection into Egg reservoir model saturated with oil. Snap-shots at 5\% (top-left), 25\% (top-right), $50\%$  (bottom-left) and 65\% PVI (bottom-right). Eight injection wells are indicated in blue, and four production wells in red.} 
 \label{fig::ex6egg2ph}
 \end{figure}

\begin{figure}
 \centering
 \includegraphics[width=\textwidth]{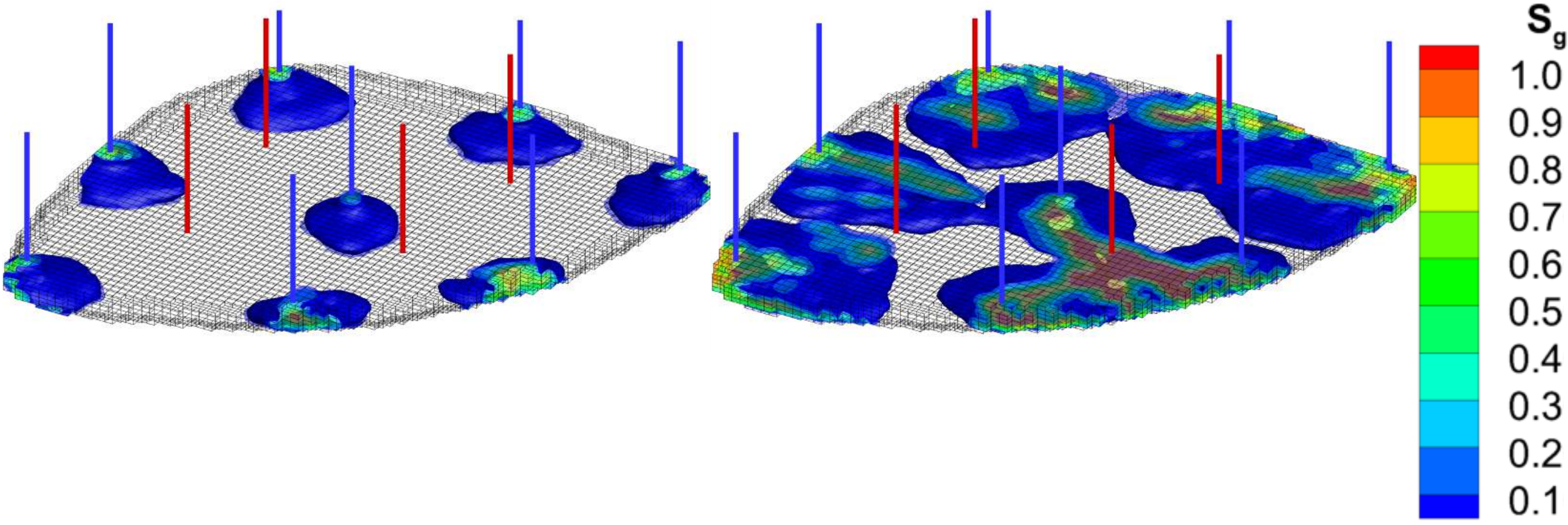}
 \caption{Example 6: Same as Figure~\ref{fig::ex6egg2ph} but for CO$_{2}$ injection into Egg reservoir saturated with oil and water (three-phase flow problem). Snap-shots at 5\% (left) and 25\% PVI (right).  Note that these correspond to a 31\% lower effective pore volume. } 
 \label{fig::ex6egg3ph}
 \end{figure}

	 Instead of modeling the same immiscible two-phase flow problem, we combine the grid and realistic petrophysical properties from the Egg model with a challenging `benchmark' compositional problem from our own work. We consider CO$_{2}$ injection into a Brazilian oil which we modeled before \citep{moortgatIV} at conditions where CO$_{2}$ is supercritical and denser than the oil, and the CO$_{2}$-oil mixtures are near the critical point. All parameters, except the grid, permeability and porosity, are the same as in our earlier work \citep{moortgatIV}, and we also include Fickian diffusion in this example.
	  CO$_{2}$ is injected at a constant rate of $5\%$ PV/yr from 8 injectors that perforate the full depth of the formation ($7$ layers), represented by a total of $56$ grid cells. Constant pressure production wells are defined on the outer boundaries of the domain, and are represented by one element in the top and one in the bottom of each of the $4$ producers indicated in Figures~\ref{fig::ex6egg2ph}--\ref{fig::ex6egg3ph} (a total of 8 grid cells). The grid has a total of $18,553$ active grid cells.
	 
	 Figure~\ref{fig::ex6egg2ph} shows gas saturations at $5$, $25$, $50$, and $65\%$ PVI for two-phase gas-oil flow. The panels show how dense CO$_{2}$, injected from the $8$ wells in blue, segregates to the bottom even in this relatively thin formation. The high permeability channels in the Egg model enhance flow towards some of the producers (in red), while impeding flow to, for instance, the bottom-left producer. We also simulate the same problem, but as a three-phase compositional problem with a 31\% connate water saturation and allowing for CO$_{2}$ dissolution into the aqueous phase and the resulting volume swelling. Figure~\ref{fig::ex6egg3ph} shows that the results are quite similar.

\section{Conclusions}
In this work, we study for the first time an attractive combination of higher-order finite element methods for compressible, compositional three-phase flow on unstructured 3D grids. The main strengths of our IMPEC MHFE-DG scheme are summarized below.
\begin{enumerate}
\item Both the DG and MHFE methods work as well on unstructured grids as they do on structured ones and with similar complexity of implementation. The DG transport update can capture discontinuities in fluid properties, e.g., across phase boundaries, layers, and fractures, with low numerical dispersion. The MHFE approach provides accurate, globally continuous, velocity fields, even on highly heterogeneous and anisotropic permeability fields. The only geometry/grid dependence of the MHFE-DG implementation consists of coefficients that are computed in a preprocessing step.
\item Our new vertex-based MHFE-DG method is more accurate than previous face-based implementations. The reason is that \textit{multiple} values are upwinded across each face, which transfers information regarding not only the magnitude of phase properties, but also their \textit{gradients}.  
\item These finite element methods are remarkably insensitive to mesh orientation and quality, which we demonstrated for grids made of tetrahedra, prisms, and general hexahedra in 3D, as well as triangles and general quadrilaterals in 2D.
\end{enumerate}
The first item has been demonstrated on unstructured grids in the past for incompressible immiscible flow, but not for compressible compositional flow. Compositional flow is a considerably more complicated problem due to the higher degree of non-linearity. 

To build confidence in the algorithm, we presented a wide range of numerical experiments in Sections~\ref{sec::exs}. The examples demonstrate that simulations of multicomponent multiphase flow with gravity-driven counter-current-flow and phase behavior yield identical results on all considered grid types, and for a range of processes (Example 1). As expected, we find that the MHFE method exhibits no grid orientation effects (Example 2), and can accommodate a {\color{black}full} permeability tensor (Example 3). For transport we showed that by using a multi-linear DG approximation we can achieve a higher convergence rate than for element-wise constant approximations. This is critical in 3D, where the CPU cost of grid refinement by a factor $\delta f$ scales with $\delta f^{4}$ for an IMPEC method {\color{black}(one factor of $\delta f$ is due to the CFL time-step constraint and could be eliminated in fully implicit schemes)}. 
Using even higher orders of DG may further improve accuracy, but involves a more subtle trade-off than for immiscible incompressible flow models, because of the additional (CPU expensive) phase-split computations.  
Combinations of different grid types were considered in Example 5, and two- and three-phase flow with Fickian diffusion were  modeled for realistic geostatistical benchmark problems in terms of petrophysical properties in Example 6.

{
\section*{Appendix A: Phase-Split Computations}\label{app:A}
Details of the phase stability and phase-split computations for compositional three-phase flow are discussed in \cite{moortgatIII}. 
The basic algorithm proceeds along the following steps
\begin{enumerate}
\item For a given mixture composition $z_i$, temperature and pressure, perform a phase stability analysis to determine whether a single-phase state is stable. The phase stability analysis is based on the tangent-plane distance and finds the phase state with the minimum Gibbs free energy. If the the fluid is stable, $x_{i,\alpha} = z_i$ and $c_\alpha = c$ and no further calculations need to be performed. If the phase is unstable, we proceed to a two-phase phase-split computation.
\item The phase-split calculations determine the phase compositions and amounts such that the fugacities of each component are equal in all phases. 
We have found that the most numerically stable and efficient approach (fewest iterations) is to carry out these calculations in terms of the logarithm of equilibrium ratios: $\ln K_i = \ln \phi_{i,\alpha_1} - \ln \phi_{i,\alpha_2}$, in which $\phi_{i,\alpha}$ are the fugacity coefficients and $\alpha_1$, $\alpha_2$ are two of the three possible phases. The molar fractions of each phase, $\beta_\alpha$ are found from the Rachford-Rice equation: $\sum_i (x_{i,\alpha_1} - x_{i,\alpha_2}) = 0$.
The equilibrium ratios and phase fractions $\beta_\alpha$ are found iteratively. First, successive substitution iterations (SSI) are used until both $K_i$ and $\beta_\alpha$ have converged to a tolerance of typically $10^{-4}$. After this switch criterion, Newton-Raphson (NR) iterations are used until the error is $<10^{-10}$. The NR method requires the computation of the inverse of a Jacobian matrix of derivatives of the aforementioned governing equations with respect to $\ln K_i$ and $\beta_\alpha$. However, while the SSI has linear converge, NR shows quadratic convergence and only requires a few iterations.
\item Once a two-phase solution is found, we perform a second phase-stability analysis to determine whether the two-phase state is stable, or whether a three-phase state has a lower Gibbs free energy. This can be done by testing the stability of either of the two phases found in the previous step. If the two-phase state is stable, phase compositions can be computed from $z_i$, $K_i$ and $\beta_\alpha$. If not, a three-phase split computation is performed. 
\item The three-phase split computation, while computationally more challenging, is mathematically similar to the two-phase split. We solve for two sets of equilibrium ratios from $\ln K_{i,\alpha_1} = \ln \phi_{i,\alpha_1} - \ln \phi_{i,\alpha_2}$ and $\ln K_{i,\alpha_3} = \ln \phi_{i,\alpha_3} - \ln \phi_{i,\alpha_2}$, and two Rachford-Rice equations. The same combination of SSI and NR with the same switch criterion and final tolerance is used for the iterative solution procedure. 
\end{enumerate}
The full (stand-alone) phase stability and phase-split algorithm is very computationally expensive. In the context of reservoir simulations, many of the computational steps can be avoided by spatial and temporal information. For example, whenever equilibrium ratios and phase amounts from a previous time-step are available, these can be used as initial guesses for a phase-split calculation. In $>99\%$ of cases the phase-split computation with such a guess converge in a few iterations, avoiding the need for a phase-stability analysis. 
Also, if a grid cell and all its neighboring grid cells where in single-phase in the previous time-step, it is safe to assume that this grid cell will remain in single-phase at the next time-step (for fluid injection, not depletion, and for an IMPES scheme with a time-step constraint that essentially says that a fluid cannot propagate through more than one grid cell in one time-step). A similar argument can be used for grid cells saturated with water and oil, when an injection gas front is still far away. The phase-split computations are further optimized by adaptively adjusting the switch criterion: if the average number of SSI iterations is large, while only one or two NR iterations are used, we increase the tolerance (switch to NR earlier). Conversely, if the number of NR iterations is high or NR fails due to inadequate initial guesses, the tolerance is decreased (more SSI iterations before NR).

\section*{Appendix B:}
In this appendix we provide the complete expressions for the second-order Discontinuous Galerkin update on two-dimensional triangles and rectangles, and three-dimensional prisms, tetrahedra, and hexahedra. Our formulation allows for any combination of these element types within a single grid.

To solve \eqref{eq::DGdiscr} we invert the matrix $\int_K \phi_N \phi_{N^\prime}\mathrm{d}\mathbf{x} $ analytically and 
arrive at the generic expression
\begin{eqnarray}\label{eq::dgupdate} 
&& (c_i)^{n+1}_{K,N} = (c_i)^{n}_{K,N} + \frac{\Delta t}{\phi |K|} \sum_\alpha [(V_i)^n_{\alpha} - (E_i)^n_{\alpha} + (F_i)_K^n]
\end{eqnarray}
for each component $i$ at each node $N$.
$(V_i)^n_{\alpha}$ and $(E_i)^n_{\alpha} $ collect all the terms in the volume and surface integrals in \eqref{eq::DGdiscr}, respectively. 
 
The expressions for $(V_i)^n_{K,\alpha}$ and $(E_i)^n_{E,\alpha}$ are the same for each grid cell, each phase, and each time-step, so we drop indices $n$ and $\alpha$. As such, $c_N$ refers to the phase composition \textit{inside} grid cell $K$ at node $N$. Note that the flux index refers to edges. For the upwind phase compositions we use slightly different notation here $\widetilde{c_{E,n}}$ will refer to the phase composition upwinded with respect to the flux $q_{\alpha,E}$, at \textit{local} node $n$ of edge $E$ (i.e, $n=1,2$ for triangular and quadrilateral elements, $n=1,2,3$ for tetrahedra and two faces on a prism, and $n=1,\ldots, 4$ on the quadrilateral faces on a prism and for hexahedra).

We write in matrix notation $\vec{V} = V_{N,\alpha}$, $\vec{Q} = q_{\alpha,E}$, $\vec{c} = c_{N,\alpha}$

\newpage
\subsection*{Triangular Elements}
Figure~\ref{fig::doftrian} illustrates the degrees-of-freedom (DOF) for the MHFE-DG method. The {\color{black}vertex}-based DOF for the DG transport update are defined \textit{inside} each element $K$, and can be discontinuous across edges $E$. 

 \begin{figure}[h!]
 \centering
\includegraphics[width=0.8\textwidth]{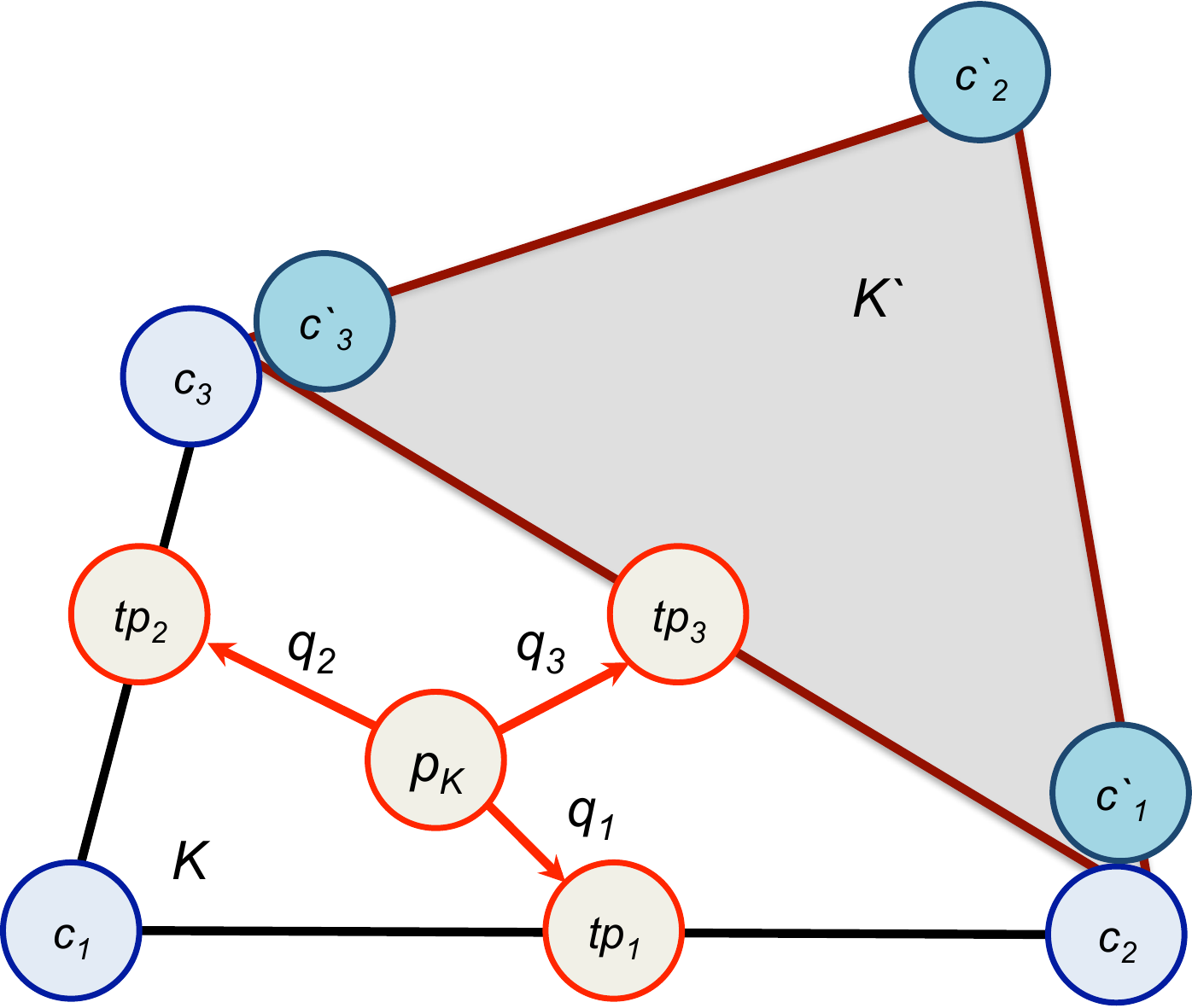} 
\caption{Degrees of freedom for MHFE-DG on a \textbf{triangular} grid element $K$ with the following notation: average pressure $p_{K}$, pressure traces on edges $tp_{E}$, edge fluxes $q_{E}$ and fluid properties at {\color{black}vertices}, e.g.~phase molar density of component $i$ at node $N$, $c_{N}$. Note that the $c_{N}$ are defined inside element $K$; the corresponding \textit{global} node has as many discontinuous values $c_{N^{\prime}}$ as elements $K^{\prime}$ sharing the same global node. One neighbouring element $K^{\prime}$ is shown to illustrated the discontinuous $c_{N}$ across edge $E_{3}$.}
 \label{fig::doftrian}
 \end{figure}
\noindent
The volume integrals in \eqref{eq::dgupdate} can be written as: $V_{N,\alpha} = \vec{c} \bar{\vec{V}}_N\vec{q}/2 $, with 
\begin{eqnarray}\nonumber
&&
\bar{\vec{V}}_{1} =
 \left( \begin{array}{ccc}  2& 2 & -2\\
 1 & 1 & -3\\
 1 & 1 & -3
   \end{array} \right), \quad 
  \bar{\vec{V}}_{2}  =
    \left( \begin{array}{ccc}  
{\color{black} 1} & {\color{black} -3} & {\color{black} 1}\\
 {\color{black}2}& {\color{black} -2 }& {\color{black} 2}\\
{\color{black} 1} & {\color{black} -3} & {\color{black} 1}
   \end{array} \right), \quad 
   \bar{\vec{V}}_{3}  =
 \left( \begin{array}{ccc}  
 -3& 1 & 1\\
 -3 & 1 & 1\\
 -2 & 2& 2
   \end{array} \right).
\end{eqnarray}
The surface integrals are given explicitly by $
E_{N,\alpha} = \bar{\vec{E}}_N\vec{q} /2
$, with 
\begin{eqnarray}\nonumber
\bar{\vec{E}}_1 &=& \left(5 \widetilde{c_{1,1}} +  \widetilde{c_{1,2}},
5 \widetilde{c_{2,1}} +  \widetilde{c_{2,2}},
-3 [\widetilde{c_{3,1}} + \widetilde{c_{3,2}}] \right),\\\nonumber
\bar{\vec{E}}_2 &=& \left(\widetilde{c_{1,1}} + 5 \widetilde{c_{1,2}},
-3[ \widetilde{c_{2,1}} +  \widetilde{c_{2,2}}],
5 \widetilde{c_{3,1}} + \widetilde{c_{3,2}}\right),\\\nonumber
\bar{\vec{E}}_3 &=& \left(-3[ \widetilde{c_{1,1}} +  \widetilde{c_{1,2}}],
\widetilde{c_{2,1}} +  5\widetilde{c_{2,2}},
\widetilde{c_{3,1}} +5 \widetilde{c_{3,2}}\right).
\end{eqnarray}

\newpage
\subsection*{Quadrilaterals}
 \begin{figure}[h!]
 \centering
\includegraphics[width=0.8\textwidth]{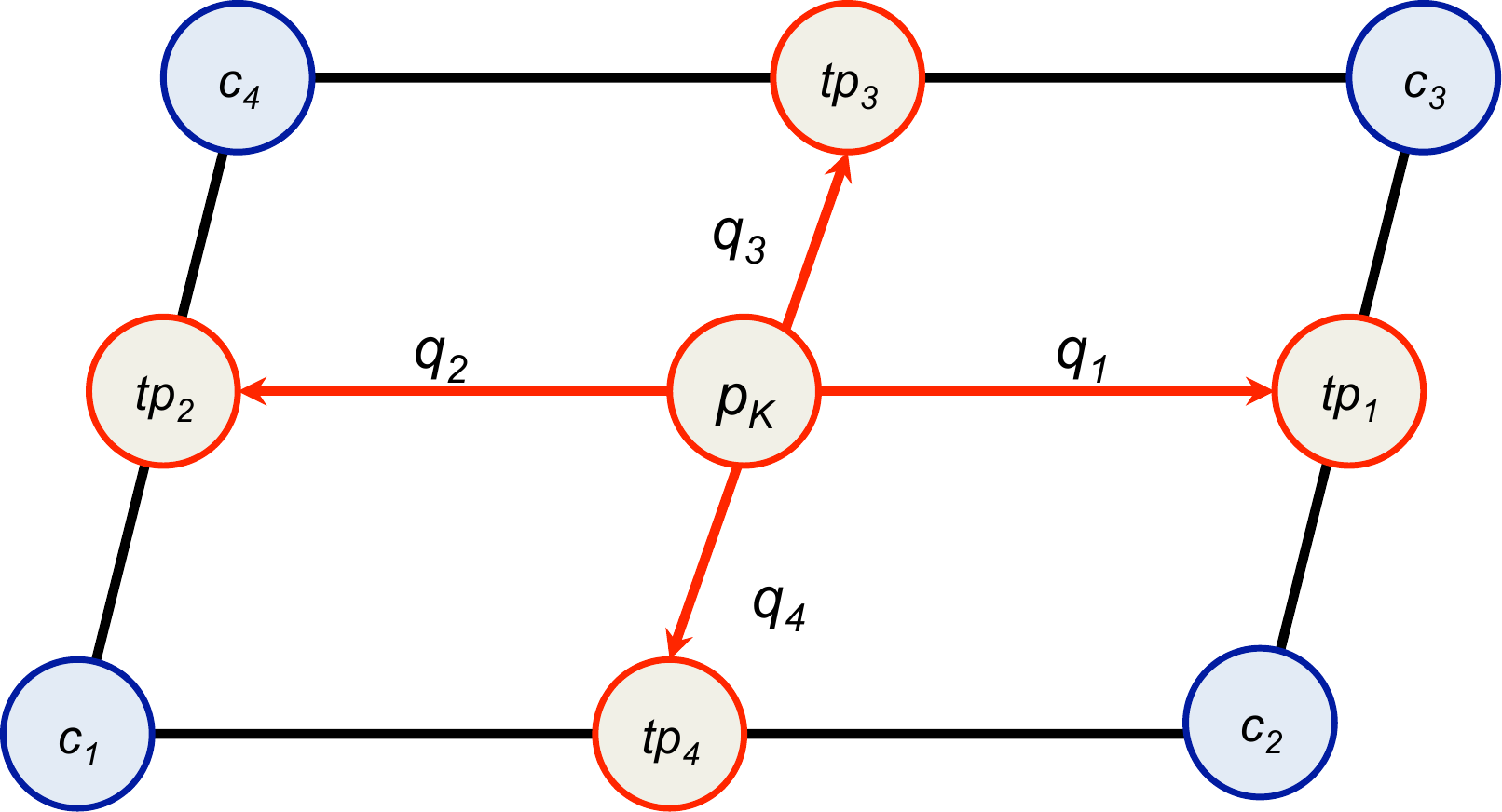} 
\caption{Degrees of freedom for MHFE-DG on \textbf{quadrilateral} grid elements. Notation as in Figure~\ref{fig::doftrian}.}
 \label{fig::dofrect}
 \end{figure}\noindent
We write the volume integrals in \eqref{eq::DGdiscr}-\ref{eq::dgupdate} as $V_{N,\alpha} = \frac{1}{9} \vec{A}^{-1}( \vec{c} \bar{\vec{V}}_N\vec{q} )
$ with
\begin{eqnarray}\nonumber
\vec{A}^{-1} &=& 
 \left( \begin{array}{cccc}  
 4 & -2 & 1 & -2 \\
 -2 & 4 & -2 & 1 \\
 1 & -2 & 4 &  -2\\
  -2 & 1 & -2 & 4 
   \end{array} \right),
\end{eqnarray}
and
\begin{eqnarray}\nonumber
&&
\bar{\vec{V}}_{1} =
 \left( \begin{array}{cccc}  
 -2 & 4 & -2 & 4 \\
 -4 & 2 & -1 & 2 \\
 -2 & 1 & -2 &  1\\
  -1 & 2 & -4 & 2 
   \end{array} \right), \quad 
  \bar{\vec{V}}_{2}  =
    \left( \begin{array}{cccc}  
 2 & -4 & -1 & 2 \\
 4 & -2 & -2 & 4 \\
 2 & -2 & -4 &  2\\
  1 & -2 & -2 & 1 
   \end{array} \right), \\ \nonumber
   &&
   \bar{\vec{V}}_{3}  =
 \left( \begin{array}{cccc}  
 1 & -2 & 1 & -2 \\
 2 & -1 & 2 & -4 \\
 4 & -2 & 4 &  -2\\
  2 & -4 & 2 & -1 
   \end{array} \right),
   \quad
    \bar{\vec{V}}_{4}  =
 \left( \begin{array}{cccc}  
 -1 & 2 & -2 & -4 \\
 -2 & 1& 1 & -2 \\
 -4 & 2 & 2 &  -1\\
  -2 & 4 & 4 & -1 
   \end{array} \right).
\end{eqnarray}
The surface integrals are computed from $
E_{N,\alpha} = \frac{2}{3}\vec{A}^{-1} (\bar{\vec{E}}_N\vec{q} )
$, with 
\begin{eqnarray}\nonumber
\bar{\vec{E}}_1 &=& \left(0, \widetilde{c_{2,1}} + 2 \widetilde{c_{2,2}}, 0, 2 \widetilde{c_{4,1}} + \widetilde{c_{4,2}} \right) ,\\\nonumber
\bar{\vec{E}}_2 &=& \left(2 \widetilde{c_{1,1}} +  \widetilde{c_{1,2}}, 0, 0,  \widetilde{c_{4,1}} + 2 \widetilde{c_{4,2}} \right) ,\\\nonumber
\bar{\vec{E}}_3 &=& \left(\widetilde{c_{1,1}} +  2 \widetilde{c_{1,2}}, 0, 2 \widetilde{c_{3,1}} + \widetilde{c_{3,2}},0 \right), \\\nonumber
\bar{\vec{E}}_4 &=& \left(0, 2 \widetilde{c_{2,1}} +  \widetilde{c_{2,2}}, \widetilde{c_{3,1}} + 2 \widetilde{c_{3,2}},0 \right). \\\nonumber
\end{eqnarray}

\newpage
\subsection*{Tetrahedra}
 \begin{figure}[h!]
 \centering
\includegraphics[width=0.8\textwidth]{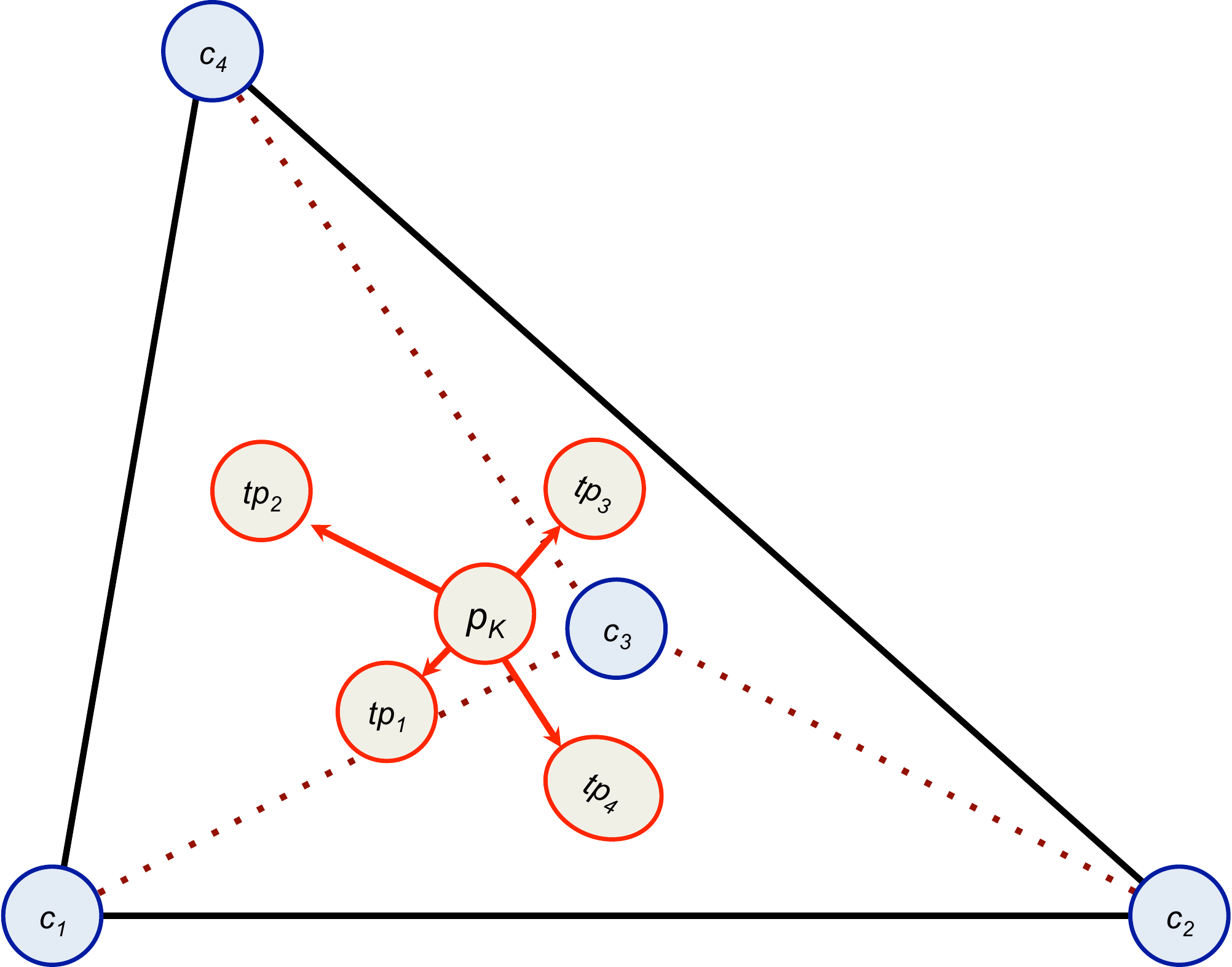} 
\caption{Degrees of freedom for MHFE-DG on \textbf{tetrahedral} grid elements. Notation as in Figure~\ref{fig::doftetra}.}
 \label{fig::doftetra}
 \end{figure}\noindent
For tetrahedra we find $V_{N,\alpha} = \vec{c} \bar{\vec{V}}_N\vec{q} $/3, with
\begin{eqnarray}\nonumber
&&
\bar{\vec{V}}_{1} =
 \left( \begin{array}{cccc}  
 2 & 2 & -3 & 2\\
 1 & 1 & -4 & 1\\
 1 & 1 & -4 & 1\\
 1 & 1 & -4 & 1\\
   \end{array} \right), \quad 
  \bar{\vec{V}}_{2}  =
    \left( \begin{array}{cccc}  
    1 & -4 & 1 & 1\\
    2 & -3 & 2 & 2\\
    1 & -4 & 1 & 1\\
    1 & -4 & 1 & 1\\
   \end{array} \right), \\ \nonumber
   &&
   \bar{\vec{V}}_{3}  =
 \left( \begin{array}{cccc}  
 -4 & 1 & 1 & 1 \\
 -4 & 1 & 1 & 1 \\
-3 & 2 & 2 & 2\\
 -4 & 1 & 1 & 1 \\
   \end{array} \right),
   \quad
    \bar{\vec{V}}_{4}  =
 \left( \begin{array}{cccc}  
 1 & 1 & 1 & -4 \\
 1 & 1 & 1 & -4 \\
 1 & 1 & 1 & -4 \\
2 & 2 & 2 & -3 
   \end{array} \right).
\end{eqnarray}
The surface integrals are given by $
E_{N,\alpha} =  \bar{\vec{E}}_N\vec{q}/3 
$, with 
\begin{eqnarray}\nonumber
\bar{\vec{E}}_1 &=& \left(
6 \widetilde{c_{1,1}} +  \widetilde{c_{1,2}} +  \widetilde{c_{1,3}} ,
6 \widetilde{c_{2,1}} +  \widetilde{c_{2,2}} +  \widetilde{c_{2,3}} ,
-4 [\widetilde{c_{3,1}} +  \widetilde{c_{3,2}} +  \widetilde{c_{3,3}}] ,
6 \widetilde{c_{4,1}} +  \widetilde{c_{4,2}} +  \widetilde{c_{4,3}}  \right), \\\nonumber
\bar{\vec{E}}_2 &=& \left(
6 \widetilde{c_{1,1}} +  \widetilde{c_{1,2}} +  \widetilde{c_{1,3}} ,
-4[ \widetilde{c_{2,1}} +  \widetilde{c_{2,2}} +  \widetilde{c_{2,3}} ],
6\widetilde{c_{3,1}} +  \widetilde{c_{3,2}} +  \widetilde{c_{3,3}} ,
6 \widetilde{c_{4,1}} +  \widetilde{c_{4,2}} +  \widetilde{c_{4,3}}  \right), \\\nonumber
\bar{\vec{E}}_3 &=& \left(
-4[ \widetilde{c_{1,1}} +  \widetilde{c_{1,2}} +  \widetilde{c_{1,3}}] ,
6 \widetilde{c_{2,1}} +   \widetilde{c_{2,2}} +  \widetilde{c_{2,3}} ,
6 \widetilde{c_{3,1}} +  \widetilde{c_{3,2}} +   \widetilde{c_{3,3}} ,
6 \widetilde{c_{4,1}} +  \widetilde{c_{4,2}} +  \widetilde{c_{4,3}}  \right), \\\nonumber
\bar{\vec{E}}_4 &=& \left(
6 \widetilde{c_{1,1}} +  \widetilde{c_{1,2}} +  \widetilde{c_{1,3}} ,
6 \widetilde{c_{2,1}} +   \widetilde{c_{2,2}} +  \widetilde{c_{2,3}} ,
6\widetilde{c_{3,1}} +  \widetilde{c_{3,2}} +  \widetilde{c_{3,3}},
-4[ \widetilde{c_{4,1}} +  \widetilde{c_{4,2}} +  \widetilde{c_{4,3}} ] \right) .
\end{eqnarray}

\newpage
\subsection*{Prismatic Elements}
 \begin{figure}[h!]
 \centering
\includegraphics[width=0.6\textwidth]{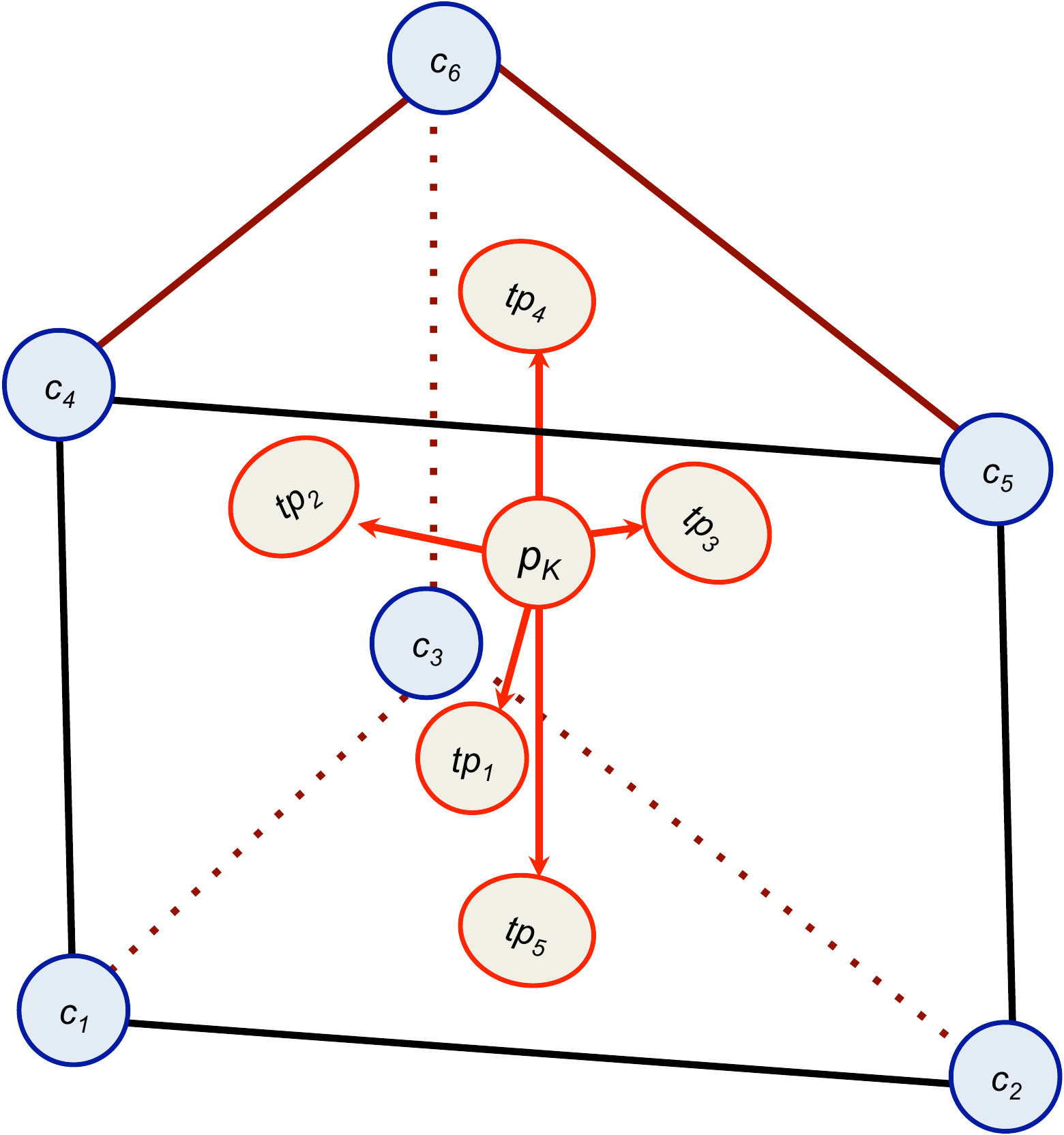} 
\caption{Degrees of freedom for MHFE-DG on \textbf{prismatic} grid elements. Notation as in Figure~\ref{fig::doftrian}.}
 \label{fig::dofprism}
 \end{figure}\noindent
Note that this is the first element type where the number of nodes and faces is not equal.

The volume integrals are $V_{N,\alpha} = \vec{c} \bar{\vec{V}}_N\vec{q}/2 $, with
\begin{eqnarray}\nonumber
&&
\bar{\vec{V}}_{1} =
 \left( \begin{array}{ccccc}  
2 & 2 & -2 & -2 & 4  \\
1 & 1 & -3 & 0 & 0 \\
1 & 1 & -3 & 0 & 0\\
  0 & 0 & 0 & -4 & 2\\
0 & 0 & 0 & 0 & 0\\
0 & 0 & 0 & 0 & 0
   \end{array} \right), \quad 
\bar{\vec{V}}_{2} =
 \left( \begin{array}{ccccc} 
 1 & -3 & 1 & 0 & 0\\
 2 & -2 & 2 & -2 & 4\\
 1 & -3 & 1 & 0 & 0 \\
 0 & 0 & 0 & 0 & 0 \\
 0 & 0 & 0 & -4 & 2\\
 0 & 0 & 0 & 0 & 0
   \end{array} \right), \quad 
   \bar{\vec{V}}_{3} =
 \left( \begin{array}{ccccc} 
 -3 & 1 & 1 & 0 & 0 \\
 -3 & 1 & 1 & 0 & 0 \\
- 2 & 2 & 2 & -2 & 4\\
0 & 0 & 0 & 0 & 0\\ 
0 & 0 & 0 & 0 & 0\\ 
0 & 0 & 0 & -4 & 2\\ 
   \end{array} \right), 
   \\\nonumber
   &&
\bar{\vec{V}}_{4} =
 \left( \begin{array}{ccccc}  
 0 & 0 & 0 & 2 & -4\\
 0 & 0 & 0 & 0 & 0\\ 
0 & 0 & 0 & 0 & 0\\ 
2 & 2 & -2 & 4 & -2\\
1 & 1& -3 & 0 & 0 \\
1 & 1 & -3 & 0 & 0 \\
   \end{array} \right), \quad 
\bar{\vec{V}}_{5} =
 \left( \begin{array}{ccccc}  
 0 & 0 & 0 & 0 & 0\\ 
 0 & 0 & 0 & 2 & -4\\
  0 & 0 & 0 & 0 & 0\\ 
1 & -3 & 1 & 0 & 0 \\
2 & -2 & 2 & 4 & -2 \\
1 & -3 & 1 & 0 & 0 \\
   \end{array} \right), \quad 
   \bar{\vec{V}}_{6} =
 \left( \begin{array}{ccccc}  
  0 & 0 & 0 & 0 & 0\\ 
 0 & 0 & 0 & 0 & 0\\ 
 0 & 0 & 0 & 2 & -4\\
 -3 & 1 & 1 & 0 & 0\\
  -3 & 1 & 1 & 0 & 0\\
 2 & -2 & -2 & -4 & 2 \\
    \end{array} \right), 
\end{eqnarray}
And the surface integrals satisfy $
E_{N,\alpha} =  \bar{\vec{E}}_N\vec{q}/2 
$, with 
\begin{eqnarray}\nonumber
\bar{\vec{E}}_1 &=& \left(
5 \widetilde{c_{1,1}} +  \widetilde{c_{1,3}} ,
5 \widetilde{c_{2,1}} +  \widetilde{c_{2,3}}  ,
-3 [\widetilde{c_{3,1}} +  \widetilde{c_{3,3}} ] ,
-4 \widetilde{c_{4,1}} ,
8  \widetilde{c_{5,1}}  \right), \\\nonumber
\bar{\vec{E}}_2 &=& \left(
 \widetilde{c_{1,1}} + 5 \widetilde{c_{1,3}} ,
-3[ \widetilde{c_{2,1}} + 5 \widetilde{c_{2,3}} ] ,
5 \widetilde{c_{3,1}} +  \widetilde{c_{3,3}}  ,
-4 \widetilde{c_{4,2}} ,
8  \widetilde{c_{5,2}}  \right), \\\nonumber
\bar{\vec{E}}_3 &=& \left(
-3[ \widetilde{c_{1,1}} +  \widetilde{c_{1,3}} ],
 \widetilde{c_{2,1}} + 5 \widetilde{c_{2,3}}  ,
 \widetilde{c_{3,1}} +  5\widetilde{c_{3,3}}  ,
-4 \widetilde{c_{4,3}} ,
8 \widetilde{c_{5,3}}  \right), \\\nonumber
\bar{\vec{E}}_4 &=& \left(
5 \widetilde{c_{1,2}} +  \widetilde{c_{1,4}} ,
5 \widetilde{c_{2,2}} +  \widetilde{c_{2,4}}  ,
-3 [\widetilde{c_{3,2}} +  \widetilde{c_{3,4}} ] ,
8 \widetilde{c_{4,1}} ,
-4  \widetilde{c_{5,1}}  \right), \\\nonumber
\bar{\vec{E}}_5 &=& \left(
 \widetilde{c_{1,2}} + 5 \widetilde{c_{1,4}} ,
-3[ \widetilde{c_{2,2}} + 5 \widetilde{c_{2,4}} ] ,
5 \widetilde{c_{3,2}} +  \widetilde{c_{3,4}}  ,
8 \widetilde{c_{4,3}} ,
-4  \widetilde{c_{5,3}}  \right), \\\nonumber
\bar{\vec{E}}_6 &=& \left(
-3[ \widetilde{c_{1,2}} +  \widetilde{c_{1,4}} ],
 \widetilde{c_{2,2}} + 5 \widetilde{c_{2,4}}  ,
 \widetilde{c_{3,2}} +  5\widetilde{c_{3,4}}  ,
8 \widetilde{c_{4,3}} ,
-4 \widetilde{c_{5,3}}  \right).
\end{eqnarray}

\subsection*{Hexahedra}
 \begin{figure}[h!]
 \centering
\includegraphics[width=0.8\textwidth]{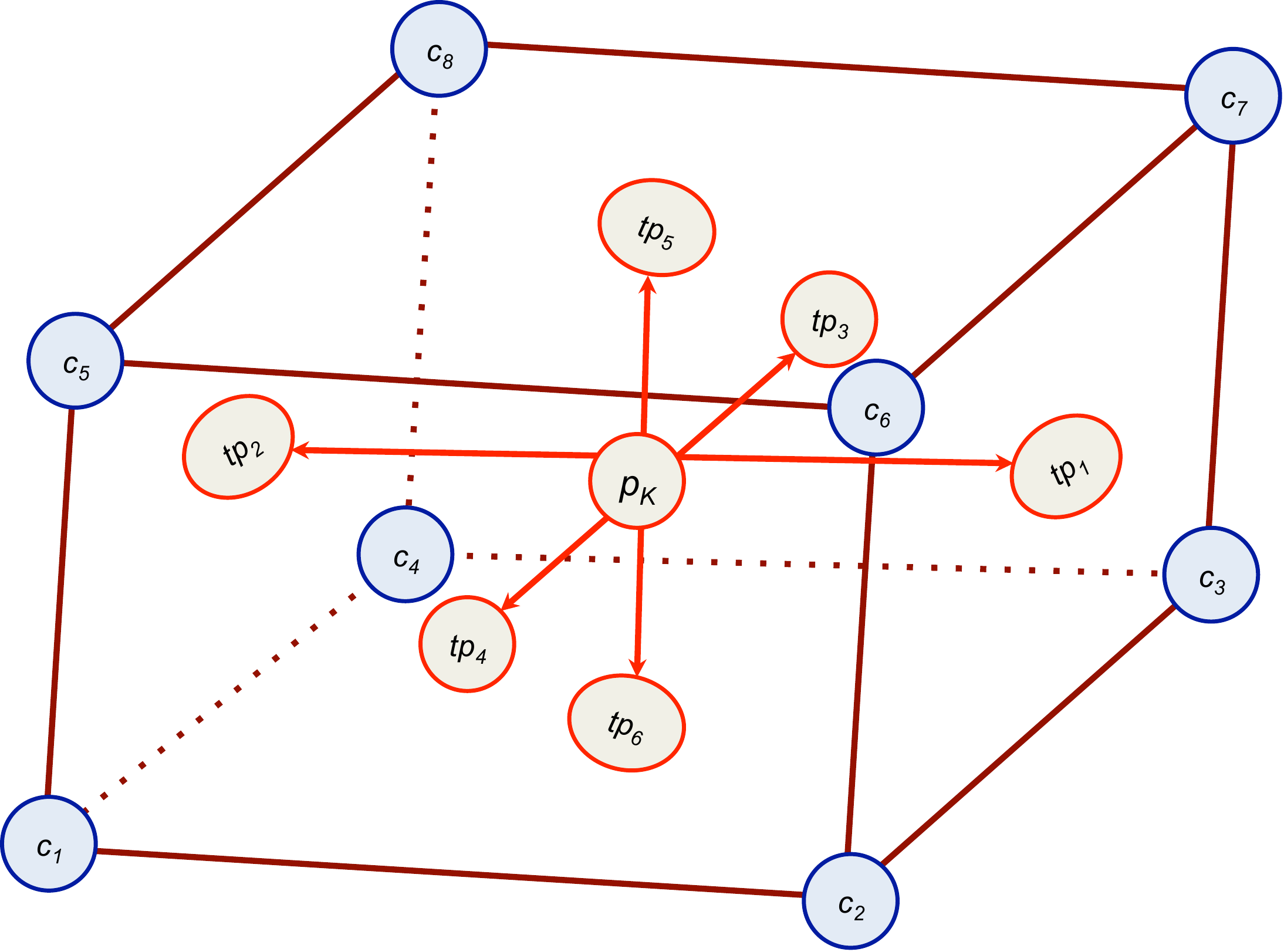} 
\caption{Degrees of freedom for MHFE-DG on \textbf{hexahedral} grid elements. Notation as in Figure~\ref{fig::doftrian}.}
 \label{fig::dofcube}
 \end{figure}
The volume integrals are given by $V_{N,\alpha} = \vec{c} \bar{\vec{V}}_N\vec{q} $ with $N = 1, \ldots 8$ and six edges $E$.

\begin{eqnarray}\nonumber
&&
\bar{\vec{V}}_{1} =
 \left( \begin{array}{cccccc}  
 -1 & 2 & -1 & 2 & -1 & 2 \\
 -2 & 1 & 0 & 0 & 0 & 0 \\
 0 & 0 & 0 & 0 & 0 & 0 \\
 0 & 0 & -2 & 1 & 0 & 0 \\
  0 & 0 & 0 & 0 & -2 & 1 \\
 0 & 0 & 0 & 0 & 0 & 0 \\
 0 & 0 & 0 & 0 & 0 & 0 \\
 0 & 0 & 0 & 0 & 0 & 0 
   \end{array} \right), \quad 
   \bar{\vec{V}}_{2} =
 \left( \begin{array}{cccccc}  
1 & -2 & 0 & 0 & 0 & 0 \\
2 & -1 & -1 & 2 & -1 & 2\\
0 & 0 & -2 & 1 & 0 &  0\\
 0 & 0 & 0 & 0 & 0 & 0 \\
 0 & 0 & 0 & 0 & 0 & 0 \\
  0 & 0 & 0 & 0 & -2 & 1\\
 0 & 0 & 0 & 0 & 0 & 0 \\
 0 & 0 & 0 & 0 & 0 & 0 
    \end{array} \right), \\\nonumber
    &&
\bar{\vec{V}}_{3} =
 \left( \begin{array}{cccccc}  
  0 & 0 & 0 & 0 & 0 & 0 \\
0 & 0 & 1 & -2 & 0 & 0 \\
2 & -1 & 2 & -1 & -1 & 2 \\
1 & -2 & 0 & 0 & 0 & 0 \\
 0 & 0 & 0 & 0 & 0 & 0 \\
 0 & 0 & 0 & 0 & 0 & 0 \\
0 & 0 & 0 & 0 & -2 & 1\\
 0 & 0 & 0 & 0 & 0 & 0 
   \end{array} \right), \quad 
   \bar{\vec{V}}_{4} =
 \left( \begin{array}{cccccc}  
 0 & 0 & 1 & -2 & 0 & 0\\
  0 & 0 & 0 & 0 & 0 & 0 \\
-2 & 1 & 0 & 0 & 0 & 0 \\
-1 & 2 & 2 & - 1 & -1 & 2 \\
  0 & 0 & 0 & 0 & 0 & 0 \\
  0 & 0 & 0 & 0 & 0 & 0 \\
  0 & 0 & 0 & 0 & 0 & 0 \\
 0 & 0 & 0 & 0 & -2 & 1 
    \end{array} \right), \\\nonumber
    &&
\bar{\vec{V}}_{5} =
 \left( \begin{array}{cccccc}  
   0 & 0 & 0 & 0 &1 & -2\\
    0 & 0 & 0 & 0 & 0 & 0 \\
  0 & 0 & 0 & 0 & 0 & 0 \\
  0 & 0 & 0 & 0 & 0 & 0 \\
-1 & 2 &-1 & 2 & 2 & -1\\
 -2 & 1 & 0 & 0 & 0 & 0 \\
   0 & 0 & 0 & 0 & 0 & 0 \\
0 & 0 & -2 & 1 & 0 & 0
   \end{array} \right), \quad 
   \bar{\vec{V}}_{6} =
 \left( \begin{array}{cccccc} 
   0 & 0 & 0 & 0 & 0 & 0 \\
   0 & 0 & 0 & 0 & 1 & -2\\
    0 & 0 & 0 & 0 & 0 & 0 \\
  0 & 0 & 0 & 0 & 0 & 0 \\
  1 & -2  & 0 & 0 & 0 & 0\\
 2 & -1 & -1 & 2 & 2 & -1 \\
 0 & 0 & -2 & 1 & 0 & 0\\
    0 & 0 & 0 & 0 & 0 & 0 
    \end{array} \right), \\\nonumber
    &&
\bar{\vec{V}}_{7} =
 \left( \begin{array}{cccccc}  
    0 & 0 & 0 & 0 & 0 & 0 \\
  0 & 0 & 0 & 0 & 0 & 0 \\
   0 & 0 & 0 & 0 & {\color{black}\mathbf{1}} & {\color{black}-2} \\
      0 & 0 & 0 & 0 & 0 & 0 \\
  0 & 0 & 0 & 0 & 0 & 0 \\
  0 & 0 & 1 & -2 & 0 & 0 \\
  2 & -1 & 2 & -1 & 2 & -1 \\
  1 & -2 & 0 & 0 & 0 & 0
   \end{array} \right), \quad 
   \bar{\vec{V}}_{8} =
 \left( \begin{array}{cccccc} 
   0 & 0 & 0 & 0 & 0 & 0 \\
      0 & 0 & 0 & 0 & 0 & 0 \\
   0 & 0 & 0 & 0 & 0 & 0 \\
   0 & 0 & 0 & 0 &1 & -2 \\
   0 & 0 & 1 & -2 & 0 & 0 \\
      0 & 0 & 0 & 0 &   0 & 0 \\
   -2 & 1 & 0 & 0 &   0 & 0 \\
-1 & 2 & 2 & -1 & 2 & -1
    \end{array} \right).
    \end{eqnarray}
The surface integrals are calculated from $
E_{N,\alpha} =  2 \bar{\vec{E}}_N\vec{q} 
$ with 
\begin{eqnarray}\nonumber
\bar{\vec{E}}_1 &=& \left(
- \widetilde{c_{1,1}},
2 \widetilde{c_{2,1}},
{\color{black}- \widetilde{c_{3,2}}},
2\widetilde{c_{4,1}},
-\widetilde{c_{5,1}},
2 \widetilde{c_{6,1}} \right),\\\nonumber
\bar{\vec{E}}_2 &=& \left(
2\widetilde{c_{1,1}},
- \widetilde{c_{2,1}},
- \widetilde{c_{3,1}},
2\widetilde{c_{4,2}},
-\widetilde{c_{5,2}},
2 \widetilde{c_{6,2}} \right),\\\nonumber
\bar{\vec{E}}_3 &=& \left(
2\widetilde{c_{1,2}},
- \widetilde{c_{2,2}},
2 \widetilde{c_{3,1}},
-\widetilde{c_{4,2}},
-\widetilde{c_{5,3}},
2 \widetilde{c_{6,3}} \right),\\\nonumber
\bar{\vec{E}}_4 &=& \left(
-\widetilde{c_{1,2}},
2 \widetilde{c_{2,2}},
2 \widetilde{c_{3,2}},
-\widetilde{c_{4,1}},
-\widetilde{c_{5,4}},
2 \widetilde{c_{6,4}} \right),\\\nonumber
\bar{\vec{E}}_5 &=& \left(
-\widetilde{c_{1,3}},
2 \widetilde{c_{2,3}},
-\widetilde{c_{3,4}},
\widetilde{c_{4,3}},
2\widetilde{c_{5,1}},
- \widetilde{c_{6,1}} \right),\\\nonumber
\bar{\vec{E}}_6 &=& \left(
2\widetilde{c_{1,3}},
-\widetilde{c_{2,3}},
-\widetilde{c_{3,3}},
2\widetilde{c_{4,4}},
2\widetilde{c_{5,2}},
- \widetilde{c_{6,2}} \right),\\\nonumber
\bar{\vec{E}}_7 &=& \left(
2\widetilde{c_{1,4}},
-\widetilde{c_{2,4}},
2\widetilde{c_{3,3}},
-\widetilde{c_{4,4}},
2\widetilde{c_{5,3}},
- \widetilde{c_{6,3}} \right),\\\nonumber
\bar{\vec{E}}_8 &=& \left(
-\widetilde{c_{1,4}},
2\widetilde{c_{2,4}},
2\widetilde{c_{3,4}},
-\widetilde{c_{4,3}},
2\widetilde{c_{5,4}},
- \widetilde{c_{6,4}} \right).\\\nonumber
\end{eqnarray}
}

	 \begin{table}[htdp]
	 \caption{\label{table::NNNMNF}Example~1: Grids.}	
	 \begin{center}
	 \begin{tabular*}{\hsize}{@{\extracolsep{\fill}}lrrr}\hline
	 Gird &  $N_K$ & $N_N$ & $N_E$\\\hline
	Hexahedra		& $9,000$	 &   $10,659$	&         $28,580$	\\
	Prisms			& $11,240$	&   $6,754$	&         $29,734$	\\
	Tetrahedra 1		& $9,000$	 &    $2,046$	&         $19,000$	\\
	Tetrahedra 2		& $10,467$	&   $2,611$	&         $22,670$	\\
	Tetrahedra 3		& $36,000	$ &   $7,216$	&        $74,300$	\\
	 \hline
	 \end{tabular*}
	 \end{center}
	 \end{table}

	 	 \begin{table}[htdp]
	 \caption{\label{table::ex3wathexa}{\color{black}Example~4}: Convergence Analysis on Hexahedra.}	
	 \begin{center}
	 \begin{tabular*}{\hsize}{@{\extracolsep{\fill}}rrrrrrr}\hline
	 $N_K$ 			& $N_N$ 		& $N_E$ 		& $h$ (mm)	& $L_1$ error ($\times 10^{-3}$) & Convergence rate $p$ & CPU time \\\hline
	$245,760$		& $245,760$	&   $753,664$	& $1.25$		&         - 	& - 		& 36 min		\\
	$30,720$			& $34,969$	&   $96,256$	& $2.50$		&     2.55	& - 		& 3 min		\\
	$3,840$			& $4,941$		&   $12,544$	& $5.00$		&     7.25	& 1.51  	& 14 sec		\\
	$480$			& $775$		&   $1,696$	& $10.00$		&     17.54& 1.39 	& 1 sec		\\
	$60$				& $144$		&   $244$		& $20.00$		&     36.32& 1.28 	& $<1$ sec	\\\hline
	 \end{tabular*}
	 \end{center}
	 \end{table}

	 	 \begin{table}[htdp]
	 \caption{\label{table::ex3wattetra}{\color{black}Example~4}: Convergence Analysis on Tetrahedra.}	
	 \begin{center}
	 \begin{tabular*}{\hsize}{@{\extracolsep{\fill}}rrrrrrr}\hline
	 $N_K$ 			& $N_N$ 		& $N_E$ 		& $h$ (mm)	& $L_1$ error ($\times 10^{-3}$)& Convergence rate $p$ & CPU time \\\hline
	$184,320$ 		& $34,969$	&   $376,832$	& $1.376$		&         - 	& - 	   & 3 min		\\	
	$23,040$			& $4,941$		&   $48,128$	& $2.752$		&     1.89	& - 	   & 14 sec		\\
	$2,880$			& $775$		&   $6,272$	& $5.503$		&     4.92 	& 1.38  & 8 sec		\\
	$360$			& $144$		&   $848$		& $11.007$	&    16.90 	& 1.58  & $<1$ sec	\\
	 \hline
	 \end{tabular*}
	 \end{center}
	 \end{table}
 
 \clearpage

  \clearpage

   \clearpage

\clearpage

\section*{References}
\begin{thebibliography}{84}
\expandafter\ifx\csname natexlab\endcsname\relax\def\natexlab#1{#1}\fi
\providecommand{\bibinfo}[2]{#2}
\ifx\xfnm\relax \def\xfnm[#1]{\unskip,\space#1}\fi
\bibitem[{Fung et~al.(2013)Fung, Ding, and Dogru}]{larry2013}
\bibinfo{author}{L.~S. Fung}, \bibinfo{author}{X.~Y. Ding},
  \bibinfo{author}{A.~H. Dogru},
\newblock \bibinfo{title}{Using unstructured grids for modeling densely-spaced
  complex wells in field-scale reservoir simulation},
\newblock in: \bibinfo{booktitle}{Paper SPE-17062-MS presented at the 6th
  International Petroleum Technology Conference, Beijing, China, Mar 26-28
  (2013)}, pp. \bibinfo{pages}{1--14}.
\bibitem[{Fung and Dogru(2008)}]{larry2008}
\bibinfo{author}{L.~S.~K. Fung}, \bibinfo{author}{A.~H. Dogru},
\newblock \bibinfo{title}{Parallel unstructured-solver methods for simulation
  of complex giant reservoirs},
\newblock \bibinfo{journal}{{SPE J.}} \bibinfo{volume}{{13}}
  (\bibinfo{year}{{2008}}) \bibinfo{pages}{{440--446}}.
\bibitem[{Aavatsmark et~al.(1996)Aavatsmark, Barkve, B{\o}e, and
  Mannseth}]{Aavatsmark19962}
\bibinfo{author}{I.~Aavatsmark}, \bibinfo{author}{T.~Barkve},
  \bibinfo{author}{{\O}.~B{\o}e}, \bibinfo{author}{T.~Mannseth},
\newblock \bibinfo{title}{Discretization on non-orthogonal, quadrilateral grids
  for inhomogeneous, anisotropic media},
\newblock \bibinfo{journal}{J. Comput. Phys.} \bibinfo{volume}{127}
  (\bibinfo{year}{1996}) \bibinfo{pages}{2 -- 14}.
\bibitem[{Wu and Parashkevov(2009)}]{wu2009}
\bibinfo{author}{X.~H. Wu}, \bibinfo{author}{R.~R. Parashkevov},
\newblock \bibinfo{title}{Effect of grid deviation on flow solutions},
\newblock \bibinfo{journal}{{SPE J.}} \bibinfo{volume}{{14}}
  (\bibinfo{year}{{2009}}) \bibinfo{pages}{{67--77}}.
\bibitem[{Aavatsmark et~al.(1998)Aavatsmark, Barkve, B{\o}e, and
  Mannseth}]{Aavatsmark1998}
\bibinfo{author}{I.~Aavatsmark}, \bibinfo{author}{T.~Barkve},
  \bibinfo{author}{{\O}.~B{\o}e}, \bibinfo{author}{T.~Mannseth},
\newblock \bibinfo{title}{{Discretization on Unstructured Grids for
  Inhomogeneous, Anisotropic Media. Part I: Derivation of the Methods}},
\newblock \bibinfo{journal}{{SIAM J. Sci. Comput.}} \bibinfo{volume}{{19}}
  (\bibinfo{year}{{1998}}) \bibinfo{pages}{{1700--1716}}.
\bibitem[{Edwards(2002)}]{Edwards2002}
\bibinfo{author}{M.~G. Edwards},
\newblock \bibinfo{title}{Unstructured, control-volume distributed, full-tensor
  finite volume schemes with flow based grids},
\newblock \bibinfo{journal}{Comput. Geosci.} \bibinfo{volume}{6}
  (\bibinfo{year}{2002}) \bibinfo{pages}{433--452}.
\bibitem[{Edwards and Zheng(2010)}]{Edwards2010}
\bibinfo{author}{M.~G. Edwards}, \bibinfo{author}{H.~Zheng},
\newblock \bibinfo{title}{Quasi-positive families of continuous {Darcy}-flux
  finite volume schemes on structured and unstructured grids},
\newblock \bibinfo{journal}{{J. Comput. Appl. Math.}} \bibinfo{volume}{{234}}
  (\bibinfo{year}{{2010}}) \bibinfo{pages}{{2152--2161}}.
\bibitem[{Edwards et~al.(2011)Edwards, Zheng, Lamine, and
  Pal}]{Edwards2011tensor}
\bibinfo{author}{M.~G. Edwards}, \bibinfo{author}{H.~Zheng},
  \bibinfo{author}{S.~Lamine}, \bibinfo{author}{M.~Pal},
\newblock \bibinfo{title}{Continuous elliptic and multi-dimensional hyperbolic
  {Darcy}-flux finite-volume methods},
\newblock \bibinfo{journal}{{Comput. Fluids}} \bibinfo{volume}{{46}}
  (\bibinfo{year}{{2011}}) \bibinfo{pages}{{12--22}}.
\bibitem[{Edwards and Zheng(2011)}]{Edwards2011tensorb}
\bibinfo{author}{M.~G. Edwards}, \bibinfo{author}{H.~Zheng},
\newblock \bibinfo{title}{Quasi m-matrix multifamily continuous {Darcy}-flux
  approximations with full pressure support on structured and unstructured
  grids in three dimensions},
\newblock \bibinfo{journal}{{SIAM J. Sci. Comput.}} \bibinfo{volume}{{33}}
  (\bibinfo{year}{{2011}}) \bibinfo{pages}{{455--487}}.
\bibitem[{Lamine and Edwards(2013)}]{Sadok2013}
\bibinfo{author}{S.~Lamine}, \bibinfo{author}{M.~G. Edwards},
\newblock \bibinfo{title}{Higher order cell-based multidimensional upwind
  schemes for flow in porous media on unstructured grids},
\newblock \bibinfo{journal}{{Comput. Methods in Appl. M.}}
  \bibinfo{volume}{{259}} (\bibinfo{year}{{2013}}) \bibinfo{pages}{{103--122}}.
\bibitem[{Lamine and Edwards(2010)}]{Sadok2010}
\bibinfo{author}{S.~Lamine}, \bibinfo{author}{M.~G. Edwards},
\newblock \bibinfo{title}{Multidimensional upwind convection schemes for flow
  in porous media on structured and unstructured quadrilateral grids},
\newblock \bibinfo{journal}{{J. Comput. Appl. Math.}} \bibinfo{volume}{{234}}
  (\bibinfo{year}{{2010}}) \bibinfo{pages}{{2106--2117}}.
\bibitem[{Eymard et~al.(2010)Eymard, Gallou\"et, and Herbin}]{Eymard4}
\bibinfo{author}{R.~Eymard}, \bibinfo{author}{T.~Gallou\"et},
  \bibinfo{author}{R.~Herbin},
\newblock \bibinfo{title}{Discretization of heterogeneous and anisotropic
  diffusion problems on general nonconforming meshes sushi: a scheme using
  stabilization and hybrid interfaces},
\newblock \bibinfo{journal}{IMA J. Numer. Anal.} \bibinfo{volume}{30}
  (\bibinfo{year}{2010}) \bibinfo{pages}{1009--1043}.
\bibitem[{Droniou and Eymard(2006)}]{Eymard5}
\bibinfo{author}{J.~Droniou}, \bibinfo{author}{R.~Eymard},
\newblock \bibinfo{title}{A mixed finite volume scheme for anisotropic
  diffusion problems on any grid},
\newblock \bibinfo{journal}{Numer. Math.} \bibinfo{volume}{105}
  (\bibinfo{year}{2006}) \bibinfo{pages}{35--71}.
\bibitem[{Karimi-Fard et~al.(2004)Karimi-Fard, Durlofsky, and
  Aziz}]{karimifractures2004}
\bibinfo{author}{M.~Karimi-Fard}, \bibinfo{author}{L.~Durlofsky},
  \bibinfo{author}{K.~Aziz},
\newblock \bibinfo{title}{An efficient discrete-fracture model applicable for
  general-purpose reservoir simulators},
\newblock \bibinfo{journal}{{SPE J.}} \bibinfo{volume}{{9}}
  (\bibinfo{year}{{2004}}) \bibinfo{pages}{{227--236}}.
\bibitem[{Monteagudo and Firoozabadi(2007)}]{Monteagudo2007}
\bibinfo{author}{J.~E.~P. Monteagudo}, \bibinfo{author}{A.~Firoozabadi},
\newblock \bibinfo{title}{Control-volume model for simulation of water
  injection in fractured media: Incorporating matrix heterogeneity and
  reservoir wettability effects},
\newblock \bibinfo{journal}{{SPE J.}} \bibinfo{volume}{{12}}
  (\bibinfo{year}{{2007}}) \bibinfo{pages}{{355--366}}.
\bibitem[{Schmid et~al.(2013)Schmid, Geiger, and Sorbie}]{Geiger2013}
\bibinfo{author}{K.~S. Schmid}, \bibinfo{author}{S.~Geiger},
  \bibinfo{author}{K.~S. Sorbie},
\newblock \bibinfo{title}{Higher order {FE-FV} method on unstructured grids for
  transport and two-phase flow with variable viscosity in heterogeneous porous
  media},
\newblock \bibinfo{journal}{{J. Comput. Phys.}} \bibinfo{volume}{{241}}
  (\bibinfo{year}{{2013}}) \bibinfo{pages}{{416--444}}.
\bibitem[{Aavatsmark(2002)}]{AavatsmarkIntro}
\bibinfo{author}{I.~Aavatsmark},
\newblock \bibinfo{title}{An introduction to multipoint flux approximations for
  quadrilateral grids},
\newblock \bibinfo{journal}{{Comput. Geosci.}} \bibinfo{volume}{{6}}
  (\bibinfo{year}{{2002}}) \bibinfo{pages}{{405--432}}.
\bibitem[{Kozdon et~al.(2011)Kozdon, Mallison, Gerritsen, and
  Chen}]{kozdon2011}
\bibinfo{author}{J.~Kozdon}, \bibinfo{author}{B.~Mallison},
  \bibinfo{author}{M.~Gerritsen}, \bibinfo{author}{W.~Chen},
\newblock \bibinfo{title}{Multidimensional upwinding for multiphase transport
  in porous media},
\newblock \bibinfo{journal}{{SPE J.}} \bibinfo{volume}{{16}}
  (\bibinfo{year}{{2011}}) \bibinfo{pages}{{263--272}}.
\bibitem[{Batista~Fernandes et~al.(2013)Batista~Fernandes, Marcondes, and
  Sepehrnoori}]{Sepehrnoori2013}
\bibinfo{author}{B.~R. Batista~Fernandes}, \bibinfo{author}{F.~Marcondes},
  \bibinfo{author}{K.~Sepehrnoori},
\newblock \bibinfo{title}{Investigation of several interpolation functions for
  unstructured meshes in conjunction with compositional reservoir simulation},
\newblock \bibinfo{journal}{{Numer. Heat Tr. A-Appl}} \bibinfo{volume}{{64}}
  (\bibinfo{year}{{2013}}) \bibinfo{pages}{{974--993}}.
\bibitem[{Nordbotten et~al.(2007)Nordbotten, Aavatsmark, and
  Eigestad}]{Nordbotten2007}
\bibinfo{author}{J.~Nordbotten}, \bibinfo{author}{I.~Aavatsmark},
  \bibinfo{author}{G.~Eigestad},
\newblock \bibinfo{title}{Monotonicity of control volume methods},
\newblock \bibinfo{journal}{Num. Math.} \bibinfo{volume}{106}
  (\bibinfo{year}{2007}) \bibinfo{pages}{255--288}.
\bibitem[{Aavatsmark et~al.(2008)Aavatsmark, Eigestad, Mallison, and
  Nordbotten}]{AavatsmarkMonot}
\bibinfo{author}{I.~Aavatsmark}, \bibinfo{author}{G.~T. Eigestad},
  \bibinfo{author}{B.~T. Mallison}, \bibinfo{author}{J.~M. Nordbotten},
\newblock \bibinfo{title}{A compact multipoint flux approximation method with
  improved robustness},
\newblock \bibinfo{journal}{{Numer. Meth. Part D E}} \bibinfo{volume}{{24}}
  (\bibinfo{year}{{2008}}) \bibinfo{pages}{{1329--1360}}.
\bibitem[{Younes et~al.(2013)Younes, Fahs, and Belfort}]{Younes2013}
\bibinfo{author}{A.~Younes}, \bibinfo{author}{M.~Fahs},
  \bibinfo{author}{B.~Belfort},
\newblock \bibinfo{title}{Monotonicity of the cell-centred triangular {MPFA}
  method for saturated and unsaturated flow in heterogeneous porous media},
\newblock \bibinfo{journal}{{J. Hydro.}} \bibinfo{volume}{{504}}
  (\bibinfo{year}{{2013}}) \bibinfo{pages}{{132--141}}.
\bibitem[{Aavatsmark et~al.(2010)Aavatsmark, Eigestad, Heimsund, Mallison,
  Nordbotten, and Qian}]{AavatsmarkConvergence}
\bibinfo{author}{I.~Aavatsmark}, \bibinfo{author}{G.~T. Eigestad},
  \bibinfo{author}{B.~O. Heimsund}, \bibinfo{author}{B.~T. Mallison},
  \bibinfo{author}{J.~M. Nordbotten}, \bibinfo{author}{E.~Qian},
\newblock \bibinfo{title}{A new finite-volume approach to efficient
  discretization on challenging grids},
\newblock \bibinfo{journal}{{SPE J}} \bibinfo{volume}{{15}}
  (\bibinfo{year}{{2010}}) \bibinfo{pages}{{658--669}}.
\bibitem[{Matringe et~al.(2009)Matringe, Juanes, and Tchelepi}]{matringeMPFA}
\bibinfo{author}{S.~F. Matringe}, \bibinfo{author}{R.~Juanes},
  \bibinfo{author}{H.~A. Tchelepi},
\newblock \bibinfo{title}{A new mixed finite element and its related finite
  volume discretization on general hexahedral grids},
\newblock \bibinfo{journal}{Mech. Sol. Struct. Fluids} \bibinfo{volume}{12}
  (\bibinfo{year}{2009}) \bibinfo{pages}{77--87}.
\bibitem[{Salama et~al.(2013)Salama, Sun, and El~Amin}]{Amgad}
\bibinfo{author}{A.~Salama}, \bibinfo{author}{S.~Sun}, \bibinfo{author}{M.~F.
  El~Amin},
\newblock \bibinfo{title}{A multipoint flux approximation of the steady-state
  heat conduction equation in anisotropic media},
\newblock \bibinfo{journal}{{J. Heat Trans.}} \bibinfo{volume}{{135}}
  (\bibinfo{year}{{2013}}) \bibinfo{pages}{041302}.
\bibitem[{Ag\'elas and Masson(2008)}]{Eymard2}
\bibinfo{author}{L.~Ag\'elas}, \bibinfo{author}{R.~Masson},
\newblock \bibinfo{title}{Convergence of the finite volume {MPFA O} scheme for
  heterogeneous anisotropic diffusion problems on general meshes},
\newblock \bibinfo{journal}{Comptes Rendus Mathematique} \bibinfo{volume}{346}
  (\bibinfo{year}{2008}) \bibinfo{pages}{1007--1012}.
\bibitem[{Eymard et~al.(2012)Eymard, Guichard, Herbin, and Masson}]{Eymard1}
\bibinfo{author}{R.~Eymard}, \bibinfo{author}{C.~Guichard},
  \bibinfo{author}{R.~Herbin}, \bibinfo{author}{R.~Masson},
\newblock \bibinfo{title}{Vertex-centred discretization of multiphase
  compositional {D}arcy flows on general meshes},
\newblock \bibinfo{journal}{Comput. Geosci.} \bibinfo{volume}{16}
  (\bibinfo{year}{2012}) \bibinfo{pages}{987--1005}.
\bibitem[{Jackson et~al.(2013)Jackson, Gomes, Mostaghimi, Percival, Tollit,
  Pavlidis, Pain, El-Sheikh, Muggeridge, and Blunt}]{peyman}
\bibinfo{author}{M.~D. Jackson}, \bibinfo{author}{J.~L. Gomes},
  \bibinfo{author}{P.~Mostaghimi}, \bibinfo{author}{J.~Percival},
  \bibinfo{author}{B.~S. Tollit}, \bibinfo{author}{D.~Pavlidis},
  \bibinfo{author}{C.~C. Pain}, \bibinfo{author}{A.~H. El-Sheikh},
  \bibinfo{author}{A.~H. Muggeridge}, \bibinfo{author}{M.~J. Blunt},
\newblock \bibinfo{title}{Reservoir modeling for flow simulation using
  surfaces, adaptive unstructured meshes, and control-volume-finite-element
  methods},
\newblock in: \bibinfo{booktitle}{Paper SPE-163633-MS presented at the SPE
  Reservoir Simulation Symposium, 18-20 February (2013), The Woodlands, Texas},
  pp. \bibinfo{pages}{1--20}.
\bibitem[{Arnold(1982)}]{arnold82}
\bibinfo{author}{D.~Arnold},
\newblock \bibinfo{title}{An interior penalty finite element method with
  discontinuous elements},
\newblock \bibinfo{journal}{SIAM J. Numer. Anal.} \bibinfo{volume}{19}
  (\bibinfo{year}{1982}) \bibinfo{pages}{742--760}.
\bibitem[{Babu\u{s}ka(1973)}]{babuska73}
\bibinfo{author}{I.~Babu\u{s}ka},
\newblock \bibinfo{title}{The finite element method with penalty},
\newblock \bibinfo{journal}{Math. Comp.} \bibinfo{volume}{27}
  (\bibinfo{year}{1973}) \bibinfo{pages}{221--228}.
\bibitem[{Babu\u{s}ka and Zlamal(1973)}]{babuskazlamal73}
\bibinfo{author}{I.~Babu\u{s}ka}, \bibinfo{author}{M.~Zlamal},
\newblock \bibinfo{title}{Nonconforming elements in the finite element method
  with penalty},
\newblock \bibinfo{journal}{SIAM J. Numer. Anal.} \bibinfo{volume}{10}
  (\bibinfo{year}{1973}) \bibinfo{pages}{863--875}.
\bibitem[{Wheeler(1978)}]{wheeler}
\bibinfo{author}{M.~F. Wheeler},
\newblock \bibinfo{title}{An elliptic collocation finite element method with
  interior penalties},
\newblock \bibinfo{journal}{SIAM J. Numer. Anal.} \bibinfo{volume}{15}
  (\bibinfo{year}{1978}) \bibinfo{pages}{152--161}.
\bibitem[{Brezzi and Fortin(1991)}]{brezzifortini91}
\bibinfo{author}{F.~Brezzi}, \bibinfo{author}{M.~Fortin}, \bibinfo{title}{Mixed
  and Hybrid Finite Element Methods}, \bibinfo{publisher}{Springer-Verlag},
  \bibinfo{address}{New York}, \bibinfo{year}{1991}.
\bibitem[{Cockburn et~al.(2000)Cockburn, Karniadakis, and
  Shu}]{cockburnshubook}
\bibinfo{author}{B.~Cockburn}, \bibinfo{author}{G.~E. Karniadakis},
  \bibinfo{author}{C.~E. Shu}, \bibinfo{title}{{Discontinuous Galerkin Methods,
  Theory, Computation, and Applications}},
  \bibinfo{publisher}{Springer-Verlag}, \bibinfo{address}{Berlin},
  \bibinfo{year}{2000}.
\bibitem[{Dawson et~al.(2004)Dawson, Sun, and Wheeler}]{dawsonsunwheeler}
\bibinfo{author}{C.~Dawson}, \bibinfo{author}{S.~Sun}, \bibinfo{author}{M.~F.
  Wheeler},
\newblock \bibinfo{title}{Compatible algorithms for coupled flow and
  transport},
\newblock \bibinfo{journal}{Comput. Meth. Appl. M.} \bibinfo{volume}{193}
  (\bibinfo{year}{2004}) \bibinfo{pages}{2565--1029}.
\bibitem[{Riviere et~al.(2001)Riviere, Wheeler, and Girault}]{rivierewheeler}
\bibinfo{author}{B.~Riviere}, \bibinfo{author}{M.~Wheeler},
  \bibinfo{author}{V.~Girault},
\newblock \bibinfo{title}{A priori error estimates for finite element methods
  based on discontinuous approximation spaces for elliptic problems},
\newblock \bibinfo{journal}{SIAM J. Numer. Anal.} \bibinfo{volume}{39}
  (\bibinfo{year}{2001}) \bibinfo{pages}{902--931}.
\bibitem[{Girault et~al.(2008)Girault, Sun, Wheeler, and Yotov}]{giraultshuuy}
\bibinfo{author}{V.~Girault}, \bibinfo{author}{S.~Sun}, \bibinfo{author}{M.~F.
  Wheeler}, \bibinfo{author}{I.~Yotov},
\newblock \bibinfo{title}{Coupling discontinuous {Galerkin} and mixed finite
  element discretizations using mortar finite elements},
\newblock \bibinfo{journal}{SIAM J. Numer. Anal.} \bibinfo{volume}{46}
  (\bibinfo{year}{2008}) \bibinfo{pages}{949--979}.
\bibitem[{Sun and Wheeler(2005{\natexlab{a}})}]{sunwheeler2005a}
\bibinfo{author}{S.~Sun}, \bibinfo{author}{M.~F. Wheeler},
\newblock \bibinfo{title}{Discontinuous {Galerkin} methods for coupled flow and
  reactive transport problems},
\newblock \bibinfo{journal}{Appl. Numer. Math.} \bibinfo{volume}{52}
  (\bibinfo{year}{2005}{\natexlab{a}}) \bibinfo{pages}{273--298}.
\bibitem[{Sun and Wheeler(2005{\natexlab{b}})}]{sunwheeler2005b}
\bibinfo{author}{S.~Sun}, \bibinfo{author}{M.~F. Wheeler},
\newblock \bibinfo{title}{$l^2 (h^1)$ norm a posteriori error estimation for
  discontinuous {Galerkin} approximations of reactive transport problems},
\newblock \bibinfo{journal}{J. Sci. Comput.} \bibinfo{volume}{22}
  (\bibinfo{year}{2005}{\natexlab{b}}) \bibinfo{pages}{501--530}.
\bibitem[{Sun and Wheeler(2005{\natexlab{c}})}]{sunwheeler2005c}
\bibinfo{author}{S.~Sun}, \bibinfo{author}{M.~F. Wheeler},
\newblock \bibinfo{title}{Symmetric and nonsymmetric discontinuous {Galerkin}
  methods for reactive transport in porous media},
\newblock \bibinfo{journal}{SIAM J. Numer. Anal.} \bibinfo{volume}{43}
  (\bibinfo{year}{2005}{\natexlab{c}}) \bibinfo{pages}{195--219}.
\bibitem[{Sun and Wheeler(2006{\natexlab{a}})}]{sunwheeler2006a}
\bibinfo{author}{S.~Sun}, \bibinfo{author}{M.~F. Wheeler},
\newblock \bibinfo{title}{Analysis of discontinuous {Galerkin} methods for
  multicomponent reactive transport problems},
\newblock \bibinfo{journal}{Comput. Math. Appl.} \bibinfo{volume}{52}
  (\bibinfo{year}{2006}{\natexlab{a}}) \bibinfo{pages}{637--650}.
\bibitem[{Sun and Wheeler(2006{\natexlab{b}})}]{sunwheeler2006b}
\bibinfo{author}{S.~Sun}, \bibinfo{author}{M.~F. Wheeler},
\newblock \bibinfo{title}{Anisotropic and dynamic mesh adaptation for
  discontinuous {Galerkin} methods applied to reactive transport},
\newblock \bibinfo{journal}{Comput. Method. Appl. M.} \bibinfo{volume}{195}
  (\bibinfo{year}{2006}{\natexlab{b}}) \bibinfo{pages}{3382--3405}.
\bibitem[{Sun and Wheeler(2006{\natexlab{c}})}]{sunwheeler2006c}
\bibinfo{author}{S.~Sun}, \bibinfo{author}{M.~F. Wheeler},
\newblock \bibinfo{title}{A dynamic, adaptive, locally conservative and
  nonconforming solution strategy for transport phenomena in chemical
  engineerin},
\newblock \bibinfo{journal}{Chem. Eng. Commun.} \bibinfo{volume}{193}
  (\bibinfo{year}{2006}{\natexlab{c}}) \bibinfo{pages}{1527--1545}.
\bibitem[{Sun and Wheeler(2007)}]{sunwheeler2007a}
\bibinfo{author}{S.~Sun}, \bibinfo{author}{M.~F. Wheeler},
\newblock \bibinfo{title}{Discontinuous {Galerkin} methods for simulating
  bioreactive transport of viruses in porous media},
\newblock \bibinfo{journal}{Adv. Water Resour.} \bibinfo{volume}{30}
  (\bibinfo{year}{2007}) \bibinfo{pages}{1696--1710}.
\bibitem[{Arnold et~al.(2002)Arnold, Brezzi, Cockburn, and
  Marini}]{Arnoldreview}
\bibinfo{author}{D.~Arnold}, \bibinfo{author}{F.~Brezzi},
  \bibinfo{author}{B.~Cockburn}, \bibinfo{author}{L.~Marini},
\newblock \bibinfo{title}{{Unified analysis of discontinuous Galerkin methods
  for elliptic problems}},
\newblock \bibinfo{journal}{{SIAM J. Numer. Anal.}} \bibinfo{volume}{{39}}
  (\bibinfo{year}{{2002}}) \bibinfo{pages}{{1749--1779}}.
\bibitem[{Peraire and Persson(2008)}]{Persson2008}
\bibinfo{author}{J.~Peraire}, \bibinfo{author}{P.~O. Persson},
\newblock \bibinfo{title}{{The compact discontinuous Galerkin (CDG) method for
  elliptic problems}},
\newblock \bibinfo{journal}{{SIAM. J. Sci. Comput.}} \bibinfo{volume}{{30}}
  (\bibinfo{year}{{2008}}) \bibinfo{pages}{{1806--1824}}.
\bibitem[{Nguyen et~al.(2009{\natexlab{a}})Nguyen, Peraire, and
  Cockburn}]{Nguyen2009}
\bibinfo{author}{N.~C. Nguyen}, \bibinfo{author}{J.~Peraire},
  \bibinfo{author}{B.~Cockburn},
\newblock \bibinfo{title}{{An implicit high-order hybridizable discontinuous
  Galerkin method for nonlinear convection-diffusion equations}},
\newblock \bibinfo{journal}{{J. Comput. Phys.}} \bibinfo{volume}{{228}}
  (\bibinfo{year}{{2009}}{\natexlab{a}}) \bibinfo{pages}{{8841--8855}}.
\bibitem[{Nguyen et~al.(2009{\natexlab{b}})Nguyen, Peraire, and
  Cockburn}]{Nguyen2009a}
\bibinfo{author}{N.~C. Nguyen}, \bibinfo{author}{J.~Peraire},
  \bibinfo{author}{B.~Cockburn},
\newblock \bibinfo{title}{{An implicit high-order hybridizable discontinuous
  Galerkin method for linear convection-diffusion equations}},
\newblock \bibinfo{journal}{{J. Comput. Phys.}} \bibinfo{volume}{{228}}
  (\bibinfo{year}{{2009}}{\natexlab{b}}) \bibinfo{pages}{{3232--3254}}.
\bibitem[{Ag\'elas et~al.(2010)Ag\'elas, Di~Pietro, Eymard, and
  Masson}]{Eymard3}
\bibinfo{author}{L.~Ag\'elas}, \bibinfo{author}{D.~A. Di~Pietro},
  \bibinfo{author}{R.~Eymard}, \bibinfo{author}{R.~Masson},
\newblock \bibinfo{title}{An abstract analysis framework for nonconforming
  approximations of anisotropic heterogeneous diffusion},
\newblock \bibinfo{journal}{Int. J. Finite Vol.} \bibinfo{volume}{7}
  (\bibinfo{year}{2010}) \bibinfo{pages}{1--29}.
\bibitem[{Di~Pietro and Ern(2011)}]{Eymard6}
\bibinfo{author}{D.~A. Di~Pietro}, \bibinfo{author}{A.~Ern},
  \bibinfo{title}{{Mathematical Aspects of Discontinuous {G}alerkin Methods
  (Vol. 69)}}, \bibinfo{publisher}{Springer Science and Business Media},
  \bibinfo{address}{Berlin}, \bibinfo{year}{2011}.
\bibitem[{Persson(2013)}]{persson2013}
\bibinfo{author}{P.-O. Persson},
\newblock \bibinfo{title}{{A sparse and high-order accurate line-based
  discontinuous Galerkin method for unstructured meshes}},
\newblock \bibinfo{journal}{{J. Comput. Phys.}} \bibinfo{volume}{{233}}
  (\bibinfo{year}{{2013}}) \bibinfo{pages}{{414--429}}.
\bibitem[{Scovazzi et~al.(2013)Scovazzi, Huang, Collis, and Yin}]{Scovazzi2013}
\bibinfo{author}{G.~Scovazzi}, \bibinfo{author}{H.~Huang},
  \bibinfo{author}{S.~S. Collis}, \bibinfo{author}{J.~Yin},
\newblock \bibinfo{title}{{A fully-coupled upwind discontinuous Galerkin method
  for incompressible porous media flows: High-order computations of viscous
  fingering instabilities in complex geometry}},
\newblock \bibinfo{journal}{{J. Comput. Phys.}} \bibinfo{volume}{{252}}
  (\bibinfo{year}{{2013}}) \bibinfo{pages}{{86--108}}.
\bibitem[{Chavent and Jaffre(1986)}]{chaventjaffre}
\bibinfo{author}{G.~Chavent}, \bibinfo{author}{J.~Jaffre},
  \bibinfo{title}{{Mathematical Models and Finite Elements for Reservoir
  Simulation. Studies in Mathematics and its Applications.}},
  \bibinfo{publisher}{Elsevier}, \bibinfo{address}{North-Holland},
  \bibinfo{year}{1986}.
\bibitem[{Mose et~al.(1994)Mose, Siegel, Ackerer, and Chavent}]{mosemfe}
\bibinfo{author}{R.~Mose}, \bibinfo{author}{P.~Siegel},
  \bibinfo{author}{P.~Ackerer}, \bibinfo{author}{G.~Chavent},
\newblock \bibinfo{title}{Application of the mixed hybrid finite-element
  approximation in a groundwater-flow model -- {Luxury} or necessity},
\newblock \bibinfo{journal}{{Water Resour. Res.}} \bibinfo{volume}{{30}}
  (\bibinfo{year}{{1994}}) \bibinfo{pages}{{3001--3012}}.
\bibitem[{Hoteit et~al.(2002)Hoteit, Erhel, Mose, Philippe, and
  Ackerer}]{hoteitreliability}
\bibinfo{author}{H.~Hoteit}, \bibinfo{author}{J.~Erhel},
  \bibinfo{author}{R.~Mose}, \bibinfo{author}{B.~Philippe},
  \bibinfo{author}{P.~Ackerer},
\newblock \bibinfo{title}{Numerical reliability for mixed methods applied to
  flow problems in porous media},
\newblock \bibinfo{journal}{Comput. Geosci.} \bibinfo{volume}{6}
  (\bibinfo{year}{2002}) \bibinfo{pages}{161--194}.
\bibitem[{Douglas et~al.(1983)Douglas, Ewing, and Wheeler}]{wheeler1983}
\bibinfo{author}{J.~Douglas}, \bibinfo{author}{R.~Ewing},
  \bibinfo{author}{M.~Wheeler},
\newblock \bibinfo{title}{A time-discretization procedure for a mixed finite
  element approximation of miscible displacement in porous media},
\newblock \bibinfo{journal}{{Numer. Anal.}} \bibinfo{volume}{{17}}
  (\bibinfo{year}{{1983}}) \bibinfo{pages}{{249--265}}.
\bibitem[{Darlow et~al.(1984)Darlow, Ewing, and Wheeler}]{wheeler1984}
\bibinfo{author}{B.~Darlow}, \bibinfo{author}{R.~Ewing},
  \bibinfo{author}{M.~Wheeler},
\newblock \bibinfo{title}{Mixed finite-element method for miscible displacement
  problems in porous media},
\newblock \bibinfo{journal}{{SPE J.}} \bibinfo{volume}{{24}}
  (\bibinfo{year}{{1984}}) \bibinfo{pages}{{391--398}}.
\bibitem[{Hoteit and Firoozabadi(2005)}]{hoteit2005}
\bibinfo{author}{H.~Hoteit}, \bibinfo{author}{A.~Firoozabadi},
\newblock \bibinfo{title}{Multicomponent fluid flow by discontinuous {Galerkin}
  and mixed methods in unfractured and fractured media},
\newblock \bibinfo{journal}{Water Resour. Res.} \bibinfo{volume}{41}
  (\bibinfo{year}{2005}) \bibinfo{pages}{1--15}.
\bibitem[{Mikysta and Firoozabadi(2010)}]{jiri}
\bibinfo{author}{J.~Mikysta}, \bibinfo{author}{A.~Firoozabadi},
\newblock \bibinfo{title}{Implementation of higher-order methods for robust and
  efficient compositional simulation},
\newblock \bibinfo{journal}{J. Comput. Phys.} \bibinfo{volume}{229}
  (\bibinfo{year}{2010}) \bibinfo{pages}{2898--2913}.
\bibitem[{Edwards(2006)}]{Edwards2006}
\bibinfo{author}{M.~G. Edwards},
\newblock \bibinfo{title}{Higher-resolution hyperbolic-coupled-elliptic
  flux-continuous {CVD} schemes on structured and unstructured grids in 3-d},
\newblock \bibinfo{journal}{Int. J. Num. Methods Fluids} \bibinfo{volume}{51}
  (\bibinfo{year}{2006}) \bibinfo{pages}{1079--1095}.
\bibitem[{Hoteit and Firoozabadi(2008)}]{hoteit2008}
\bibinfo{author}{H.~Hoteit}, \bibinfo{author}{A.~Firoozabadi},
\newblock \bibinfo{title}{An efficient numerical model for incompressible
  two-phase flow in fractured media},
\newblock \bibinfo{journal}{Adv. Water Resour.} \bibinfo{volume}{31}
  (\bibinfo{year}{2008}) \bibinfo{pages}{891--905}.
\bibitem[{Hoteit and Firoozabadi(2006{\natexlab{a}})}]{hoteit2006a}
\bibinfo{author}{H.~Hoteit}, \bibinfo{author}{A.~Firoozabadi},
\newblock \bibinfo{title}{Compositional modeling by the combined discontinuous
  {Galerkin} and mixed methods},
\newblock \bibinfo{journal}{{SPE J.}} \bibinfo{volume}{{11}}
  (\bibinfo{year}{2006}{\natexlab{a}}) \bibinfo{pages}{{19--34}}.
\bibitem[{Hoteit and Firoozabadi(2006{\natexlab{b}})}]{hoteit2006b}
\bibinfo{author}{H.~Hoteit}, \bibinfo{author}{A.~Firoozabadi},
\newblock \bibinfo{title}{Compositional modeling of discrete-fractured media
  without transfer functions by the discontinuous {Galerkin} and mixed
  methods},
\newblock \bibinfo{journal}{{SPE J.}} \bibinfo{volume}{{11}}
  (\bibinfo{year}{2006}{\natexlab{b}}) \bibinfo{pages}{{341--352}}.
\bibitem[{Hoteit and Firoozabadi(2008)}]{hoteitcapillary}
\bibinfo{author}{H.~Hoteit}, \bibinfo{author}{A.~Firoozabadi},
\newblock \bibinfo{title}{Numerical modeling of two-phase flow in heterogeneous
  permeable media with different capillarity pressures},
\newblock \bibinfo{journal}{{Adv. in Water Res.}} \bibinfo{volume}{{31}}
  (\bibinfo{year}{2008}) \bibinfo{pages}{{56--73}}.
\bibitem[{Moortgat et~al.(2011)Moortgat, Sun, and Firoozabadi}]{moortgatII}
\bibinfo{author}{J.~Moortgat}, \bibinfo{author}{S.~Sun},
  \bibinfo{author}{A.~Firoozabadi},
\newblock \bibinfo{title}{Compositional modeling of three-phase flow with
  gravity using higher-order finite element methods},
\newblock \bibinfo{journal}{{Water Resour. Res.}} \bibinfo{volume}{{47}}
  (\bibinfo{year}{2011}).
\bibitem[{Moortgat et~al.(2012)Moortgat, Li, and Firoozabadi}]{moortgatIII}
\bibinfo{author}{J.~Moortgat}, \bibinfo{author}{Z.~Li},
  \bibinfo{author}{A.~Firoozabadi},
\newblock \bibinfo{title}{Three-phase compositional modeling of {CO2} injection
  by higher-order finite element methods with {CPA} equation of state for
  aqueous phase},
\newblock \bibinfo{journal}{Water Resour. Res.} \bibinfo{volume}{48}
  (\bibinfo{year}{2012}). \bibinfo{note}{{doi:10.1029/2011WR011736}}.
\bibitem[{Li and Firoozabadi(2009)}]{liCPA}
\bibinfo{author}{Z.~Li}, \bibinfo{author}{A.~Firoozabadi},
\newblock \bibinfo{title}{Cubic-plus-association equation of state for
  water-containing mixtures: Is ``cross association{''} necessary?},
\newblock \bibinfo{journal}{{AIChE J.}} \bibinfo{volume}{{55}}
  (\bibinfo{year}{{2009}}) \bibinfo{pages}{{1803--1813}}.
\bibitem[{Moortgat et~al.(2013)Moortgat, Firoozabadi, Li, and
  Esp\'osito}]{moortgatIV}
\bibinfo{author}{J.~Moortgat}, \bibinfo{author}{A.~Firoozabadi},
  \bibinfo{author}{Z.~Li}, \bibinfo{author}{R.~Esp\'osito},
\newblock \bibinfo{title}{{CO2} injection in vertical and horizontal cores:
  Measurements and numerical simulation},
\newblock \bibinfo{journal}{SPE J.} \bibinfo{volume}{18}
  (\bibinfo{year}{2013}). \bibinfo{note}{Doi:10.2118/135563-PA}.
\bibitem[{Moortgat and Firoozabadi(2013{\natexlab{a}})}]{moortgatV}
\bibinfo{author}{J.~Moortgat}, \bibinfo{author}{A.~Firoozabadi},
\newblock \bibinfo{title}{Three-phase compositional modeling with capillarity
  in heterogeneous and fractured media},
\newblock \bibinfo{journal}{{SPE J.}} \bibinfo{volume}{{18}}
  (\bibinfo{year}{2013}{\natexlab{a}}) \bibinfo{pages}{{1150--1168}}.
\bibitem[{Moortgat and Firoozabadi(2013{\natexlab{b}})}]{moortgatVI}
\bibinfo{author}{J.~Moortgat}, \bibinfo{author}{A.~Firoozabadi},
\newblock \bibinfo{title}{Higher-order compositional modeling of three-phase
  flow in {3D} fractured porous media based on cross-flow equilibrium},
\newblock \bibinfo{journal}{J. Comput. Phys.} \bibinfo{volume}{250}
  (\bibinfo{year}{2013}{\natexlab{b}}) \bibinfo{pages}{425--445}.
\bibitem[{Christie and Blunt(2001)}]{spe10}
\bibinfo{author}{M.~A. Christie}, \bibinfo{author}{M.~J. Blunt},
\newblock \bibinfo{title}{Tenth {SPE} comparative solution project: A
  comparison of upscaling techniques},
\newblock \bibinfo{journal}{SPE Reserv. Eval. Eng.} \bibinfo{volume}{4}
  (\bibinfo{year}{2001}) \bibinfo{pages}{308--317}.
\bibitem[{Jansen et~al.(2014)Jansen, Fonseca, Kahrobaei, Siraj, Van~Essen, and
  Van~den Hof}]{eggmodel2}
\bibinfo{author}{J.~D. Jansen}, \bibinfo{author}{R.~M. Fonseca},
  \bibinfo{author}{S.~Kahrobaei}, \bibinfo{author}{M.~M. Siraj},
  \bibinfo{author}{G.~M. Van~Essen}, \bibinfo{author}{P.~M.~J. Van~den Hof},
\newblock \bibinfo{title}{The {Egg} model; a geological ensemble for reservoir
  simulation},
\newblock \bibinfo{journal}{Geosci. Data J.} \bibinfo{volume}{1}
  (\bibinfo{year}{2014}) \bibinfo{pages}{192--195}.
\bibitem[{Jansen(2013)}]{eggmodel}
\bibinfo{author}{J.~Jansen}, \bibinfo{title}{{Egg Model}},
  \bibinfo{howpublished}{\url{http://data.3tu.nl/repository/uuid:916c86cd-3558-4672-829a-105c62985ab2}},
  \bibinfo{year}{2013}.
\bibitem[{Peng and Robinson(1976)}]{PREOS}
\bibinfo{author}{D.-Y. Peng}, \bibinfo{author}{D.~B. Robinson},
\newblock \bibinfo{title}{A new two-constant equation of state},
\newblock \bibinfo{journal}{Ind. Eng. Chem. Fundam.} \bibinfo{volume}{15}
  (\bibinfo{year}{1976}) \bibinfo{pages}{59--64}.
\bibitem[{Moortgat and Firoozabadi(2010)}]{moortgatI}
\bibinfo{author}{J.~Moortgat}, \bibinfo{author}{A.~Firoozabadi},
\newblock \bibinfo{title}{Higher-order compositional modeling with {Fickian}
  diffusion in unstructured and anisotropic media},
\newblock \bibinfo{journal}{Adv. Water Resour.} \bibinfo{volume}{33}
  (\bibinfo{year}{2010}) \bibinfo{pages}{951 -- 968}.
\bibitem[{Moortgat and Firoozabadi(2013)}]{moortgatVII}
\bibinfo{author}{J.~Moortgat}, \bibinfo{author}{A.~Firoozabadi},
\newblock \bibinfo{title}{{Fickian} diffusion in discrete-fractured media from
  chemical potential gradients and comparison to experiment},
\newblock \bibinfo{journal}{Energy Fuel} \bibinfo{volume}{27}
  (\bibinfo{year}{2013}) \bibinfo{pages}{5793--5805}.
\bibitem[{Acs et~al.(1985)Acs, Doleschall, and Farkas}]{acs}
\bibinfo{author}{G.~Acs}, \bibinfo{author}{S.~Doleschall},
  \bibinfo{author}{E.~Farkas},
\newblock \bibinfo{title}{General purpose compositional model},
\newblock \bibinfo{journal}{{SPE J.}} \bibinfo{volume}{{25}}
  (\bibinfo{year}{{1985}}) \bibinfo{pages}{{543--553}}.
\bibitem[{Watts(1986)}]{watts}
\bibinfo{author}{J.~W. Watts},
\newblock \bibinfo{title}{A compositional formulation of the pressure and
  saturation equations},
\newblock \bibinfo{journal}{SPE Reser. Eng.} \bibinfo{volume}{1}
  (\bibinfo{year}{1986}) \bibinfo{pages}{243--252}.
\bibitem[{Shahraeeni et~al.(2015)Shahraeeni, Moortgat, and
  Firoozabadi}]{shahraeeni}
\bibinfo{author}{E.~Shahraeeni}, \bibinfo{author}{J.~Moortgat},
  \bibinfo{author}{A.~Firoozabadi},
\newblock \bibinfo{title}{High resolution finite element methods for 3d
  simulation of compositionally triggered instabilities in porous media},
\newblock \bibinfo{journal}{Comput. Geosci.} \bibinfo{volume}{19}
  (\bibinfo{year}{2015}) \bibinfo{pages}{899--920}.
\bibitem[{Hoteit et~al.(2004)Hoteit, Ackerer, Mose, Erhel, and
  Philippe}]{hoteitslope}
\bibinfo{author}{H.~Hoteit}, \bibinfo{author}{P.~Ackerer},
  \bibinfo{author}{R.~Mose}, \bibinfo{author}{J.~Erhel},
  \bibinfo{author}{B.~Philippe},
\newblock \bibinfo{title}{{New two-dimensional slope limiters for discontinuous
  Galerkin methods on arbitrary meshes}},
\newblock \bibinfo{journal}{{Int. J. Numer. Methods Eng.}}
  \bibinfo{volume}{{61}} (\bibinfo{year}{{2004}})
  \bibinfo{pages}{{2566--2593}}.
\bibitem[{Moortgat(2016)}]{moortgatFinger}
\bibinfo{author}{J.~Moortgat},
\newblock \bibinfo{title}{Viscous and gravitational fingering in multiphase
  compositional and compressible flow},
\newblock \bibinfo{journal}{Adv. Water Resour.} \bibinfo{volume}{89}
  (\bibinfo{year}{2016}) \bibinfo{pages}{53--66}.
\bibitem[{Stone(1970)}]{stone1}
\bibinfo{author}{H.~Stone},
\newblock \bibinfo{title}{Probability model for estimating three-phase relative
  permeability},
\newblock \bibinfo{journal}{J. Petrol. Technol.} \bibinfo{volume}{22}
  (\bibinfo{year}{1970}) \bibinfo{pages}{214--218}.
\bibitem[{Stone(1973)}]{stone2}
\bibinfo{author}{H.~Stone},
\newblock \bibinfo{title}{Estimation of three-phase relative permeability and
  residual oil data},
\newblock \bibinfo{journal}{J. Can. Petrol. Technol.} \bibinfo{volume}{12}
  (\bibinfo{year}{1973}) \bibinfo{pages}{53}.
\bibitem[{Lie et~al.(2012)Lie, Krogstad, Ligaarden, Natvig, Nilsen, and
  Skaflestad}]{mrst}
\bibinfo{author}{K.-A. Lie}, \bibinfo{author}{S.~Krogstad},
  \bibinfo{author}{I.~S. Ligaarden}, \bibinfo{author}{J.~R. Natvig},
  \bibinfo{author}{H.~M. Nilsen}, \bibinfo{author}{B.~Skaflestad},
\newblock \bibinfo{title}{Open source matlab implementation of consistent
  discretisations on complex grids},
\newblock \bibinfo{journal}{Comput. Geosci.} \bibinfo{volume}{16}
  (\bibinfo{year}{2012}) \bibinfo{pages}{297--322}.

\end{thebibliography}
\end{document}